\documentclass[twocolumn,showpacs,preprintnumbers,amsmath,amssymb]{revtex4}

\usepackage[dvips]{graphicx}
\usepackage{dcolumn}
\usepackage{bm}

\begin{document}

\title{Theory of double-resonant Raman spectra in graphene:
intensity and line shape of defect-induced and two-phonon bands}

\author{Pedro Venezuela$^{1,2}$, Michele Lazzeri$^1$, and Francesco Mauri$^1$}

\affiliation{$^1$ IMPMC, Universit{\'e} Pierre et Marie Curie, CNRS, 4 place Jussieu, F-75252 Paris, France \\
$^2$ Instituto de F{\'\i}sica, Universidade Federal Fluminense, 24210-346 Niter{\'o}i, RJ, Brazil
}

\begin{abstract}

We calculate the double resonant (DR) Raman spectrum of graphene, and
determine the lines associated to both phonon-defect processes (such
as in the $D$ line at $\sim$~1350~cm$^{-1}$, $D'$ at
$\sim$~1600~cm$^{-1}$ and $D''$ at $\sim$~1100~cm$^{-1}$), and
two-phonons ones (such as in the $2D$, $2D'$, or $D+D''$ lines).
Phonon and electronic dispersions reproduce calculations based on
density functional theory corrected with GW.  
Electron-light, -phonon , and -defect scattering matrix elements
and the electronic linewidth are explicitly calculated.
Defect-induced processes are
simulated by considering different kind of idealized defects.  
For an excitation energy of $\epsilon_L=2.4$~eV, the agreement with
measurements is very good and calculations reproduce: the relative
intensities among phonon-defect or among two-phonon lines;
the measured small widths of the $D$, $D'$, $2D$ and $2D'$ lines;
the line shapes; the presence of small intensity lines 
in the 1800, 2000 cm$^{-1}$ range.
We determine how the spectra depend on the excitation energy,
on the light polarization, on the electronic linewidth,
on the kind of defects and on their concentration.
According to the present findings,
the intensity ratio between the $2D'$ and $2D$ lines
can be used to determine experimentally the
electronic linewidth.
The intensity ratio between the $D$ and $D'$ lines
depends on the kind of model defect, suggesting that this
ratio could possibly be used to identify the kind of defects present
in actual samples.
Charged impurities outside the graphene plane provide an almost undetectable
contribution to the Raman signal.
The present analysis reveals that, for both $D$ and $2D$
lines, the dominant DR processes are those in which electrons and
holes are both involved in the scattering, because of a destructive
quantum interference that kills processes involving only electrons or
only holes. The most important phonons belong to the
{\bf K}$\rightarrow{\bm \Gamma}$ direction ($inner$ phonons) and not
to the {\bf K}$\rightarrow${\bf M} one ($outer$ phonons), as
usually assumed.  The small $2D$ line width at
$\epsilon_L=2.4$~eV is a consequence of the interplay
between the opposite trigonal warpings of the electron and phonon
dispersions.  At higher excitation, e.g. $\epsilon_L=3.8$~eV,
the $2D$ line becomes broader and evolves in an asymmetric double peak structure.

\end{abstract}

\pacs{78.30.-j,78.67.Wj,81.05.ue}

\maketitle


\section{Introduction}

Raman spectroscopy is one of the most important experimental
techniques for the characterization of graphitic materials. In
particular, for graphene, this technique provides information about
the number of layers~\cite{ferrari,gupta06},
doping~\cite{lazzeri06PRL,pisana07,yan07},
disorder~\cite{chen09,lucchese,ni10} and phonon
properties~\cite{mafra}.

Lowest-oder Raman processes correspond to the scattering with a zero momentum phonon
({\bf q=0}).  The Raman $G$ line in graphene and
graphite ($\sim$1582~cm$^{-1}$) is associated with the E$_{2g}$ phonon at
${\bm \Gamma}$ and it is a lowest-order process.  Graphene and graphite
present other lines, due to higher order processes, which are usually
interpreted in terms of the so called double resonance (DR) mechanism
~\cite{thomsen00}.  The DR mechanism is used to interpret two distinct
kind of phenomena.  The first is the excitation of a phonon with
momentum {\bf q}$\ne${\bf 0} due to the presence of defects in the
sample.  This process, called {\it defect-induced}, is not allowed in
a purely crystalline sample (without defects) because of momentum
conservation. In graphene and graphite, it gives rise to the well
studied $D$ line at $\sim$~1350~cm$^{-1}$ and also to less intense
lines such as the $D'$ ($\sim$~1600~cm$^{-1}$), and the $D''$
($\sim$~1100~cm$^{-1}$~\cite{lucchese,erlon}).  The second process
corresponds to the excitation of two phonons with opposite momenta
{\bf q} and {\bf -q}.  This process, called {\it two-phonon}, can be
observed in purely crystalline samples since the momentum is conserved
and gives rise to the very intense $2D$ line at $\sim$~2700~cm$^{-1}$
(which is an overtone of the $D$ line) and, for instance, to
the $D+D''$ and $2D'$ lines at $\sim$~2450~cm$^{-1}$ and
$\sim$~3200~cm$^{-1}$.  The lines related to DR defect-induced and
two-phonon processes have a remarkable property: they are dispersive,
i.e. their positions change with excitation energy.

It has been shown experimentally~\cite{ferrari,gupta06} that the $2D$
line in graphene changes in shape, width and position with number of
layers.  Later, the phonon dispersion of graphene, near the Dirac {\bf
K} points, was probed by measurements~\cite{mafra} of the $2D$ and
$D+D''$ lines as a function of the excitation energies.  Usually,
Raman experiments are performed in graphene layers that were deposited
or grown over a substrate.  However, experimental measurements of the
$G$ and $2D$ lines have also been performed for free-standing graphene
monolayers~\cite{berciaud}.  Lucchese {\it et. al}~\cite{lucchese} and
Martins Ferreira {\it et. al}~\cite{erlon} have studied the evolution
of the Raman spectra for mono and multi-layer graphene with increasing
disorder, showing that the intensity of the $D$ line, which is absent
in pristine graphene, increases when disorder is induced in the sample
up to a maximum value where it begins to decrease.  On the other hand,
the $2D$ line intensity is maximum for pristine graphene and it
decreases with increasing disorder.

Frequencies, intensities and linewidths of all DR Raman bands may be determined
by the calculation of the Raman cross section~\cite{martinfalicov}.
Several excellent theoretical works already appeared on the topic
providing an overall good understanding of the situation.  However,
the many different approximations used by different authors (e.g.
constant electron-phonon matrix elements, resonant phonons are assumed
to be on some high symmetry line, in some cases the electronic
dispersion is conic, the electronic life-time is a parameter, etc.)
and the several debates still going on lead the sensation that
something is missing.
Thomsen and Reich\cite{thomsen00} and Kurti {\it et. al}~\cite{kurti02}
studied the $D$ line for graphite and carbon nanotubes, respectively. Also,
Narula and Reich\cite{narula} studied the $D$ and $2D$ Raman lines in
graphene and graphite. 
In these works~\cite{thomsen00,kurti02,narula} the scattering matrix elements
(electron-light, electron-phonon and electron-defect)
are assumed to be constants
and the electronic linewidth is a parameter set ot a fixed value.
Basko~\cite{basko} has studied the two-phonon
and four-phonon Raman bands in graphene under the assumption of conical bands,
which is valid only in the limit of small excitation energies, not suitable
for most experimental data available in the literature. Also, his work is
limited to disorder-free graphene. Park {\it et. al}~\cite{park} have
studied the two-phonon processes in single, double and triple layer graphene,
making the assumption of conical bands and limiting their work to
disorder-free graphene.

In this context, some questions are currently debated.
For instance, according to previous theoretical works~\cite{thomsen00,kurti02,narula},
phonons in the {\bf K}$\rightarrow${\bf M} direction of the Brillouin zone 
should give the most important contribution to the $D$ line intensity.
However, recent works~\cite{daniela,mohr,huang10,frank11,yoon11} have argued that the phonons in the 
{\bf K}$\rightarrow{\bm \Gamma}$ direction should be more important.
Other open questions refer to the processes more relevant for the DR Raman spectra.
In some Raman processes only the electrons are scattered, while in other processes
both electrons and holes are scattered simultaneously.
Some authors claim that, at least for the $2D$ line, this last kind of processes should
be dominant because they are associated to a triple resonance~\cite{basko07}.
On the other hand, several authors perform their studies considering only
electron-electron processes, as 
in the seminal work by Thomsen and Reich~\cite{thomsen00}.

Besides, several fundamental questions are almost untouched.
So far, the DR mechanism has been basically used to give an overall description
of the physics and to determine which are the excited phonons.
Can the DR theory be used to obtain a quantitative description of the
intensities of the Raman lines?
Can the DR theory be used to obtain a quantitative description of the
shape and of the width of the Raman lines?
The most studied Raman lines, the $2D$ and the $D$ ones,
present a relatively narrow linewidth similar to the one
of the $G$ line (which is not due to DR).
This fact is very surprising and, indeed, the theoretical approaches used so far
were not able to reproduce the observed small width of these lines.
Which are the missing ingredients?
Is this a consequence of the approximations used so far, or, on the contrary,
is this a limit of the perturbative approach inherent to the DR theory?
Finally, the $D$ line is activated by disorder and is routinely used to probe
the quality of the samples of graphitic materials.
However, which kind of defects activate the $D$ line is not known. 
For instance, do neutral impurities, vacancies and charged defects
affect the $D$ line in the same way?
Which kind of defects are probed by measuring different defect-activates lines?
Does Raman spectroscopy probe the defects which mostly influence electronic transport?

Here, as a first step to answer these questions,
we calculate the double resonant Raman spectrum of graphene,
considering both defect-induced and two-phonons processes,
trying to provide a computational method overcoming the most
common approximations used in literature.
Calculations are done using the standard approach based on
the golden rule generalized to the perturbative fourth-order~\cite{thomsen00}.
The electronic summation is performed all over the two dimensional Brillouin zone
and all the possible phonons (with any wavevector) are considered.
The phonon dispersion is obtained from fully ab-initio calculations based on
density functional theory (DFT) corrected with GW.
Electronic structure calculations are based on a tight binding approach
in which the parameters are fitted to reproduce DFT+GW calculations.
The electronic lifetime is calculated explicitly and
the defect-induced processes are simulated by considering
three different kind of ideal model defects.

Sec.~\ref{sec_method} describes the computational method;
Sec.~\ref{sec_results} describes and discusses the results;
Sec.~\ref{sec_conclusions} resumes the main conclusions of the paper.

\section{Method}
\label{sec_method}
This section describes the method used to compute the DR Raman spectra.
Sec.~\ref{sec2a} gives the general framework and provides the equations
to obtain double resonant Raman spectra in graphene within the perturbative approach.
The other subsections describe the details to obtain the quantities used in the actual
implementation.
In particular,
Sec.~\ref{sec_el_ph_disp} describes the electronic and phononic band dispersions;
Sects.~\ref{sec_el-ph_scatt},~\ref{sec_el-light},~\ref{sec_el-def_scatt}
describe the electron-phonon, electron-light and electron-defect scattering matrix elements;
Sec.~\ref{sec_el-life}
describes the calculation of the electronic linewidth.

\subsection{Double resonant Raman intensity}
\label{sec2a}

In vibrational Raman, the spectrum usually consists in well defined lines
associated with emission (Stokes) or absorption (anti-Stokes) of a phonon.
Here, only Stokes processes are considered.
Note also that the $G$ line (lowest-order excitation of the E$_{2g}$ ${\bm \Gamma}$
phonon) is not described by the present formalism and is, thus, not present
in the calculated spectra.
Within the DR scheme~\cite{thomsen00}, the light-electron and electron-phonon interactions,
as well as
the defect-induced electron-electron scattering
are treated at the first order in perturbation theory.
The Raman cross section $I$ of the light scattered by a crystal
is obtained from the golden rule generalized to the fourth-order~\cite{martinfalicov}:
\begin{widetext}
\begin{equation}
I\propto\sum_f
\left|
\sum_{A,B,C}
\frac{{\cal M}_{fC}{\cal M}_{CB}{\cal M}_{BA}{\cal M}_{Ai}}
     {(\epsilon_i-\epsilon_C-i\frac{\gamma^C}{2})
      (\epsilon_i-\epsilon_B-i\frac{\gamma^B}{2})
      (\epsilon_i-\epsilon_A-i\frac{\gamma^A}{2})}
\right|^2
\delta(\epsilon_i-\epsilon_f),
\label{eq1}
\end{equation}
\end{widetext}
where $\epsilon_i$ is the energy of the initial state which consists in
a quantum of light with energy $\epsilon_L=\hbar\omega_L$ (the laser energy)
and in which the crystal is in the ground state.
The sum is performed on intermediate virtual states $A,B,C$, with energy
$\epsilon_A$, $\epsilon_B$, $\epsilon_C$,
which are described by electronic and phononic excitation of the crystal.
$\epsilon_f$ is the energy of the final state $f$, in which
the electronic degrees of freedom of the crystal are in the ground state,
one or two phonons with total energy $\hbar\omega_p$ have been excited, and
a quantum of light with energy $\epsilon_L-\hbar\omega_p$ has been emitted.
$\delta$ is the Dirac distribution.
$\gamma^A$, $\gamma^B$, $\gamma^C$
are the inverse of the lifetimes of the electronic excitations of the virtual
states $A$, $B$, $C$, respectively.
${\cal M}_{JK}$ are first-order scattering matrix elements between the states $J$ and $K$.
So far, no attempts have been reported to go beyond the approximation inherent to Eq.~\ref{eq1},
for graphitic materials.
Note that within the present approach, the $G$ line (which in literature is usually
referred to as a ``first-order'' process) is a third-order process.

The processes described by Eq.~\ref{eq1} are in general associated to lines
which are much weaker than ``first-order'' Raman lines.
Graphene and graphite are notable exceptions.
During the intermediate virtual transition the energy is not necessarily conserved
and the three denominators of Eq.~\ref{eq1} are in general different from zero.
However, in graphene and graphite two or more of the denominators of Eq.~\ref{eq1}
can be equal to zero simultaneously.
In literature this is called double-resonance condition, and
can be associated to Raman lines which have an intensity
comparable to that of lower-order processes (the $G$ line).

In the DR Raman scattering,
the process ${\cal M}_{Ai}$ in Eq.~\ref{eq1}
corresponds to the absorption of light by creation of an electron-hole pair
in the $\pi/\pi^*$ bands.
Then, the carriers are scattered twice before recombination
(${\cal M}_{BA}$ and ${\cal M}_{CB}$ in Eq.~\ref{eq1}).
For temperatures typically
present in Raman measurements in graphene, only Stoke processes (phonon emission) are relevant.
Thus, in one possible case, one scattering event is due to collision with a  defect
and the other to the creation of a phonon ({\it phonon-defect} process).
In a second possible case,
both scattering events are due to creation of phonons ({\it two-phonon} process).
Finally, the process ${\cal M}_{fC}$ in Eq.~\ref{eq1}
corresponds to the recombination of the carriers by light emission.
We define $I^{pd}_{{\bf q}\nu}$ as the probability to excite a phonon {\bf -q}$\nu$,
with momentum {\bf -q}, branch index $\nu$ and energy $\hbar\omega_{\bf -q}^\nu$
through a {\it phonon-defect} process.
$I^{pp}_{{\bf q}\nu\mu}$ is the probability to excite the two phonons
{\bf -q}$\nu$ and {\bf q}$\mu$ through a {\it two-phonon} process.
The Raman intensity as a function of the frequency $\omega$ of the scattered light is
proportional to
\begin{widetext}
\begin{eqnarray}
I(\omega)&=&
\frac{1}{N_q}\sum_{{\bf q},\nu}I^{pd}_{{\bf q}\nu}
\delta(\omega_L-\omega-\omega_{\bf -q}^\nu)
[n(\omega_{\bf -q}^\nu)+1]+
\frac{1}{N_q}\sum_{{\bf q},\nu,\mu}I^{pp}_{{\bf q}\nu\mu}
\delta(\omega_L-\omega-\omega_{\bf -q}^\nu-\omega_{\bf q}^\mu)
[n(\omega_{\bf -q}^\nu)+1][n(\omega_{\bf q}^\mu)+1]
, \label{eq2} \\
I^{pd}_{{\bf q}\nu}    &=& N_d \left|
\frac{1}{N_{k}}\sum_{{\bf k},\alpha}
K^{pd}_\alpha ({\bf k},{\bf q},\nu) \right|^2
~~~~;~~~~
I^{pp}_{{\bf q}\nu\mu} =    \left|
\frac{1}{N_{k}}\sum_{{\bf k},\beta}
K^{pp}_\beta ({\bf k},{\bf q},\nu,\mu) \right|^2 \label{eq3}.
\end{eqnarray}
\end{widetext}
The sum in Eq.~\ref{eq2} is performed on a uniform grid of $N_q$ 
phonon wavevectors {\bf q} in the Brillouin zone and on all
the branch indexes $\nu$ and $\mu$.
In the limit $N_{\bf q}\rightarrow\infty$,
$\delta(\omega)$ is the Dirac distribution.
$n(\omega)$ is the Bose-Einstein occupation.
In Eq.~\ref{eq3}, $N_d$ is the average number of defects in the unit cell.
$I^{pd}\propto N_d$, because we assume that the contributions of defects
on different sites add up incoherently.
The first sum in Eq.~\ref{eq3} is performed on a uniform grid of $N_k$ 
electronic wavevectors {\bf k}.
$\alpha$ and $\beta$ are labels running on the eight different possible processes that
we call $ee1,ee2,hh1,hh2,eh1,eh2,he1,he2$, which are represented diagrammatically in Fig.~\ref{fig1}.
The reader might be familiar with an alternative representation of the processes, reported in 
Fig.~\ref{diagr_bis}.
Expressions for the DR scattering amplitudes $K$ are given in the appendix.
Here we report, as examples, $K_{ee1}^{pd}$ and $K_{ee1}^{pp}$:
\begin{widetext}
\begin{eqnarray}
K_{ee1}^{pd} ({\bf k},{\bf q},\nu) =
\frac{
\langle {\bf k}          \pi   |D_{out}                 | {\bf k}          \pi^* \rangle
\langle {\bf k}          \pi^* |H_D                     | {\bf k}+{\bf q},\pi^* \rangle
\langle {\bf k}+{\bf q},\pi^* |\Delta H_{{\bf q},\nu} | {\bf k}          \pi^* \rangle
\langle {\bf k}          \pi^* |D_{in}                  | {\bf k}          \pi   \rangle
}
{
(\epsilon_L-\epsilon^{\pi^*}_{\bf k}+\epsilon^{\pi}_{\bf k}-\hbar\omega_{{\bf -q}}^\nu
-i\frac{\gamma^C_{\bf k}}{2})
(\epsilon_L-\epsilon^{\pi^*}_{\bf k+q}+\epsilon^{\pi}_{\bf k}-\hbar\omega_{{\bf -q}}^\nu
-i\frac{\gamma^B_{\bf k}}{2})
(\epsilon_L-\epsilon^{\pi^*}_{\bf k}+\epsilon^{\pi}_{\bf k}
-i\frac{\gamma^A_{\bf k}}{2})
}, \label{eq4} \\
K_{ee1}^{pp}({\bf k},{\bf q},\nu,\mu) =
\frac{
\langle {\bf k}          \pi   |D_{out}                  | {\bf k}          \pi^* \rangle
\langle {\bf k}          \pi^* |\Delta H_{{\bf -q},\mu} | {\bf k}+{\bf q},\pi^* \rangle
\langle {\bf k}+{\bf q},\pi^* |\Delta H_{{\bf q},\nu}  | {\bf k}          \pi^* \rangle
\langle {\bf k}          \pi^* |D_{in}                   | {\bf k}          \pi   \rangle
}
{
(\epsilon_L-\epsilon^{\pi^*}_{\bf k}+\epsilon^{\pi}_{\bf k}-\hbar\omega_{{\bf -q}}^\nu-\hbar\omega_{{\bf q}}^\mu
-i\frac{\gamma^C_{\bf k}}{2})
(\epsilon_L-\epsilon^{\pi^*}_{\bf k+q}+\epsilon^{\pi}_{\bf k}-\hbar\omega_{{\bf -q}}^\nu
-i\frac{\gamma^B_{\bf k}}{2})
(\epsilon_L-\epsilon^{\pi^*}_{\bf k}+\epsilon^{\pi}_{\bf k}
-i\frac{\gamma^A_{\bf k}}{2})
}. \label{eq5}
\end{eqnarray}
\begin{figure}[ht]
\includegraphics[width=16.0cm]{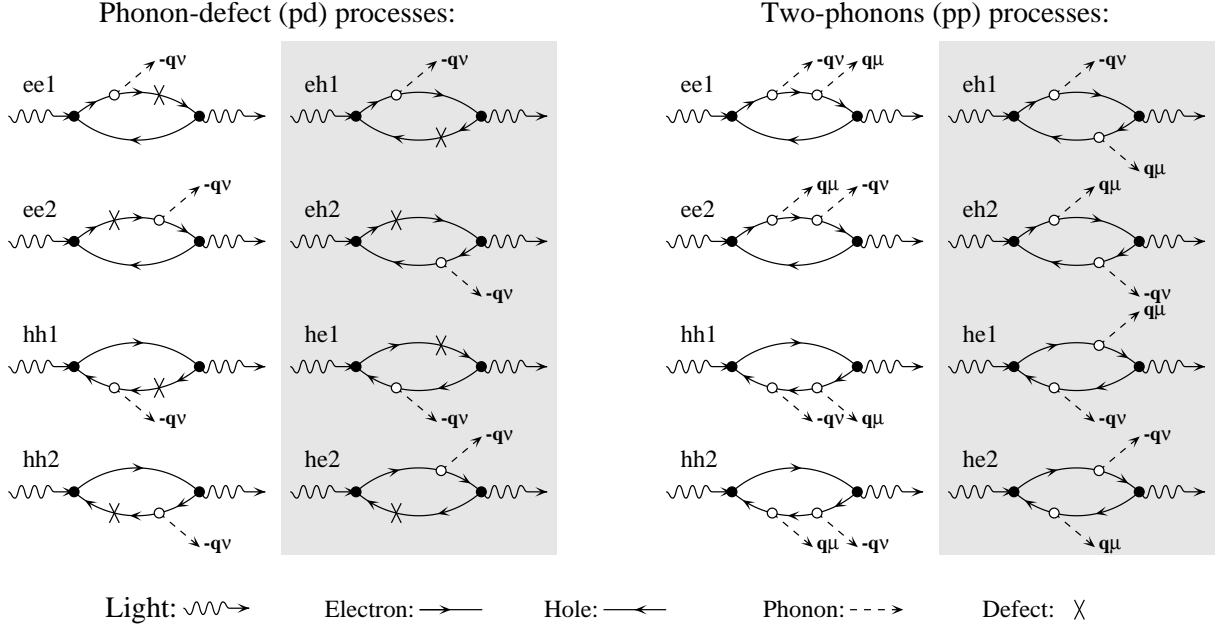}
\caption{Goldstone diagrams for the double resonant Raman processes considered in this
work.
In this manuscript, the term ``$ab$ processes'' refers to the processes highlighted
by the gray area ($eh1$, $eh2$, $he1$, and $he2$).
The other processes are referred to as ``$aa$ processes''.
The largest part of the Raman intensity is due to the $ab$ processes.
The reader might be familiar with an alternative representation of the processes,
reported in Fig.~\ref{diagr_bis}.
}
\label{fig1}
\end{figure}
\end{widetext}

Eq.~\ref{eq4} corresponds to the phonon-defect diagram $ee1$ in Fig.~\ref{fig1}.
Initially, the excitation laser
creates an electron-hole pair with momentum ${\bf k}$. Thus, using the notation of Eq.~\ref{eq1},
$\mathcal{M}_{Ai}=\langle \pi^*{\bf k}|D_{in} |\pi{\bf k}\rangle$, where
$|{\bf k}\pi\rangle$ and $|{\bf k}\pi^*\rangle$ are the electronic
occupied and empty states and $D_{in}$ is the operator coupling
the incident electromagnetic wave with the crystal.
$\epsilon_i=\epsilon_L$ and $\epsilon_A = \epsilon^{\pi^*}_{\bf k}-\varepsilon^{\pi}_{\bf k}$,
being $\epsilon^{\pi}_{\bf k}$
the energy of $|{\bf k}\pi\rangle$.
Secondly, the excited electron
is scattered into a ${\bf k}+{\bf q}$ state
by emitting a phonon with momentum {\bf -q}. Thus,
$\mathcal{M}_{BA}=\langle {\bf k}+{\bf q}, \pi^* |\Delta H_{{\bf q},\nu} |{\bf k}\pi^*\rangle$,
being $\Delta H_{{\bf q},\nu}$ the electron-phonon coupling operator. Now,
$\varepsilon_B=\epsilon^{\pi^*}_{\bf k+q}-\epsilon^{\pi}_{\bf k}+\hbar\omega_{{\bf -q}}^\nu$.
The third step in the process $K^{pd}_{ee1}$ is the scattering of the
${\bf k}+{\bf q}$ electron by a defect back to the
${\bf k}$ state. Thus,
$\mathcal{M}_{CB}=\langle {\bf k}\pi^*|H_D|{\bf k}+{\bf q},\pi^*
\rangle$, being $H_D$ the defect scattering operator and
$\epsilon_C = \epsilon^{\pi^*}_{\bf k}-\epsilon^{\pi}_{\bf k}+\hbar\omega_{\bf -q}^\nu$.
Finally, the electron and hole recombine vertically in the ${\bf k}$-state, by emitting light.
Thus,
$\mathcal{M}_{fC}= \langle {\bf k}\pi|D_{out}|{\bf k}\pi^*\rangle$,
being $D_{out}$ the operator coupling the emitted photon with the crystal.
The broadening energies $\gamma_{\bf k}$ in the denominators of
the DR amplitudes $K$ (e.g. in Eqs.~\ref{eq4}, ~\ref{eq5})
are the inverse of the corresponding electronic lifetimes
(see Sec.~\ref{sec_el-life}).

Eq.~\ref{eq5} corresponds to the phonon-phonon diagram $ee1$ in Fig.~\ref{fig1}.
The first two step are the same as in the previous paragraph, while
in the third step, the ${\bf k}+{\bf q}$ electron is scattered into a {\bf k} electron,
by emitting the phonon with momentum {\bf q}$\mu$. Thus,
$\mathcal{M}_{CB}=\langle {\bf k} \pi^* |\Delta H_{{\bf -q},\mu} |{\bf k+q}\pi^*\rangle$
and $\epsilon_C = \epsilon^{\pi^*}_{\bf k}-\epsilon^{\pi}_{\bf k}+\hbar\omega_{\bf -q}^\nu
+\hbar\omega_{\bf q}^\mu$.
The fourth step is the same as before.
Finally, 
for graphene and graphite, the diagrams of Fig.~\ref{fig1} are sometimes schematized
with a different notation. For a comparison see Fig.~\ref{diagr_bis}.

\begin{figure}[ht]
\includegraphics[width=7.5cm]{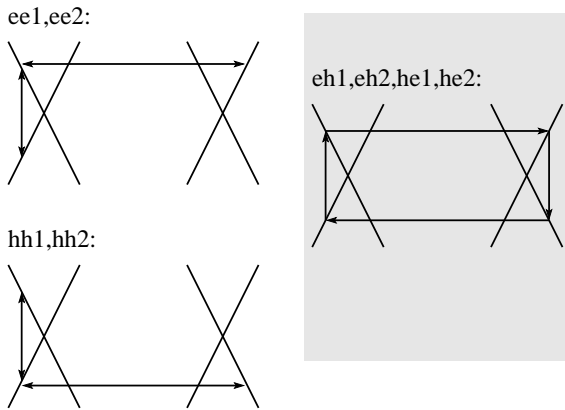}
\caption{
An alternative representation (customary for graphene and graphite)
of the processes associated to the diagrams of Fig.~\ref{fig1}.
The crosses represent the electronic dispersion near the conic region.
The vertical arrows represent the electron/hole creation and recombination.
The horizontal arrows represent the scattering with a defect or with a phonon.
For simplicity we show only the processes involving a phonon with momentum
along the {\bf K}-{\bf M} line.
In this manuscript, the term ``$ab$ processes'' refers to the processes highlighted
by the gray area ($eh1$, $eh2$, $he1$, and $he2$).
The other processes are referred to as ``$aa$ processes''.
}
\label{diagr_bis}
\end{figure}

The sums in Eq.~\ref{eq2} are performed on a uniform grid of 120$\times$120 {\bf q} points
(randomly shifted with respect to the origin)
and $\delta(\omega)$ is a Lorentzian distribution with 8~cm$^{-1}$
full width at half maximum.
The results will be plotted as a function of the Raman shift $\omega_L-\omega$.
The sums in Eq.~\ref{eq3}, are performed on grids of {\bf k} points
which are sufficiently large to ensure convergence. Depending on
the value of $\gamma^0_{\bf k}$ uniform grids between 480$\times$480 and
840$\times$840 {\bf k} points are used.
In Eq.~\ref{eq2},
we consider $\hbar\omega_{\bf q}^\nu\gg K_B T$ and, thus, $n(\omega_{\bf q}^\nu)\sim 0$.
Unless otherwise specified, the intensities are normalized to the maximum value
of the 2D peak.
In the following four sub-sections (and in App.~\ref{app_TB}), we
describe the model to obtain the DR scattering amplitudes $K$.

\subsection{Electron and phonon dispersion}
\label{sec_el_ph_disp}

\begin{figure}[ht]
\includegraphics[width=7.0cm]{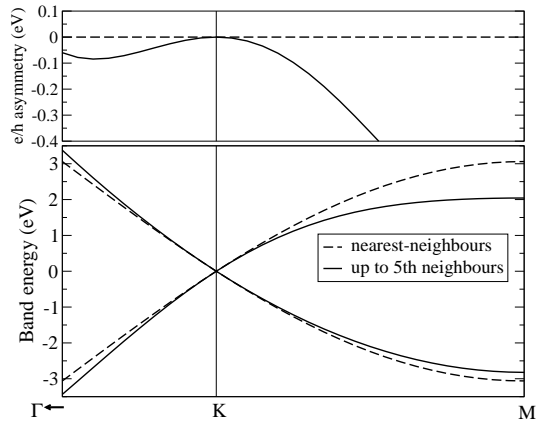}
\caption{
Graphene electronic dispersion obtained with the 5-neighbors tight-binding described
in the text (solid line). For a comparison we also show the dispersion obtained with
the 1st neighbors TB having the same Fermi velocity at {\bf K} (dashed line).
The electron/hole (e/h) asymmetry is defined as $\epsilon_{\bf k}^{\pi^*}+\epsilon_{\bf k}^\pi$
and is constant for the 1st neighbors TB.
}
\label{band}
\end{figure}

The electronic structure, $\epsilon^\alpha_{\bf k}$ and $|{\bf k},\alpha\rangle$,
is obtained from a tight binding (TB) model with
one orthonormalized $p_z$ orbital per site and interactions up to fifth neighbors
(details are in App. ~\ref{app_el-struct}).
We use
$t_1 = -3.40$ eV,
$t_2 = 0.33$ eV,
$t_3 = -0.24$ eV,
$t_4 = 0.12$ eV and
$t_5 = 0.09$ eV,
where $t_i$ is the i-th neighbor hopping parameter.
The resulting electronic dispersion is shown in Fig.~\ref{band}.
These TB parameters were obtained following~\cite{paola}:
first, the $t_i$ are fitted to
density-functional theory (DFT) electronic band dispersion
to reproduce the $\pi-\pi^*$ bands along the
${\bm \Gamma}$-{\bf K}-{\bf M} line;
then, all the $t_i$ are rescaled by +18\% in order to reproduce
the $\pi$ band slope near {\bf K} from GW calculations,
which are in excellent agreement with
angle-resolved photoemission spectra
 (ARPES) measurements on graphite
~\cite{gruneis}.

We remark that, in the present context, a good description of the
trigonal warping of the $\pi$-bands cone is very relevant, since the
actual shape of the trigonal warping determines the {\bf q} vectors of
the phonons associated to the $D$ line.
The present 5-neighbors TB can reproduce very well the
trigonal warping as obtained from DFT.
On the contrary, by using  a 1st-neighbors TB model, the trigonal
warping is underestimated.
Another relevant characteristic which is badly described by small-neighbors
TBs, but which is well described by the present 5-neighbors TB,
is the electron/hole asymmetry, $\epsilon_{\bf k}^{\pi^*}+\epsilon_{\bf k}^\pi$.
This quantity depends on the {\bf k} direction and has values of the order of the electronic broadening
(see Sec.~\ref{sec_el-life}):
e.g. for the states in resonance with a laser of 2.4 eV, the asymmetry is about 40 and 100 meV along the 
{\bf K}-${\bm \Gamma}$ and the {\bf K-M} direction, respectively (Fig.~\ref{band}).
On the contrary, in a 1st-neighbors TB model, the e/h asymmetry is {\bf k} independent and it is 
equal to zero.

\begin{figure}[ht]
\includegraphics[width=8.5cm]{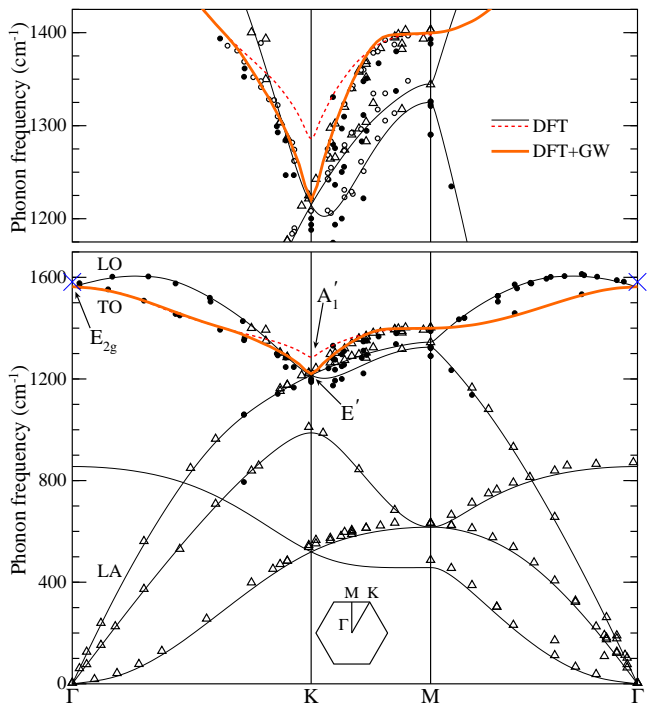}
\caption{
(Color online) Calculated graphene phonon dispersion from DFT (lines)
vs. IXS measurements on graphite from ~\cite{maultzsch04} (filled dots),
~\cite{mohr07} (triangles), and ~\cite{gruneis09} (open dots).
The highest optical branch near {\bf K} is ``corrected''
to include GW effects following~\cite{michele,gruneis09},
and is plotted with a thicker gray (red) line.
The dashed line is the same branch calculated from standard DFT,
without GW correction.
The cross at ${\bm \Gamma}$ is the measured Raman $G$ line frequency
in graphite (1582~cm$^{-1}$)
}
\label{fig2}
\end{figure}

Phonon dispersions, $\omega_{\bf q}^\nu$, are obtained from ab-initio
 DFT calculations~\cite{dfpt}
corrected with GW as in~\cite{michele,gruneis09}.
In particular, first we computed the DFT phonon dispersion,
then we ``correct'' the dispersion of the highest optical branch near {\bf K}
(the branch which is TO near ${\bm \Gamma}$ and which is associated with
the A$_1'$ mode at {\bf K}, see Fig.~\ref{fig2})
by rescaling the phonon self-energy contribution to the dynamical matrix
consistently with the GW calculated electron-phonon coupling and electronic
$\pi$ band dispersion~\cite{michele}.
Calculations are done for graphene with the same
computational details of~\cite{gruneis09}.
In ~\cite{gruneis09}, the rescaling factor is a constant, $r^{GW}=1.61$, all over the BZ
and the phonons are studied just in the neighborhood of {\bf K}.
Here, in order to obtain a phonon dispersion all over the BZ,
the rescaling factor, $r^{GW}_{\bf q}$, depends on {\bf q}.
$r^{GW}_{\bf q}=r^{GW}$ near {\bf K} and  smoothly drops to one
elsewhere:
\begin{equation}
r^{GW}_{\bf q}=1+(r^{GW}-1)\frac{1}{2}erfc\left(\frac{|{\bf q-K^n}|\frac{a_0}{2\pi}-0.2}{0.05}\right),
\end{equation}
being $a_0$ the graphene lattice constant and ${\bf K^n}$ the nearest vector to {\bf q}
among those equivalent to {\bf K}.
The GW correction associated to $r^{GW}$ changes
the phonon slope of the highest optical branch near {\bf K}
by almost +60\% (with respect to DFT)
providing a much better agreement with measurements for graphite (Fig.~\ref{fig2}).
The precise value of the phonon dispersion near {\bf K} is essential in the
present context, since it determines the dependence of the
$D$ peak dispersion as a function of the exciting laser energy~\cite{maultzsch04prb}.

Finally, notice that the present DFT calculations reproduce very well the experimental
phonon dispersion from inelastic x-ray scattering (IXS)
 of ~\cite{maultzsch04} of the highest optical branch near ${\bm \Gamma}$.
We can thus assume that the DFT frequency for the E$_{2g}$ ${\bm \Gamma}$ mode
(1561~cm$^{-1}$) is a precise fit of the IXS measurements.
The 1561~cm$^{-1}$ value is however 1.3\% smaller than the measured frequency of the
$G$ Raman line of graphite which is 1582~cm$^{-1}$ (the corresponding infra red mode is 1586~cm$^{-1}$).
This discrepancy between Raman and IXS measurements in graphite is so far unexplained.

\subsection{Electron-phonon scattering}
\label{sec_el-ph_scatt}

The electron-phonon scattering matrix elements $\Delta H_{{\bf q},\nu}$
are obtained from TB (explicit expressions are given in App.~\ref{app_el-phon})
and depend on the parameter $\eta_1$, defined
as the derivative of the nearest-neighbors hopping parameter
with respect to the bond length.
The present approach neglects the derivative of the hopping
parameters (with respect to the atomic positions) for hopping
computed for second and more distant neighbors.
This approximation reproduces very well the {\bf k} and {\bf q} dependence of the
electron-phonon matrix elements for electronic states with {\bf k} near {\bf K}
and for optical phonons with {\bf q} near ${\bm \Gamma}$ or near {\bf K}.
This was already verified in~\cite{piscanec} by direct comparison with
DFT calculations.

We define the average square of
$\sqrt{2M\omega_{{\bf q}\nu}/\hbar}~\Delta H_{{\bf q},\nu}$
between $\pi$ and $\pi^*$ at {\bf K}
as $\langle D^2_{\bm \Gamma}\rangle_F$ for the
E$_{2g}$ phonon at ${\bm \Gamma}$.
$\langle D^2_{\bf K}\rangle_F$ is the analogous quantity for
the A$'_1$ phonon at {\bf K}.
From Eqs.~\ref{eqb4},~\ref{eqb5} from App.~\ref{app_el-struct},
after some algebra, 
$\langle D^2_{\bm \Gamma}\rangle_F = 9/4 (\eta_1)^2$
and
$\langle D^2_{\bf K}\rangle_F = 9/2 (\eta_1)^2$
($\eta_1$ is defined in the previous paragraph
and the notation is consistent with ~\cite{michele}).
It follows that, within TB,
$\langle D^2_{\bf K}\rangle_F/\langle D^2_{\bm \Gamma}\rangle_F=2$
(that is, this ratio does not depend on the actual value of 
the TB parameter $\eta_1$).
This last relation is well reproduced by DFT calculations, within LDA or GGA,
but not by GW ones (see Table I of ~\cite{michele}).
As a consequence, a single value for $\eta_1$ could be used to
describe reasonably well the DFT electron-phonon interaction
for phonons in all the Brillouin zone.
On the contrary, we need two distinct values for $\eta_1$,
$\eta_1^{\bm \Gamma}=5.25$~eV/\AA~ and $\eta_1^{\bf K}=6.55$~eV\AA,
to reproduce the GW value of
$\langle D^2_{\bf K}\rangle_F$ and
$\langle D^2_{\bm \Gamma}\rangle_F$, respectively,
from Table I of ~\cite{michele}.
Here we will use $\eta_1=\eta_1^{\bm \Gamma}$ for phonons near ${\bm \Gamma}$
(those associated to the  $D'$ and $2D'$ lines), and
$\eta_1=\eta_1^{\bf K}$ for phonons near ${\bf K}$
($D$, $2D$, and $D+D''$).
A change of $\eta_1^{\bm \Gamma}$ and $\eta_1^{\bf K}$ values
will affect the present calculations as an uniform intensity scaling of some peaks with respect to others.

\subsection{Electron-light scattering}
\label{sec_el-light}

Explicit expressions for the $D_{in}$ and $D_{out}$ matrix elements
are given in App.~\ref{app_el-light}.
We assume that the polarization of the incoming and scattered light
are on the graphene ($x,y$) plane.
The computed Raman intensity $I_{i,o}$ depends on two indexes determined by
the polarization of the incident ($i=x,y$)
and of the scattered light ($o=x,y$).
The polarizations are chosen so as to reproduce different kind of Raman experiments.
In the {\it parallel polarization} case,
the incident and scattered light are parallel polarized  and $I_\parallel=I_{xx}+I_{yy}$.
In the {\it transverse polarization} case,
the incident and scattered light are perpendicularly polarized and $I_\perp=I_{xy}+I_{yx}$.
If the light is not polarized $I_{unpol}=I_{xx}+I_{yy}+I_{xy}+I_{yx}$.
Unless specified differently calculations are done in the non-polarized case.
In Sec.~\ref{sec_polariz}, 
the effects of parallel and transverse light polarizations are discussed.

\subsection{Electron-defect scattering}
\label{sec_el-def_scatt}

Defect scattering is treated within the Born approximation.
Namely, the defect scattering operator $H_D$ is the difference between the TB
Hamiltonian in presence of the defect and that of the defect free
system. $H_D$ is determined by considering three distinct kind of defects.

i) The {\it on-site defects}: defects that change the value of the on-site TB parameter by
$\delta V_0$.

ii) The {\it hopping defects}: change the value of one of the first-neighbor hopping
TB parameters by $\delta t_1$.

iii) The {\it Coulomb defects}: charged impurities adsorbed at a distance $h$ from
the graphene sheet that interact with graphene with a Coulomb potential.
Following \cite{dassarma}, we consider an environment dielectric constant $\kappa$ = 2.5.

We remark that these are very simplified prototypical models and that
a realistic description of a given type of impurity, which is beyond the
present scope, will result in a combination of these three kind of perturbations.
However, it is reasonable to expect that the present three models describe the
most important characteristics of certain kind of defects.
For instance, the on-site defect is the most simple description of
an hydrogen atom bound to a carbon atom in the graphene sheet.
Hopping defects are any defects that lead to deformations of the carbon-carbon bonds in graphene.
A Coulomb defect describes
any charged atom or molecule adsorbed over the graphene sheet.
Explicit expressions of the three defect scattering operators $H_D$
are given in App.~\ref{app_el-def}.
The three models are characterized by the parameters
$\delta V_0$, $\delta t_1$, and $h$, whose values will be specified in the discussion.
The results will be expressed as a function of the defect concentration
$n_d=N_d/A_0$, where $A_0=\sqrt{3}/2a_0^2$ is the graphene unit-cell area, being
$a_0=2.46$~\AA~the graphene lattice spacing.

Note that the Raman intensity of the defect-induced lines (e.g. $D$, $D'$, and $D''$)
is proportional to the average number of defects in the unit cell, $N_d$ (Eq.~\ref{eq3}).
This is because the scattering from defects on different sites is considered as incoherent,
which is reasonable for low defect-concentrations.
In particular, 
for on-site and hopping defects, the defect-induced intensities are proportional to
$\alpha_{on}=n_d(\delta V_0)^2$ and to $\alpha_{hopp}=n_d(\delta t_1)^2$,
being $n_d$ the defect concentration.
Through the text, we will specify the value of these parameters,
in order to make meaningful the
comparison of the defect-induced line intensities with those of the
phonon-phonon lines (e.g. $2D$, $2D'$, and $D'+D''$).

\subsection{Electronic linewidth}
\label{sec_el-life}

An electronic state $|{\bf k}\alpha\rangle$ ($\alpha=\pi^*$~or~$\pi$)
has a finite life-time $\tau^\alpha_{\bf k}$
(which is associated to a line broadening energy $\gamma^\alpha_{\bf k}=\hbar/\tau^\alpha_{\bf k}$)
because the electronic states interact, e.g.,  with phonons and with defects.
The broadening energies $\gamma_{\bf k}$ in the denominators of
the DR amplitudes $K$ ( e.g. in Eqs.~\ref{eq4}, ~\ref{eq5})
are the sum of the broadenings of the corresponding electronic states.
As examples, in both Eqs.~\ref{eq4}, ~\ref{eq5},
$\gamma^A_{\bf k} = \gamma^{\pi^*}_{\bf k} + \gamma^\pi_{\bf k}$,~
$\gamma^B_{\bf k} = \gamma^{\pi^*}_{\bf k+q} + \gamma^\pi_{\bf k}$, and
$\gamma^C_{\bf k} = \gamma^{\pi^*}_{\bf k} + \gamma^\pi_{\bf k}$.
For $\alpha=\pi^*$~or~$\pi$, $\gamma^\alpha_{\bf k}$
is the full-width at half maximum of the
electron/hole spectral function as measured, e.g., by ARPES.

We consider $\gamma$ as the sum of two contributions
\begin{equation}
\gamma^\alpha_{\bf k} = \gamma^{\alpha(ep)}_{\bf k} + \gamma^{\alpha(D)}_{\bf k}.
\label{eq7}
\end{equation}
The first is due to electron-phonon scattering.
It is an intrinsic broadening (present in perfectly crystalline samples)
and, according to the Golden rule, is
\begin{eqnarray}
\gamma^{\alpha(ep)}_{\mathbf{k}}&=&\frac{2\pi}{N_q}\sum_{\mathbf{q},\nu}
|\langle {\bf k}+\mathbf{q},\alpha|\Delta H_{{\bf q},\nu} | {\bf k},\alpha\rangle|^2 \nonumber \\
&&\times \delta(\varepsilon^{\alpha}_{\bf k}-\varepsilon^{\alpha}_{\bf k+q}-\hbar\omega_{\bf -q}^\nu),
\label{eq_b11}
\end{eqnarray}
where $\alpha$ refers to $\pi$ or $\pi^*$ bands,
the sum is performed on a uniform grid of $N_q$ ${\bf q}$ points in the Brillouin zone
and on all the phonon branches $\nu$.
A good approximation of $\gamma^{\alpha(ep)}$
is obtained by considering conic bands
($|\epsilon|=\hbar v_F k$, being $v_F$ the Fermi velocity)
and only the two phonons E$_{2g}$ at ${\bm \Gamma}$ and A$'_1$ at {\bf K},
with energies $\hbar\omega_{\bm \Gamma}$ and $\hbar\omega_{\bf K}$.
By defining
$\langle g^2_{\bm \Gamma}\rangle=\sqrt{\hbar/(2M\omega_{\bm \Gamma})}\langle D^2_{\bm \Gamma}\rangle_F$
and
$\langle g^2_{\bf K}\rangle=\sqrt{\hbar/(2M\omega_{\bf K})}\langle D^2_{\bf K}\rangle_F$
(see Sec.~\ref{sec_el-ph_scatt}),
Eq.~\ref{eq_b11} becomes:
\begin{eqnarray}
&&\gamma^{\alpha(ep)}_{conic}=\frac{\pi}{2}
\left[
2\langle g^2_{\bm \Gamma}\rangle N_\alpha(|\epsilon|-\hbar\omega_{\bm \Gamma})
+\langle g^2_{\bf K}\rangle N_\alpha(|\epsilon|-\hbar\omega_{\bf K})
\right] \nonumber \\
&&N_\alpha(\epsilon) = \frac{\sqrt{3}}{\pi}\left(\frac{a_0}{\hbar v_F}\right)^2
|\epsilon|\theta(|\epsilon|),
\end{eqnarray}
where $N_\alpha$ is the electronic density of states of the $\alpha=\pi$~or~$\pi^*$
band, being $a_0$ the lattice spacing and $\theta(x)$ the Heaviside step function.
Using the parameters of the present work,
$N_\alpha(\epsilon)=0.07908 {\rm eV}^{-2} |\epsilon|\theta(|\epsilon|)$
and for $|\epsilon|>0.196$~eV
\begin{equation}
\gamma^{\alpha(ep)}_{conic} = 41.89 (|\epsilon|-0.1645)~{\rm meV},
\label{gamma_conic}
\end{equation}
where $\epsilon$ is expressed in eV.

The second contribution in Eq.~\ref{eq7}
is due to electron-defect elastic scattering.
It is extrinsic (it is induced by the presence of impurities and depends on the sample quality)
and is
\begin{equation}
\gamma^{\alpha(D)}_{\bf k} = N_d \frac{2\pi}{N_k'}
\sum_{{\bf k'}}
|\langle {\bf k'},\alpha|H_D | {\bf k},\alpha\rangle|^2
\delta(\epsilon^\alpha_{\bf k}-\epsilon^\alpha_{\bf k'}),
\label{eq_b12}
\end{equation}
where the sum is performed on a uniform grid of $N_k'$ ${\bf k'}$ points in the Brillouin zone.
The electron-defect scattering operator $H_D$ is defined as in Sec.~\ref{sec_el-def_scatt} and
App.~\ref{app_el-def} and depends on the considered kind of defect.
$N_d$ is the average number of defects in the unit cell.

\begin{figure}[ht]
\includegraphics[width=7.8cm]{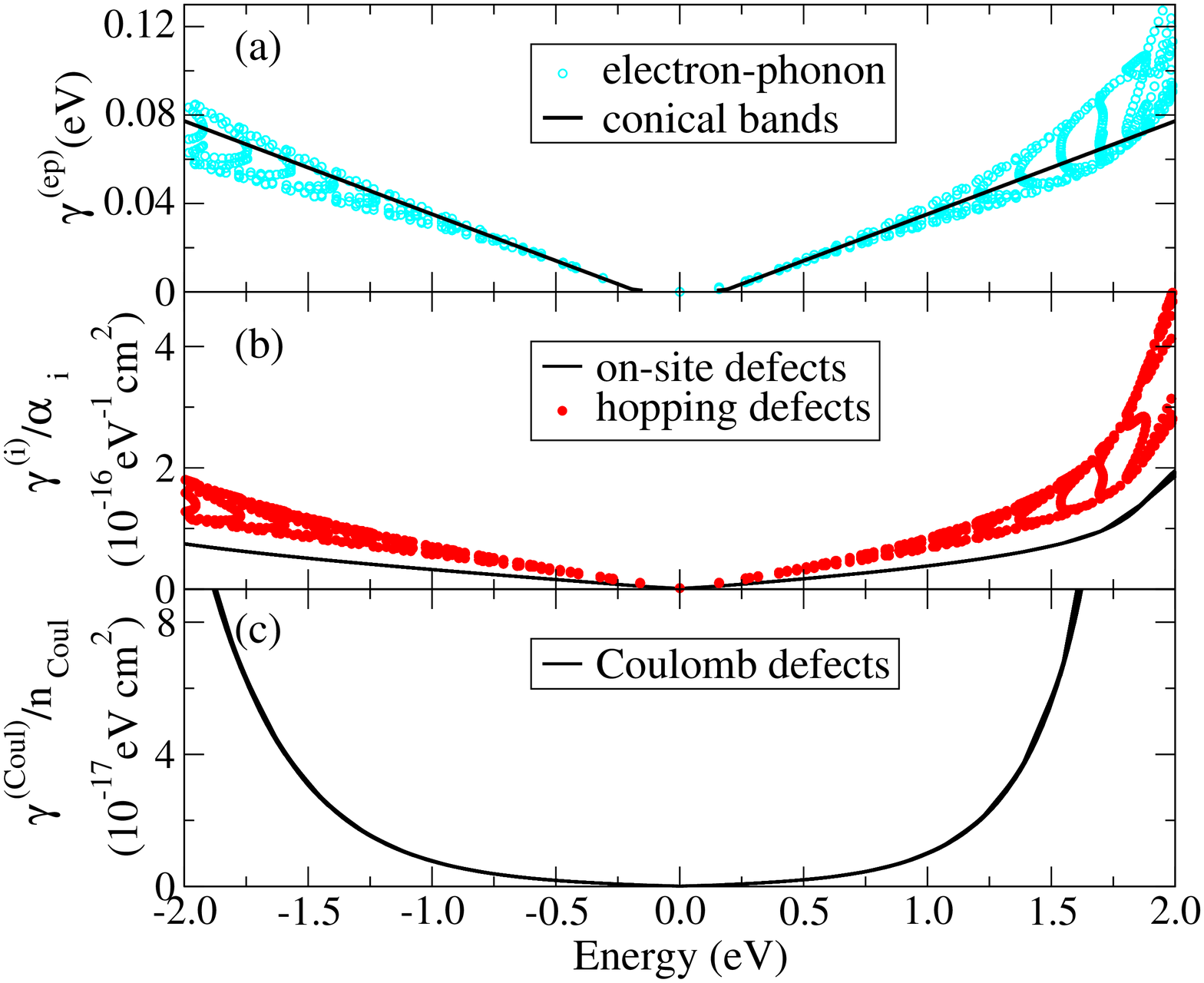}
\caption{(Color online) Electronic linewidth as a function of energy. (a)
Contribution of electron-phonon scattering to the electronic linewidth, $\gamma^{(ep)}$,
compared to conical bands results (Eq.~\ref{gamma_conic}).
(b) Contribution of on-site and hopping impurity scattering to the electronic linewidth.
$\gamma^{(on)}$ is proportional to $\alpha_{on} = n_d(\delta V_0)^2$ and
$\gamma^{(hopp)}$ is proportional to $\alpha_{hopp} = n_d(\delta t_1)^2$
(Sec.~\ref{sec_el-def_scatt}).
We, thus, plot $\gamma^{(i)}/\alpha_i$, where the
label ``i'' refers to ``on'' (on-site defect) or to ``hopp'' (hopping defect).
(c) Contribution of Coulomb impurity scattering to the electronic linewidth.
$n_{Coul}$ is the Coulomb impurity concentration.
The distance between graphene and the charged impurity $h$ = 0.27~nm
(see the discussion in Sec.~\ref{type}).
}
\label{broadening}
\end{figure}

Fig.~\ref{broadening} shows $\gamma^{(ep)}$ and $\gamma^{(D)}$ for the three
kind of defects we considered ($\gamma^{(D)}=\gamma^{(on)}$, 
$\gamma^{(D)}=\gamma^{(hopp)}$, or $\gamma^{(D)}=\gamma^{(Coul)}$).
The $\gamma$ in Fig.~\ref{broadening} are calculated with Eqs.~\ref{eq_b11},~\ref{eq_b12}
and are plotted as a function of the energy of the corresponding electronic state
($\epsilon^{\pi^*}_{\bf k}$ or $\epsilon^{\pi}_{\bf k}$).
$\gamma^{(ep)}$ is compared with the conic-band results of Eq.~\ref{gamma_conic}.
As expected, the two results are similar for energies smaller than 1~eV.

$\gamma^{(on)}$ and $\gamma^{(Coul)}$ are univocally determined by the energy
and, in Fig.~\ref{broadening}, are represented by  lines.
$\gamma^{(on)}$, in particular, is proportional to the density of states.
On the contrary, $\gamma^{(ep)}$ and  $\gamma^{(hopp)}$ display a dispersion
associated to the fact that different ${\bf k}$ electronic states with 
the same energy can have a different life-time.
However, the dispersion is relatively small, and for the present purpose they will also
be considered a function of the energy.
All the contributions ($\gamma^{(ep)}$, $\gamma^{(on)}$, $\gamma^{(hopp)}$ and $\gamma^{(Coul)}$)
increase with energy and display a noticeable asymmetry between positive and negative energies
due to the graphene electron/hole asymmetry.

In actual calculations (e.g. in Eqs. ~\ref{eq4},~\ref{eq5}) we 
neglect the dependence on {\bf k} and we use 
\begin{equation}
\gamma^A_{\bf k} = \gamma^B_{\bf k} = \gamma^C_{\bf k} = \gamma^{tot},
\label{eq_gtot}
\end{equation}
where $\gamma^{tot}$ depends only on the excitation energy $\epsilon_L$, on the kind of defect
$D$ and on its concentration $n_D$, through
\begin{equation}
\gamma^{tot}=\tilde\gamma^{(ep)}(\epsilon_L)+\tilde\gamma^{(D)}(\epsilon_L,n_D).
\label{eq13}
\end{equation}
$\tilde\gamma$ are the 
sum of the two contributions for $\pi$ an $\pi^*$ bands
in a small energy range close to half the excitation energy $\epsilon_L$.
As an example,
$\tilde\gamma^{(ep)}=\overline\gamma^{(ep)}(\epsilon_L/2)+\overline\gamma^{(ep)}(-\epsilon_L/2)$,
where $\overline\gamma^{(ep)}(\epsilon)$ is the average of $\gamma^{(ep)}$
from Fig.~\ref{broadening} at that energy, in particular,
for $\epsilon_L\gtrsim 1.0$~eV,
\begin{equation}
\tilde\gamma^{(ep)} (\epsilon_L) = (~18.88~\epsilon_L + 6.802~\epsilon_L^2~)~{\rm meV},
\end{equation}
where $\epsilon_L$ is expressed in eV.
While comparing these values with literature,
notice that $\gamma^{(tot)}$ and the $\tilde\gamma$'s
correspond to the sum of the width of electrons and holes and are, thus,
roughly two times bigger that the width of electronic states.
To give some examples,
for $\epsilon_L$ = 2.4~eV, and
for the typical defect concentrations of the present work,
$\alpha_{on}=\alpha_{hopp}=6.4\times10^{13}$~eV$^2$cm$^{-2}$,
$\tilde\gamma^{(on)}=5$~meV and $\tilde\gamma^{(hopp)}=12$~meV, and
for $n_{Coul}=10^{12}$~cm$^{-2}$, $\tilde\gamma^{(Coul)}=0.01$~meV.
On the other hand, for $\epsilon_L$ = 2.4~eV, $\tilde\gamma^{(ep)}=84$~meV
is the dominant contribution and, in several cases, we will just consider
$\gamma^{tot}\sim\tilde\gamma^{(ep)}$.
Similar values of $\gamma^{tot}\sim\tilde\gamma^{(ep)}$ have been
extracted from measurements in ~\cite{basko2}
(note that $\gamma_{e-ph}$ of~\cite{basko2} corresponds to
$\tilde\gamma^{(ep)}/4$ in the present notation).

Finally, in charged graphene a further contribution to the broadening
due to electron-electron interaction~\cite{basko2} can be relevant
when 0.06$|\epsilon_F|\gtrsim \tilde\gamma^{(ep)}/4$
where $\epsilon_F$ is the Fermi energy (see e.g. Eq.8 of~\cite{basko2}).
For electron/hole concentrations of the order of $10^{12}$~cm$^{-2}$ this contribution
is negligible and, here, it is not considered.

\section{Results and Discussion}
\label{sec_results}

This section presents the calculation of the double resonant (DR)
Raman spectra of graphene and discuss the results.
Sec.~\ref{fix} describes the overall agreement with measurements.
Sec.~\ref{sec_laser} describes the dependence of the spectra on
excitation energy and light polarization.
Sec.~\ref{sec_intensities} describes the dependence of the Raman intensities on
various parameters such as the electronic linewidth,
the excitation energy, and the defect concentration.
Sec.~\ref{type} describes the  dependence of the spectra on the type of defect.
Sec.~\ref{sec_interpretation} is dedicated to the interpretation of the results.
It is focused on some specific issues such as the determination of the
most relevant processes and phonons, the role of quantum interference,
and on the interpretation of the small width of the main DR Raman lines.

\subsection{Overall agreement with measurements}
\label{fix}
\begin{figure}[ht]
\includegraphics[width=7.8cm]{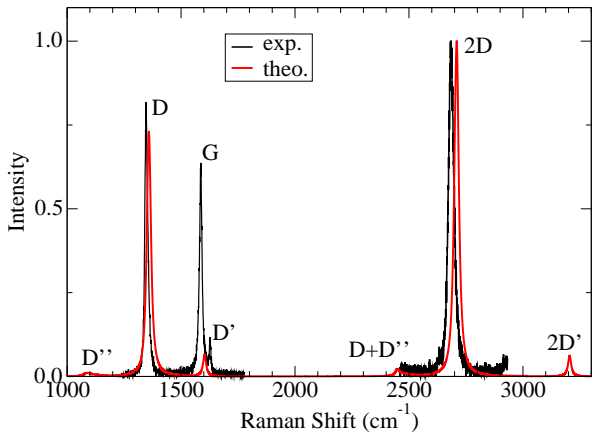}
\vspace{0.2cm}
\caption{
(Color online)
Intensity vs. Raman shift for $\varepsilon_L$ = 2.4 eV. 
Comparison of the present calculations with the measurements from~\cite{erlon}. 
Notice that our model includes
only double-resonant processes and, thus, the $G$ line is not present. 
Measurements correspond to a
defect concentration  $n_d$ = $10^{12}$ cm$^{-2}$.
Calculations are done using
$\gamma^{tot}=96$~meV, and hopping defects with
$\alpha_{hopp}=6.4\times10^{13}$~eV$^2$cm$^{-2}$. All the
intensities are normalized to the maximum value of the $2D$ line.
}
\label{fig3}
\end{figure}

\begin{figure}[ht]
\includegraphics[width=7.8cm]{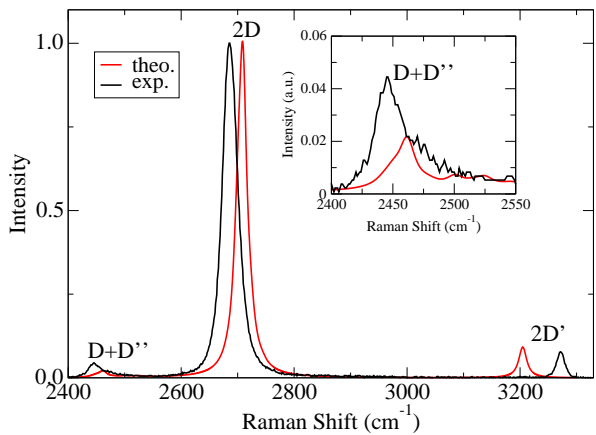}
\caption{(Color online)
Intensity vs. Raman shift for $\varepsilon_L$ = 2.4 eV. 
Comparison of the present calculations with the measurements from Ref.~\cite{ferrari}.
The figure reports only two-phonon processes.
Calculations are done using $\gamma^{tot}=84$~meV.
All the
intensities are normalized to the maximum value of the $2D$ line.
The inset shows the $D$+$D''$ band in a different scale.
}
\label{2D}
\end{figure}

Figs.~\ref{fig3} and ~\ref{2D} compare the present calculations with Raman spectra
of Refs.~\cite{erlon,ferrari}, for an excitation energy $\epsilon_L=2.4$~eV.
In Fig.~\ref{fig3}, below 2000 cm$^{-1}$ the processes are due to phonon-defect
scattering and calculations are done considering only the hopping defects
(this choice is justified in Sec.~\ref{type}), using the parameter
$\alpha_{hopp}=6.4\times10^{13}$~eV$^2$cm$^{-2}$ (see Sec.~\ref{sec_el-def_scatt}),
which reproduces the measured ratio of the integrated areas between $D$ and $2D$ lines
of~\cite{erlon}.
Above 2000 cm$^{-1}$, all the processes are due to two-phonon scattering.
We remark that the $G$ line is a single-resonant
process which is not included in the present calculations.

The agreement between calculations and measurements is extremely good.
In particular, all the lines observed experimentally, even the small intensity ones,
are present in the
calculated spectra and the relative intensities among 
phonon-defect lines (such as the $D$ and the $D'$)
or among two-phonon lines (such as $2D$, $2D'$, or $D+D''$)
are correctly reproduced.
The most remarkable agreement relates to the line widths. Indeed,
the present model reproduces very well the measured small widths
of the $D$, $D'$, $2D$ and $2D'$ lines.
Moreover, the model reproduces quite well 
the symmetric Lorentzian shapes of the $2D$ and $2D'$ lines and the
asymmetric shape of $D+D''$ band.
We remark that, in the present model, the only parameter used to fit
the Raman data is $\alpha_{hopp}$.
This parameter determines the ratio of the $D$ vs. $2D$ intensities
but does not affect
the relative intensities among phonon-defect or among two-phonon lines,
the width of the lines, and their shape.

As far as the line frequencies are concerned, 
calculations and measurements display some small deviations
of the order of a few meV.
We remark that
the line frequencies are determined by a subtle interplay between
the phononic and electronic energy dispersions, and that
the present dispersions are obtained from state of the
art ab-initio computational methods
which correctly reproduce ARPES and IXS measurements
(Sec.~\ref{sec_el_ph_disp}).
A correction of the electronic or of the phononic dispersions,
to reproduce with more precision the Raman frequencies,
would be done at the expense of introducing fitting parameters
to the model, which is beyond the present scope.

\subsection{Dependence of the spectra on the laser}
\label{sec_laser}

This section describes the dependence of the spectra on
excitation energy and light polarization.
Excitation energies vary from
1.2 to 4.0 eV, which are energies mainly used in actual experiments.

\subsubsection{Dependence of the main lines on the excitation energy}

\begin{figure}[ht]
\includegraphics[width=7.8cm]{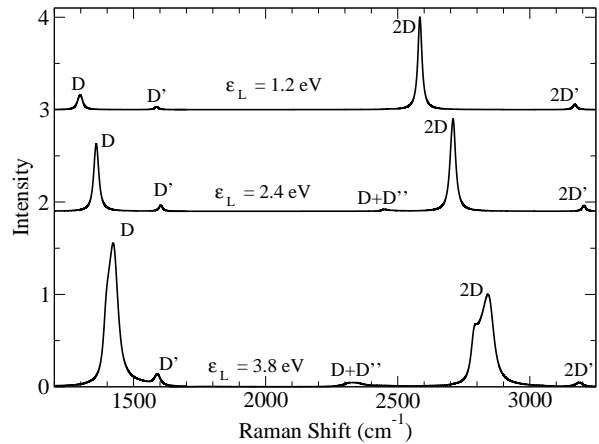}
\caption{Calculated Raman spectra
for $\varepsilon_L$ = 1.2 eV and $\gamma^{tot}$ = 32 meV,
$\varepsilon_L$ = 2.4 eV and $\gamma^{tot}$ = 84 meV,
$\varepsilon_L$ = 3.8 eV and $\gamma^{tot}$ = 170 meV.
Calculations are done using
hopping defects with $\alpha_{hopp}=6.4\times10^{13}$~eV$^2$cm$^{-2}$.
All the intensities are normalized to the corresponding $2D$ line maxima. }
\label{spectra}
\end{figure}

\begin{figure}[h]
\includegraphics[width=8.5cm]{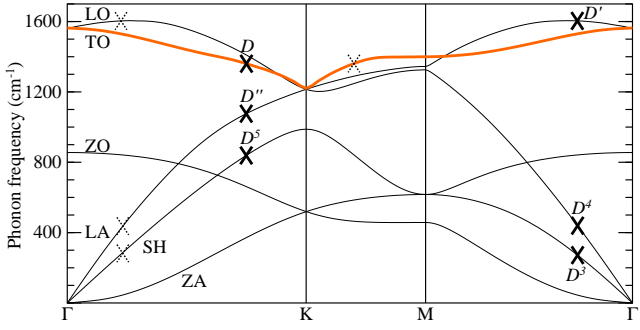}
\caption{(Color online) 
Phonon dispersion of graphene along high symmetry lines.  Bold crosses
indicate the phonons that mostly contribute to the $D$, $D'$, $D''$,
$D^3$, $D^4$ and $D^5$ Raman bands, for $\epsilon_L=2.4$~eV.  Dotted
crosses indicate phonons that also contribute to the $D$, $D'$, $D^3$, and $D^4$
bands but with smaller intensity.
The crosses are determined from the maximum of ${\cal I}_{\bf q}$
as defined in Sec.~\ref{phon}.
}
\label{fig_attrib}
\end{figure}

\begin{figure}[h]
\includegraphics[width=7.8cm]{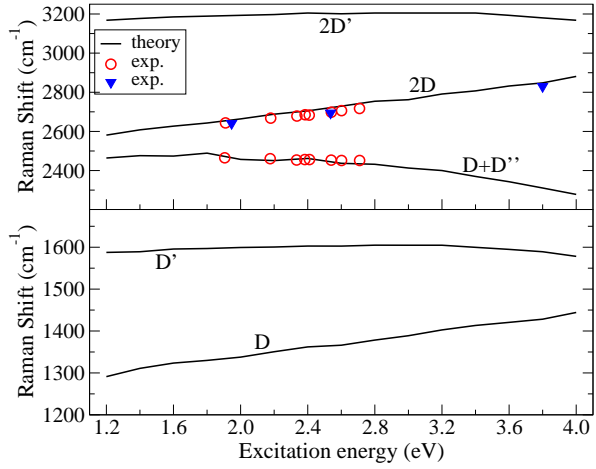}
\caption{(Color online) Raman shift as a function of excitation energy.
Upper panel: two-phonon bands.
Lower panel: disorder induced bands. Our results compared to experimental data
from  Ref.~\cite{mafra} (circles) and Ref.~\cite{calizo} (triangles).  }
\label{frequency}
\end{figure}

Fig.~\ref{spectra} displays the calculated spectra of the main double resonant Raman
lines for three different excitation energies.  In all cases,
we use the electronic broadening $\gamma^{tot}=
\tilde\gamma^{(ep)}$, calculated at the corresponding excitation
energy (Sec.~\ref{sec_el-life}).  In general, 
by increasing the excitation energy,
the bands become broader and the relative intensities change.
The behavior of the $2D$ line is particularly
interesting.  At $\epsilon_L=2.4$~eV, the $2D$ line presents a Lorentzian
lineshape with a relatively small linewidth, while at $\epsilon_L=3.8$~eV,
it is much broader showing two components with smaller, 2$D^-$,
and higher, 2$D^+$ Raman shifts,
as discussed in detail in Sec.~\ref{sec_2Dwidth}.  Here, we just
remark that the presence of a small width $2D$ line with Lorentzian
shape is commonly used to detect a graphene monolayer in samples
containing flakes with a different number of graphene layers
~\cite{ferrari}.  According to Fig.~\ref{spectra}, this kind of
experiment makes sense only when it is done at $\epsilon_L\lesssim
2.4$~eV, but not at higher excitation energies.

Fig.~\ref{fig_attrib} shows the wavevector and the
branch of the high symmetry phonons which mostly contribute to the
DR graphene lines, for $\epsilon_L=2.4$~eV.  
The figure display the phonons associated with the single-phonon Raman
lines $D$, $D'$, $D''$, $D^3$, $D^4$ and $D^5$, where
$D^3$, $D^4$ and $D^5$ refer to the small intensity lines of
Fig.~\ref{small}.  The $D$ line is associated to the phonon branch
affected by the Kohn anomaly (thick grey line in
Fig.~\ref{fig2}). This branch, near ${\bm \Gamma}$, becomes almost
transverse (TO).  The $D'$ line is associated to the branch which,
near ${\bm \Gamma}$, is almost longitudinal (LO).  The two-phonon
bands, such as the $2D$, $2D'$ and $D+D''$ are associated with the
emission of two phonons which, in the scale of Fig.~\ref{fig_attrib}, are
almost indistinguishable from those of the $D$, $D'$, and $D''$ lines.

Fig.~\ref{frequency} shows the calculated shift of the main Raman lines
as a function of the excitation energy, $\epsilon_L$.  The Raman shift
of the $D$ and $2D$ lines increases with increasing laser energy.  The
$D'$ Raman shift does not show a monotonic behavior but it does not
change significantly.  The $D+D''$ Raman shift is almost constant for
$\epsilon_L$ between 1.2 and 1.8 eV, and decreases for
$\epsilon_L\gtrsim$1.8 eV.  Fig.~\ref{frequency} also shows the
experimental data from Ref.\cite{mafra} for the $2D$ and $D+D''$ lines
and from Ref.\cite{calizo} for the $2D$ line.  The good agreement with
measurements is not surprising since the dispersion of a DR line as a
function of $\epsilon_L$ is determined by the phonon dispersion and in
Ref.~\cite{michele} it was already shown that the present phonon
dispersions (obtained from DFT plus GW corrections) reproduce the
measured $D$ line shift as a function of $\epsilon_L$.  The behavior of
the shift as a function of $\epsilon_L$ is easily understood by
comparing with the phonon dispersions in Fig.~\ref{fig_attrib}.  For
instance, for the $D$ line, when the excitation energy increases, the
phonons mostly involved in the DR process move away from {\bf K}, and their
frequencies are higher.  The same reasoning explains the
behavior of the $D'$ frequency.  For the two-phonon lines,
one has to consider the frequencies of the two phonon
involved. For instance, the $2D$ line Raman
shifts are twice as large as the $D$ ones.  For the $D+D''$
line, the energy of one phonon branch increases, while the other
decreases while moving away from {\bf K}.

\subsubsection{Small intensity bands}

\begin{figure}[ht]
\includegraphics[width=7.8cm]{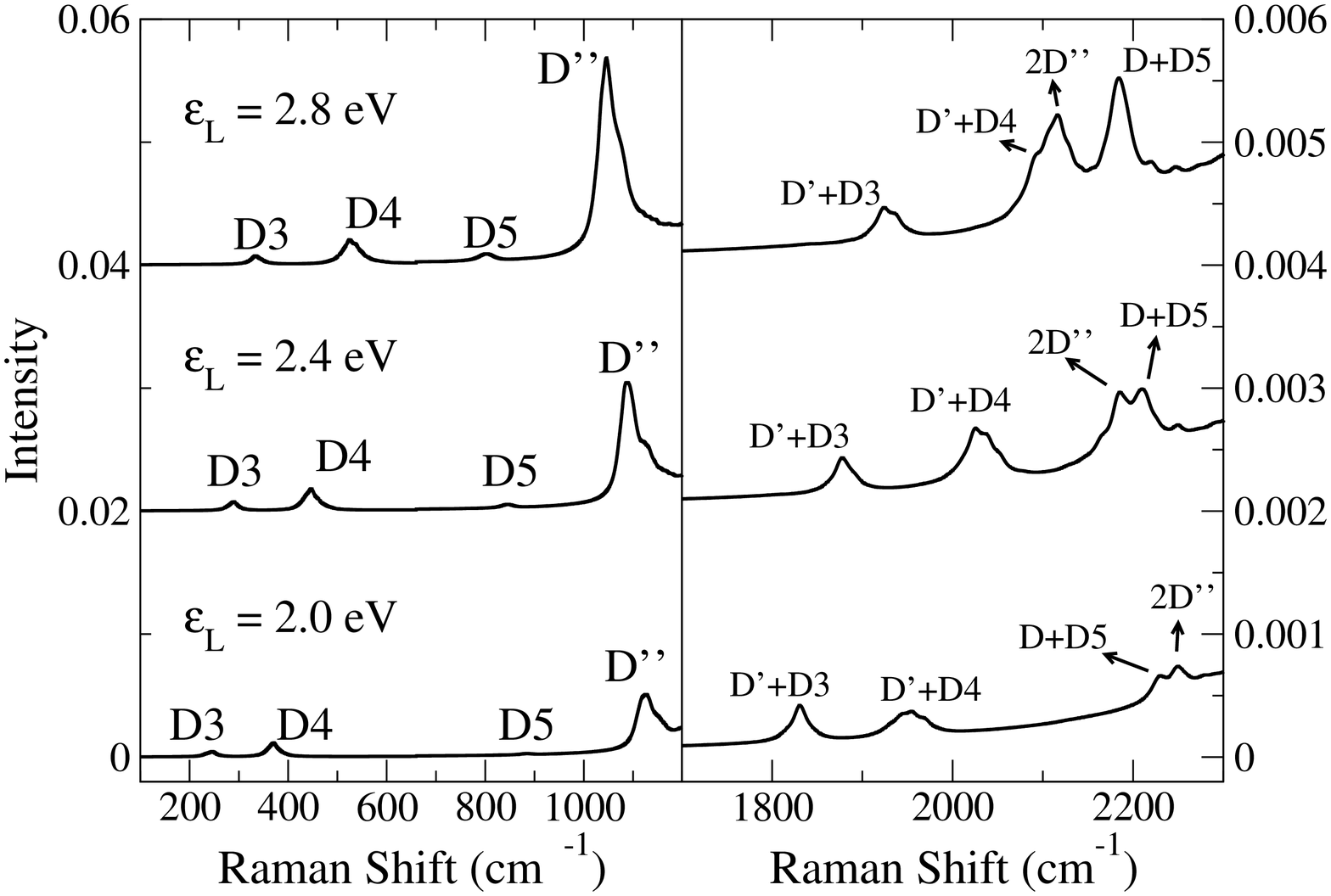}
\caption{Calculated Raman spectra for small intensity bands.
Calculations are done using
$\epsilon_L$ = 2.0 eV and $\gamma^{tot}$ = 65 meV (upper),
$\epsilon_L$ = 2.4 eV and $\gamma^{tot}$ = 84 meV (middle),
$\epsilon_L$ = 2.8 eV and $\gamma^{tot}$ = 106 meV (lower). 
We consider
hopping defects with $\alpha_{hopp}=6.4\times10^{13}$~eV$^2$cm$^{-2}$.
All the intensities are normalized to the corresponding $2D$ line maxima. }
\label{small}
\end{figure}

\begin{figure}[ht]
\includegraphics[width=8.2cm]{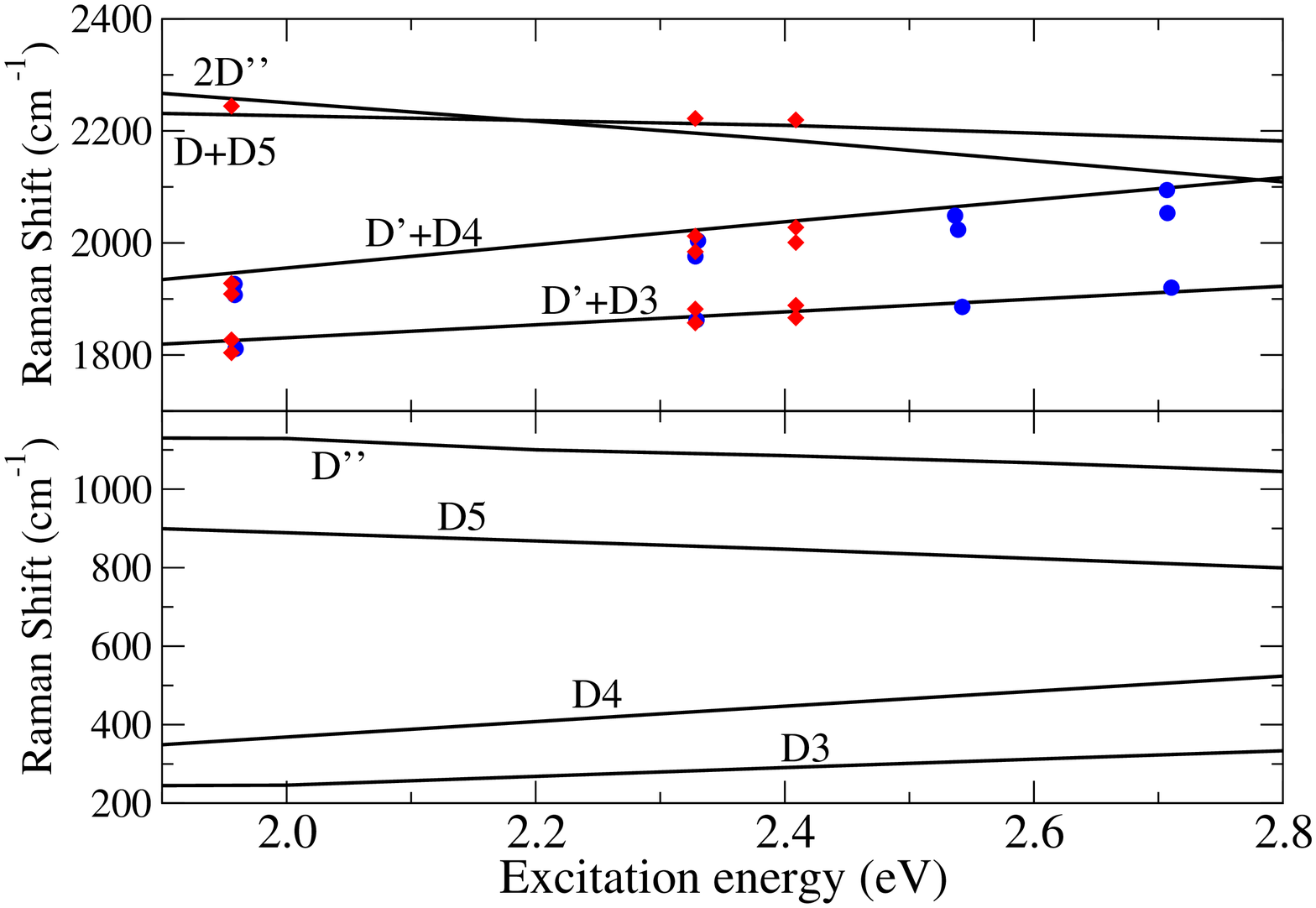}
\caption{(Color online) Raman shift vs. excitation energy for the
small intensity bands of Fig.~\ref{small}. Upper and lower panels
display results for two-phonon and defect-induced bands, respectively.
Upper panel calculations are compared
with measurements from ~\cite{cong} (dots)
and ~\cite{rao11} (diamonds).}
\label{freqsmall}
\end{figure}

The calculated spectra display some small intensity bands which are shown in
Fig.~\ref{small}.
Some of these bands are extremely weak and it is not clear whether they could
be possibly measured, on the other hand the $D''$ is observed~\cite{lucchese,erlon}
and the bands that we label as $D'+D^4$ and $D'+D^3$ have been measured
recently ~\cite{cong,rao11}.
Fig.~\ref{freqsmall} reports the shift of these small intensity bands as a function
of the excitation energy. The agreement with available measurements is good.
Fig.~\ref{fig_attrib} reports the high symmetry phonons associated
with the bands that we label as $D^3$, $D^4$, $D^5$, and $D''$.
The $D^3$ and $D^4$ bands are associated with phonons near ${\bm \Gamma}$, that
have a momentum very similar to the momentum of the phonons associated to the $D'$ line.
The $D^5$ and $D''$ bands are associated with phonons near {\bf K}, with
a momentum very similar to the momentum of the $D$ phonons.
The $D^3$, $D^4$, $D^5$, and $D''$ bands are however much weaker than the $D$ and $D'$
ones, because the electron-phonon coupling (between $\pi$ electronic bands) for those
branches, is much weaker than the one of the $D$ and $D'$ (see~\cite{piscanec}).


\subsubsection{Dependence on the light polarization}
\label{sec_polariz}

\begin{figure}[ht]
\includegraphics[width=7.8cm]{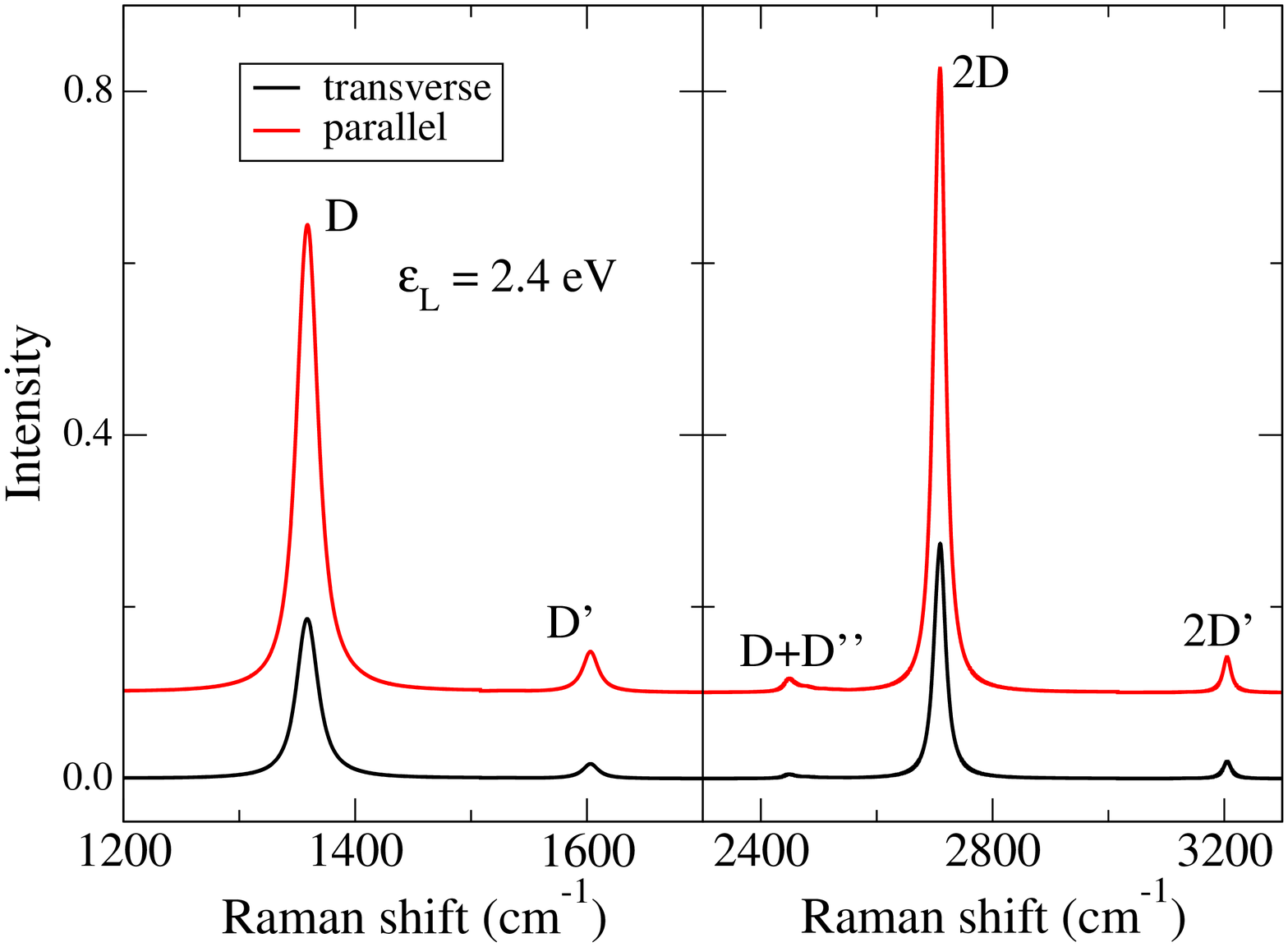}
\vspace{0.5cm}
\includegraphics[width=7.8cm]{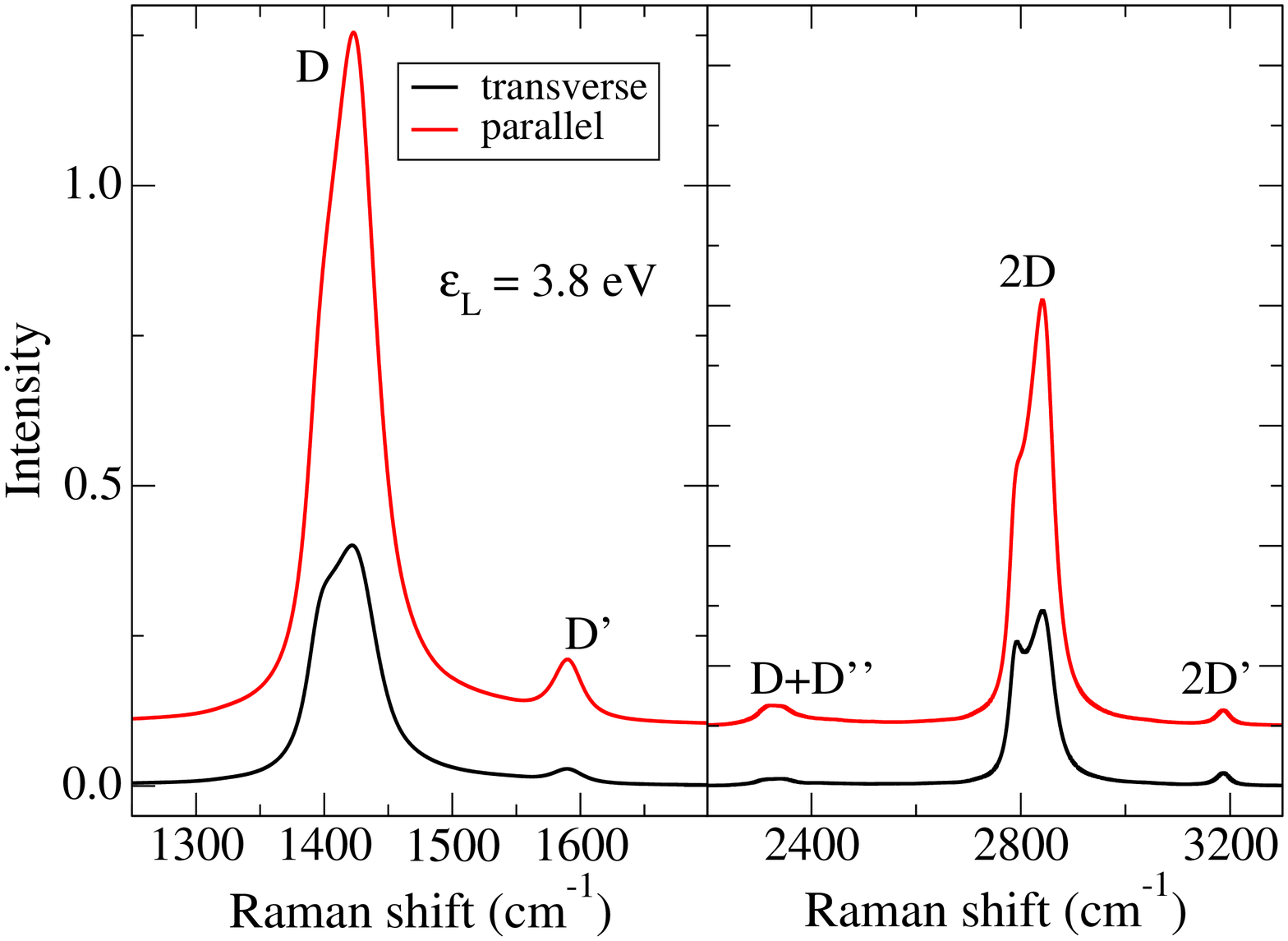}
\caption{(Color online) Comparison of calculated Raman spectra
done with different light polarizations.
Calculations are done using
$\epsilon_L$ = 2.4 eV and $\gamma^{tot}$ = 84  meV (upper plot), or
$\epsilon_L$ = 3.8 eV and $\gamma^{tot}$ = 170 meV (lower plot).
We used hopping defects with $\alpha_{hopp}=6.4\times10^{13}$~eV$^2$cm$^{-2}$.
The intensities are normalized to the corresponding $2D$ line maxima
calculated with unpolarized light.
``Parallel'' and ``transverse'' refer to $I_\parallel$ and $I_\perp$
as defined in Sec.~\ref{sec_el-light}.
}
\label{polar}
\end{figure}

So far, we have shown calculations done with unpolarized light.
We now discuss how the results are affected by the use of polarized
light.
For parallel and transverse polarizations, we calculated $I_\parallel$ and $I_\perp$ as
defined in Sec.~\ref{sec_el-light}.
Fig.~\ref{polar} compares
the results obtained for $\epsilon_L$ = 2.4 eV and $\epsilon_L$ = 3.8 eV.
The intensity in the parallel polarization case is considerably larger than
in the transverse one, as expected.
For $\varepsilon_L$ = 2.4 eV, the spectrum shape almost does not depend on the polarization
and the ratio $I_\parallel$/$I_\perp$ is about 2.7, in reasonable agreement
with measurements in graphite~\cite{reichptrs},
graphene~\cite{yoon08} and earlier theoretical predictions~\cite{basko}.
For $\epsilon_L$ = 3.8 eV, the $D$ and $2D$ bands split into two components
(see Sec.~\ref{sec_2Dwidth} for a detailed discussion) and
the intensity ratio between the two components 
depends on the polarization. For example, the intensities of the two components
of the $2D$ band, $2D^+$ and $2D^-$, are very similar within transverse polarization, while
the $2D^+$ intensity is slightly higher than the $2D^-$ one, within parallel polarization.
This finding is very remarkable since it could lead to measurable effects.

\subsection{Dependence of the Raman intensities on the various parameters}
\label{sec_intensities}

In this section we discuss how the intensity of the main DR Raman lines is
affected by the various parameters such as the electronic linewidth (Sec.~\ref{lw}),
the excitation energy (Sec.~\ref{sec_int_laser}), and
the defect concentration (Sec.~\ref{def}).
In general, the absolute value of the intensities is affected by these parameters,
however, we will mainly focus on how the ratio of the intensities of
different lines is affected, since this last quantity can be measured
more easily.

\subsubsection{Dependence on the electronic broadening}
\label{lw}

\begin{figure}[h]
\includegraphics[width=7.8cm]{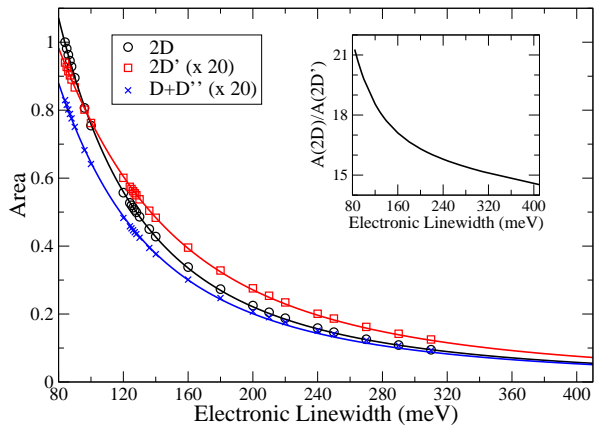}
\caption{(Color online) 
Integrated areas under the $2D$, $2D'$ and $D+D''$ lines
[A$(2D)$, A$(2D')$, and A$(D+D'')$]
as a function of the electron + hole linewidth ($\gamma^{tot}$),
for $\varepsilon_L$ = 2.4 eV.
The areas are normalized to A$(2D)$ calculated with $\gamma^{tot}=\tilde\gamma^{(ep)}$ = 84 meV.
For clarity,  A$(2D')$ and A$(D+D'')$ are multiplied by 20.
Symbols are calculations, lines are the fit from Eq.~\ref{eq_fit_area}.
Inset: A$(2D)/$A$(2D')$ ratio.
}
\label{2d_2dprime}
\end{figure}

As already discussed in Sec.~\ref{sec_el-life} the broadening
parameter $\gamma^{tot}$ (the sum of the electron and hole linewidths,
see Eq.~\ref{eq_gtot}) results from an intrinsic component (due to
electron-phonon scattering), which depends on the laser energy, and
from an extrinsic component which increases by increasing the defect
concentration.  Eventually, in charged (doped) graphene, a further
contribution due to electron-electron scattering can be relevant.  The
actual value of $\gamma^{tot}$, which depends on the defect
concentration, determines in a measurable way also the intensities of
the two-phonon lines (which are not defect induced).  Indeed,
Fig.~\ref{2d_2dprime} reports the integrated areas under the $2D$,
$2D'$ and $D+D''$ lines [A$(2D)$, A$(2D')$, and A$(D+D'')$], as a
function of $\gamma^{tot}$. The areas of these lines decrease by
increasing $\gamma^{tot}$.  In general, for all Raman lines studied
here, the intensity decreases when the electronic linewidth increases,
at fixed defect concentration.  This is because, in Eq.~\ref{eq1}, an
increase of the imaginary values $i\gamma$ tends to kill the double
resonance condition.

It is interesting to notice that also the ratio of the two areas,
A$(2D)/$A$(2D')$, depends on $\gamma^{tot}$ (inset of
Fig.~\ref{2d_2dprime}).  This result is particularly appealing since
the ratio of the two areas can be measured in a relatively easy way.
The measured value of A$(2D)/$A$(2D')$ compared to the inset of
Fig.~\ref{2d_2dprime} (which is obtained for $\epsilon_L=2.4$~eV)
could, thus, be used to determine experimentally the electron+hole
linewidth $\gamma^{tot}$ and, in particular, its components due to
defects and/or to electron-electron scattering in doped samples
(keeping in mind that for large doping the value of the
electron-phonon interaction itself is expected to
change~\cite{attaccalite10} and, thus, the inset of
Fig.~\ref{2d_2dprime} cannot be used as it is).  For
$\gamma^{tot}=\tilde\gamma^{(ep)}$ = 84 meV, which is suitable for
comparison with pristine graphene, A$(2D)/$A$(2D')$ = 21.5, in
agreement with experimental works which reported A$(2D)/$A$(2D')$ as
being 27~\cite{ferrari} and 26 $\pm$ 3 ~\cite{alzina}.

In~\cite{basko07} it has been shown that, if the electronic bands can be considered conic,
the dependence of A$(2D)$ and A$(2D')$
on $\gamma^{tot}$ should be A$=A_0/(\gamma^{tot})^2$,
where $A_0$ is a constant.
This functional form, however,  cannot be used for a quantitative description of the present results.
Indeed, the integrated areas as a function of $\gamma^{tot}$ reported in Fig.~\ref{2d_2dprime}
can be fitted by a similar, but different, law:
\begin{eqnarray}
{\rm A}(2D)   &=& 9374/((\gamma^{tot})^2+48.5^2) \nonumber \\
{\rm A}(2D')  &=& 629/((\gamma^{tot})^2+80.0^2) \nonumber \\ 
{\rm A}(D+D'')&=& 438/((\gamma^{tot})^2+59.6^2),
\label{eq_fit_area}
\end{eqnarray}
where $\gamma^{tot}$ is expressed in meV.
An explanation of the discrepancy between Eqs.~\ref{eq_fit_area}
and the model of~\cite{basko07}
(which is based on a simplified description of the electronic bands)
is probably associated to the importance of a proper inclusion
of the trigonal warping and of the electron/hole asymmetry in the description
of the electronic bands (Sec.~\ref{sec_el_ph_disp}).
Another result of ~\cite{basko07} is that
\begin{equation}
{\rm A}(2D)/{\rm A}(2D')=2(\eta^{\bf K}_1/\eta^{\bm \Gamma}_1)^4\times(\omega_{2D'}/\omega_{2D})^2.
\label{eq_basko}
\end{equation}
Eq.~\ref{eq_basko} is obtained by rewriting the equation in 
the last paragraph of~\cite{basko07} using
the notation of Sec.~\ref{sec_el-ph_scatt} and considering
$\omega_{2D}$ and $\omega_{2D'}$ are the frequencies associated with
the two Raman lines.
Indeed, for large $\gamma^{tot}$, the ratio A$(2D)/$A$(2D')$ from Eqs.~\ref{eq_fit_area}
does not depend on $\gamma^{tot}$.
However, using the parameters of the present work,
Eq.~\ref{eq_basko}, gives A$(2D)/$A$(2D')$=6.8 which is almost two times
smaller than A$(2D)/$A$(2D')$=14.7
obtained from the limit $\gamma^{tot}\rightarrow\infty$
of Eqs.~\ref{eq_fit_area}.
This second discrepancy with the model of ~\cite{basko07}
is so far unexplained, since in this limit the effect of electron-hole
asymmetry should become negligible.
We also remark that the model of ~\cite{basko07} predicts that the ratio
A$(2D)/$A$(2D')$ does not depend on the excitation energy $\epsilon_L$.
In the following we will show that, on the contrary,
A$(2D)/$A$(2D')$ strongly depends on $\epsilon_L$.

\subsubsection{Dependence on the excitation energy}
\label{sec_int_laser}

\begin{figure}[ht]
\includegraphics[width=8.0cm]{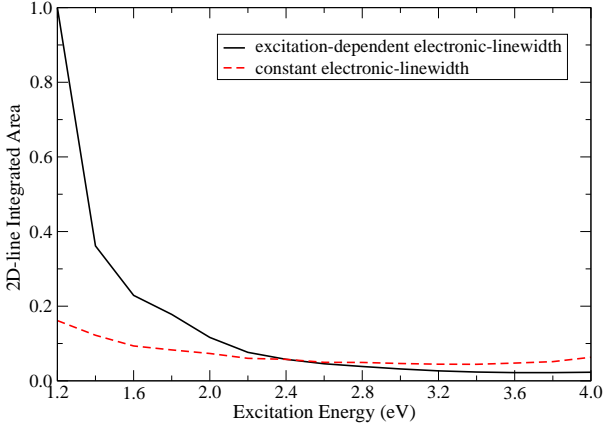}
\caption{(Color online) Integrated area under the $2D$ line as a function of the excitation energy
$\epsilon_L$. The defect concentration is zero.
The full line is obtained by including the dependence of the broadening on $\epsilon_L$,
$\gamma^{tot}=\tilde\gamma^{(ep)}(\epsilon_L)$ (see Sec.~\ref{sec_el-life}).
The dashed line is from an unrealistic simulation in which $\gamma^{tot}$ has been kept fixed to a
constant value $\gamma^{tot}=\tilde\gamma^{(ep)}(2.4~{\rm eV})=84$~meV, independent from $\epsilon_L$.
}
\label{area_laser}
\end{figure}

\begin{figure}[ht]
\includegraphics[width=7.8cm]{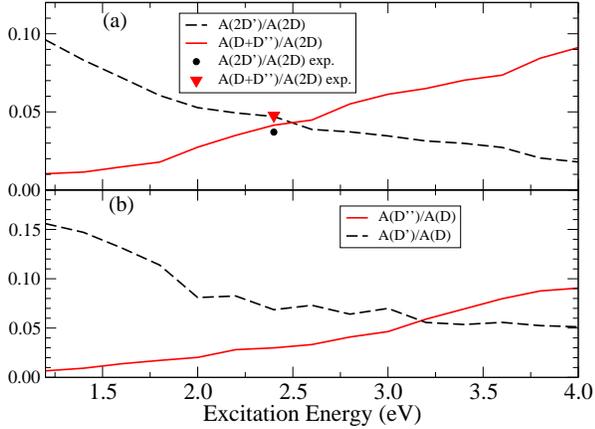}
\caption{(Color online) Ratio of the integrated areas under Raman bands as a function of excitation energy.
(a) Two-phonon bands: our results compared to experimental
data from Ref.~\cite{ferrari}. (b) Disorder induced bands from hopping impurities,
with $\alpha_{hopp}=6.4\times10^{13}$ eV$^{2}$ cm$^{-2}$.}
\label{int_laser}
\end{figure}

The intensity of the $2D$ line decreases by increasing the excitation energy $\epsilon_L$
(Fig.~\ref{area_laser}).
The most important contribution to the decrease comes from the fact that
the electron/hole broadening $\gamma^{tot}$ increases by increasing $\epsilon_L$.
This can be deduced from Fig.~\ref{area_laser} which also shows the results for a
fictitious system in which  $\gamma^{tot}$ is kept to a fixed value independent from $\epsilon_L$.
Indeed, in this second case, the dependence of A($2D$) on $\epsilon_L$ is much less
marked than in the full calculation.

Fig.~\ref{int_laser}(a) reports the calculated ratio of the integrated
areas under the bands, A$(2D')/$A($2D$) and A$(D+D'')/$A$(2D)$, as a
function of the excitation energy $\epsilon_L$.  These ratios considerably
change in the range of excitation energies of the figure.
A$(2D')/$A$(2D)$ decreases and A$(D+D'')/$A$(2D)$ increases rapidly.
The values calculated for $\epsilon_L$ = 2.4 eV compare reasonably
well with those obtained from the measurements of ~\cite{ferrari}.  In
the last paragraph of Sec.~\ref{lw} we discussed the model of
~\cite{basko07}, which was used to theoretically determine the ratio
A$(2D')/$A$(2D)$.  The simplified model of ~\cite{basko07} predicts
that the ratio A$(2D')/$A$(2D)$ does not depend on $\epsilon_L$. On
the contrary, from Fig.~\ref{int_laser}(a), this dependence is very
important.  Using Eq.~\ref{eq_basko} (which is adapted from
~\cite{basko07}) and using, for consistency, the parameters of the
present work, one obtains A$(2D')/$A$(2D)=0.15$.  
This value is significantly higher than 0.09, which we
obtain for the smallest $\epsilon_L$ of Fig.~\ref{int_laser}(a).

Fig.~\ref{int_laser}(b) reports the ratio of the integrated areas
under the defect-induced bands, A$(D')/$A$(D)$ and
A$(D'')/$A$(D)$. Here, we consider again only hopping impurities.  We
also remark that the present approach is expected to be valid in the
limit of small defect concentration.  For small excitation energies
the $D''$ band intensity is very small in comparison to the $D$
one. For larger excitation energies the $D''$ relative intensity
increases, reaching A$(D'')/$A$(D)$ = 0.09 when $\epsilon_L$ = 4.0 eV.
On the other hand, the intensity of the $D'$ band compared to the $D$
band decreases by increasing the excitation energy.  For $\epsilon_L$
up to about 3.0 eV the $D'$ band is more intense than the $D''$ band,
while for $\epsilon_L \gtrsim 3.2$~eV, the $D''$ is slightly more
intense than the $D'$.

\subsubsection{\label{def} Dependence on the defect concentration}

\begin{figure}[ht]
\includegraphics[width=7.8cm]{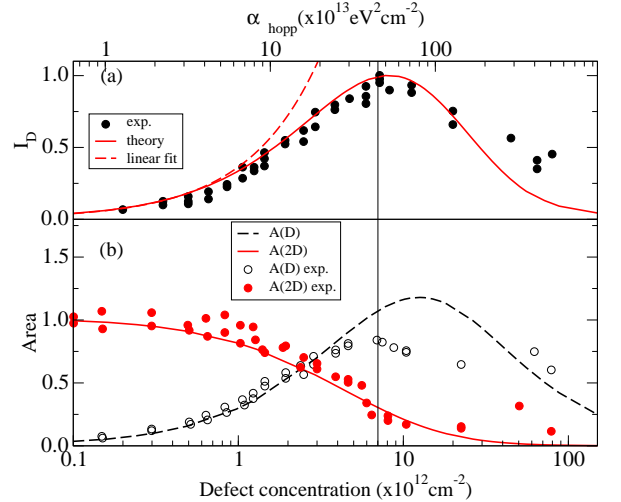}
\caption{(Color online) 
Intensity of the $D$ and $2D$ Raman lines as a function of the
defect concentration for $\epsilon_L=2.4$~eV.
Calculations are done using hopping defects and are reported as
a function of the parameter $\alpha_{hopp} = n_{d}(\delta t_1)^2$
($n_d$ is the defect concentration and $\delta t_1$ the hopping parameter),
in the upper horizontal scale.
The lower horizontal scale is obtained by considering $\delta t_1=8.0$~eV.
(a) I$_{D}$ is the maximum of the intensity of the $D$ line;
symbols are experimental data from ~\cite{lucchese}.
The dashed line is a linear fit of
the $I_D$ calculated values for $n_d <$ 5 x 10 $^{11}$ cm$^{-2}$.
Theoretical and experimental intensities have been normalized by
their maximum values.
(b) Integrated areas under $D$ and $2D$ bands, A($D$) and A($2D$).
Experimental data are from ~\cite{erlon}.
Theoretical and experimental areas are normalized by A($2D$) at minimum defect concentration. 
The vertical line indicates the defect concentration of
$7\times10^{12}$~cm$^{-2}$~
($\alpha_{hopp}=4.5\times10^{14}$~cm$^{-2}$eV$^2$)
for which the two contributions to the electronic broadening are equal:
$\tilde\gamma^{(D)}=\tilde\gamma^{(ep)}$. }
\label{intensity}
\end{figure}

We now discuss how the intensities of the Raman bands are affected by
defect concentration $n_d$.
We recall that two-phonons Raman lines (such as the $2D$) depend on $n_d$
only through the electronic broadening parameter $\gamma^{tot}$
(Eq.~\ref{eq_gtot}).
$\gamma^{tot}$ is given by the sum of an intrinsic component $\tilde\gamma^{(ep)}$
(due to the electron-phonon interaction) and an extrinsic defect-induced component
$\tilde\gamma^{(D)}$ which increases linearly by increasing $n_d$
(Eq.~\ref{eq13}).
On the other hand,
the defect-induced Raman lines (such as the $D$ line) depend on $n_d$
through two distinct mechanisms.
First, it depends on $n_d$ through $\gamma^{tot}$ as for the two-phonon lines.
Second, there is a proportionality factor between the Raman intensity and the
the number of defects in the sample
($I\propto N_d$ in Eq.~\ref{eq3}).
Basically, for a higher number of defects there are more scattering events
that can activate the defect-induce lines, which, in crystalline samples, are not Raman active.
In the following discussion, we will consider only hopping defects.
As already shown in Sects.~\ref{sec_el-def_scatt} and~\ref{sec_el-life},
the calculated Raman spectra depend on the defect concentration,  $n_{d}$,
only through the parameter $\alpha_{hopp} = n_{d}(\delta t_1)^2$,
being $\delta t_1$ the hopping parameter.

Fig.~\ref{intensity} reports the
$D$ line peak maximum (I$_D$) and the
integrated areas under the calculated $2D$ and $D$ lines, A($D$) and A($2D$),
as a function of the parameter $\alpha_{hopp}$, for $\epsilon_L=2.4$~eV.
For $\alpha_{hopp}=4.5\times10^{14}$~cm$^{-2}$eV$^2$,
the two contributions to the broadening are equal,
$\tilde\gamma^{(D)}=\tilde\gamma^{(ep)}$.
The corresponding $\alpha_{hopp}$ is indicated in
Fig.~\ref{intensity} with a vertical line.
The intensity of the $2D$ line (which corresponds to a two-phonon process)
monotonously decreases by increasing the defect concentration.
For small defect concentrations
($\alpha_{hopp}\leq 10^{14}$~cm$^{-2}$eV$^2$)
$\tilde\gamma^{(D)}\ll\tilde\gamma^{(ep)}$,
$\gamma^{tot}\sim\tilde\gamma^{(ep)}$ slightly depends on the defect concentration,
and A$(2D)$ is almost constant.
For higher defect concentrations, $\tilde\gamma^{(D)}$ becomes the
dominant contribution to $\gamma^{tot}$, which, as a consequence,
becomes more sensitive to the defect concentration.
The increase of $\gamma^{tot}$ by increasing the defect concentration
is associated to a decrease of A($2D$), because of the mechanism
discussed in Sec.~\ref{lw}.

The intensity of the $D$ line (which is a defect induced process) has a different behavior.
For low defect concentrations, it increases almost linearly, then it reaches a maximum,
and finally decreases.
This behavior results from the interplay of two competing mechanisms.
For small defect concentration $\tilde\gamma^{(D)}\ll\tilde\gamma^{(ep)}$ and
$\gamma^{tot}\sim\tilde\gamma^{(ep)}$. In this region,
the intensity is expected to increase linearly ($I\propto N_d$ in Eq.~\ref{eq3}).
Indeed, the calculated intensity is well reproduced by a linear fit up to 
$\alpha_{hopp}\leq 10^{14}$~cm$^{-2}$eV$^2$
(compare the continuous line with the dashed one in Fig.~\ref{intensity}, upper panel).
For $\alpha_{hopp} > 4.5\times10^{14}$~cm$^{-2}$eV$^2$,
the dependence of the broadening $\gamma^{tot}$ on the defect concentration
becomes the dominant mechanism,
leading to a decrease of the intensity as for the $2D$ line.
It is remarkable that the defect concentration for which 
$\alpha_{hopp}=4.5\times10^{14}$~cm$^{-2}$eV$^2$
(vertical line in Fig.~\ref{intensity})
almost coincides with the maximum 
value reached by the $D$ intensity, I$_{D}$.

Fig.~\ref{intensity} compares calculations with the intensities of
the $D$ and $2D$ measured in~\cite{erlon,lucchese} as a function of
the defect concentration.  So far, we have discussed theoretical
results as a function of $\alpha_{hopp} = n_{d}(\delta t_1)^2$.
$\alpha_{hopp}$ defines the upper horizontal scale in
Fig.~\ref{intensity}.  To make the comparison with measurements we
need to attribute a value to the hopping energy $\delta t_1$.  The
best fit to measurements is obtained for
$\delta t_1=8.0$~eV.  This value is used only to rescale the
horizontal axis of Fig.~\ref{intensity} and defines the defect
concentration as reported in the lower horizontal axis of
Fig.~\ref{intensity}.  The measured behavior as a function of the
defect concentration is well reproduce by calculations.  It is
remarkable that the same value $\delta t_1=8.0$~eV can be used to fit
equally well the $D$ and the $2D$ line data.
The value $\delta t_1=8.0$~eV is very high.
However, one should notice that in Ref.~\cite{erlon,lucchese} defects were 
induced in graphene by means of Ar$^+$ ion bombardment.  
This technique leads to the formation of Carbon
multi-vacancies in the sample. In Ref.~\cite{lucchese},
the defect average size is estimated, by means of
scanning tunnel microscopy, to be 1.85~nm.
On the contrary, the present model considers only point defects
(the hopping parameters is changed by $\delta t_1$ for a single isolated 
carbon-carbon bond).
The large value $\delta t_1=8.0$~eV is, thus, to be considered
as an effective variation of the hopping parameter that mimics
the existence of an extended defect
(a realistic description of the defect should be done by considering the
variation of the hopping parameters associated to many different neighboring
sites).
For less damaging defects, $\delta t_1$ will be smaller and the
critical defect concentration, above which the $D$ line intensity
begins to decrease, will be larger than that of Fig.~\ref{intensity}.

Finally, the behavior of the $D$ line intensity as a function of the defect
concentration has been discussed in literature using different models
~\cite{erlon,lucchese} (see also ~\cite{ferrari00}).
To make a comparison,
it can be useful to restate the present finding as follows.
According to the DR perturbative model,
the intensity of the defect-induced lines
decreases by increasing the defect concentration
when $\tilde\gamma^{(D)}$ becomes higher than $\tilde\gamma^{(ep)}$, that is
when the average length an electron/hole travels in between two scatterings events
with a defect becomes smaller than the  average length an electron/hole travels before
scattering with an optical phonon.

\subsection{Dependence of the spectra on the type of defect}
\label{type}

\begin{figure}[ht]
\includegraphics[width=7.8cm]{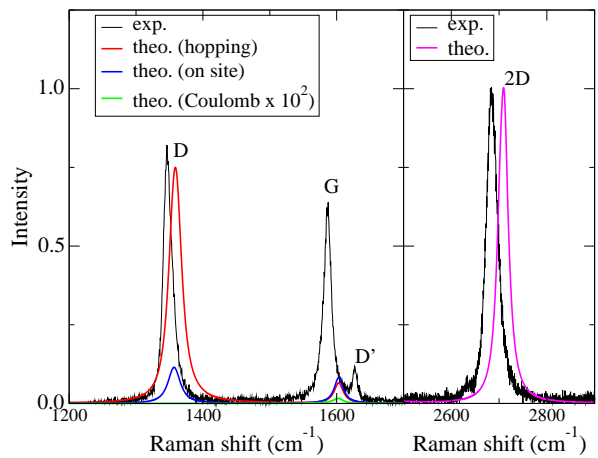}
\caption{(Color online) 
Calculated Raman spectra obtained for three different kind of defects
(hopping, on-site, and Coulomb), compared with the measurements of ~\cite{erlon}
done at $\epsilon_L=2.4$~eV.
The Raman $G$ line is not described by the present model.
Calculations are done using $\gamma^{tot}$ = 96 meV.
Other relevant parameters are given in the text.
All intensities are normalized by the corresponding $2D$ maximum.
The intensity of the Coulomb impurity spectrum is enhanced by $10^{2}$ for clarity.
}
\label{defects}
\end{figure}

Here, we discuss how the results depend on the type of defect.
Calculations were done using three different model defects 
namely, hopping defects, on-site defects, and Coulomb ones
(see Sec.~\ref{sec_el-def_scatt} for a description of the relevant parameters).
Fig.~\ref{defects} compares calculations with the measurements
from ~\cite{erlon}, which correspond to a 
defect concentration  $n_d$ = $10^{12}$ cm$^{-2}$ and $\epsilon_L=2.4$~eV.
For the hopping and on-site defects, the calculations are done using
$\alpha_{hopp}=\alpha_{on}=6.4\times10^{13}$ eV$^{2}$ cm$^{-2}$,
which, for the hopping defect, reproduces the ratio between the integrated areas of the measured
$D$ and $2D$ lines of~\cite{erlon}.
By choosing $\delta t_1=\delta V_0=8.0$~eV
(see also the discussion in Sec.~\ref{def})
, the above values of $\alpha$ 
correspond to a defect concentration $n_d$ = $10^{12}$ cm$^{-2}$.
For Coulomb impurities, the distance between the impurity and graphene is $h$ = 0.27 nm
and $n_d$ = $10^{12}$ cm$^{-2}$.

From Fig.~\ref{defects}, the hopping defect is the best model to study defect-induced
Raman processes. Indeed, contrary to the other models, the hopping defect
provides a ratio of the intensities of the $D$ and $D'$ lines which is in good agreement
with measurements.
The intensity ratio between $D$ and $D'$ strongly depends on the kind of model defect,
suggesting that this ratio could possibly be
used to experimentally determine the kind of defects present in a graphene sample.
From Fig.~\ref{defects}, we also notice that Coulomb defects (charged impurities outside the
graphene plane) provide an almost undetectable contribution to the Raman signal.
Indeed, for a defect concentration of $n_d$ = $10^{12}$ cm$^{-2}$,
the $D$ line is absent and the $D'$ intensity is
almost three orders of magnitude smaller than the experimental one.
We recall that Coulomb defects could be an important source of scattering during electronic 
transport in graphene (see ~\cite{chen08} and refs. therein).
The fact that they are not detectable by Raman spectroscopy
(which is routinely used to characterize experimentally the quality of graphene samples)
is, thus, a relevant issue which deserves some more comments.

The present simulations consider a very short graphene/impurity distance $h$, in order to
enhance the Raman signal of the Coulomb impurities. Indeed,
$h$ = 0.27~nm is the distance between K atoms and graphene planes in the
KC$_8$ intercalated graphite.
This distance corresponds to the experimental conditions of~\cite{chen08},
where K$^+$ ions are deposited on graphene.
In the case, where the impurities are charges trapped in the substrate (e.g. SiO$_2$)
a longer distance (e.g.  1~nm) is more appropriate.
It is not surprising that the  contribution of Coulomb impurities to the $D$ line
is completely negligible. Indeed, 
the Fourier transform of the Coulomb potential is maximum close to ${\bm \Gamma}$
and decays as $1/q$ far from it, Eq.~\ref{eq_b10}, and
the $D$ line is due to phonons near to the {\bf K} point and far from ${\bm \Gamma}$.
This argument, also, suggests that the $D' $ band, which is due to phonons near ${\bm \Gamma}$,
should be more sensitive to the presence of Coulomb impurities.
According to calculations, this is actually the case, however
for $\epsilon_L=2.4$~eV and $n_d=10^{12}$~cm$^{-2}$
the ratio of the integrated area ${\rm A}(D')/{\rm A}(2D) =1.5\times10^{-4}$,
meaning that the presence of a $D'$ band due to Coulomb impurities
should not be detectable.
The use of smaller energy laser increases the intensity of the $D' $ signal
since the excited phonons are nearer to ${\bm \Gamma}$.
However, for $\epsilon_L=1.2$~eV and $n_d=10^{12}$~cm$^{-2}$,
${\rm A}(D')/{\rm A}(2D) =8.0\times10^{-4}$, which is still very small.
Within the present model, ${\rm A}(D')/{\rm A}(2D)$ increases linearly by
increasing the impurity concentration, $n_d$.
$n_d$, however, cannot be higher than $10^{14}$~cm$^{-2}$,
which corresponds the density of K atoms in KC$_8$.
On the other hand, for Coulomb impurity concentrations higher than
$10^{12}$~cm$^{-2}$ doping effects should become important.
These should be associated to an increase of the electron-electron
scattering contribution to the electronic broadening~\cite{basko2},
which, in turn, will prevent the  $D'$ intensity to become detectable.
Concluding, the presence of charged impurities is not associated to a Raman
$D$ band. A $D'$ band is present, but should not be easily detectable.

\subsection{Interpretation of the results}
\label{sec_interpretation}

This section is dedicated to the interpretation of the results.
Sec.~\ref{pr} describes which are the most important processes associated to the DR.
Sec.~\ref{phon} describes which are the phonon wavevectors contributing
to each Raman band.
Sec.~\ref{angularsub} analyzes the dominant directions of the phonon wavevectors and
Sec.~\ref{sec_2Dwidth} is dedicated to the interpretation of the
small width of the main DR Raman lines.


\subsubsection{\label{pr} Dominant Processes and Interference Effects}

In this section we analyze which are the dominant processes among those
described in Fig.~\ref{fig1}.
We distinguish between two classes of processes:
processes $aa$ are those in which the two intermediate
scattering processes are associated
to both electron states or to both hole states
(namely the processes $ee1$, $ee2$, $hh1$, and $hh2$, using the notation of
Fig.~\ref{fig1});
processes $ab$ are those in which the two scattering processes are associated
one to an electron state and the other to a hole state
($eh1$, $eh2$, $he1$, and $he2$ in Fig.~\ref{fig1}).
The distinction between $aa$ and $ab$ processes holds for both
phonon-defect and two-phonon lines.

\begin{figure}[ht]
\includegraphics[width=7.8cm]{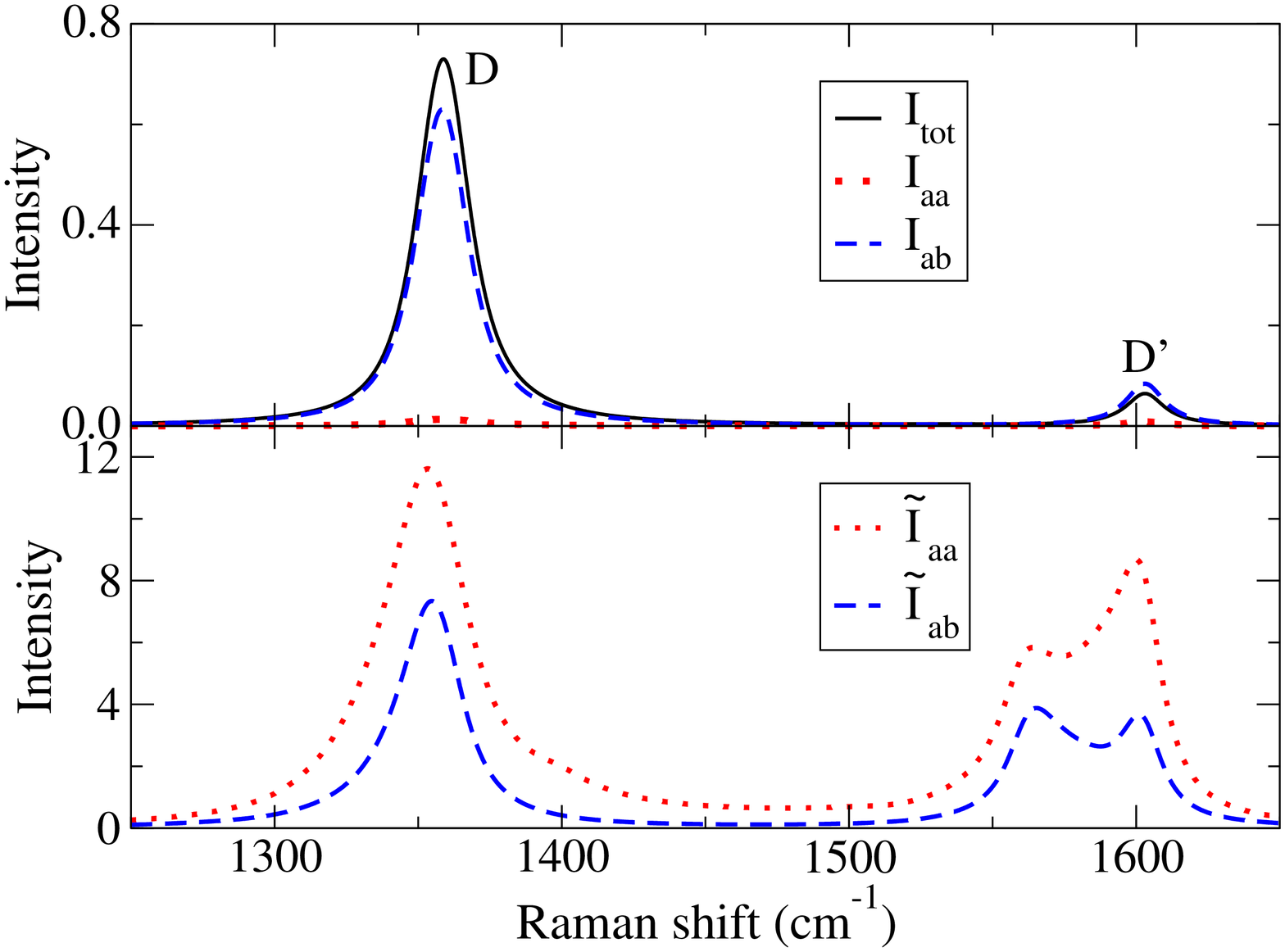}
\includegraphics[width=7.8cm]{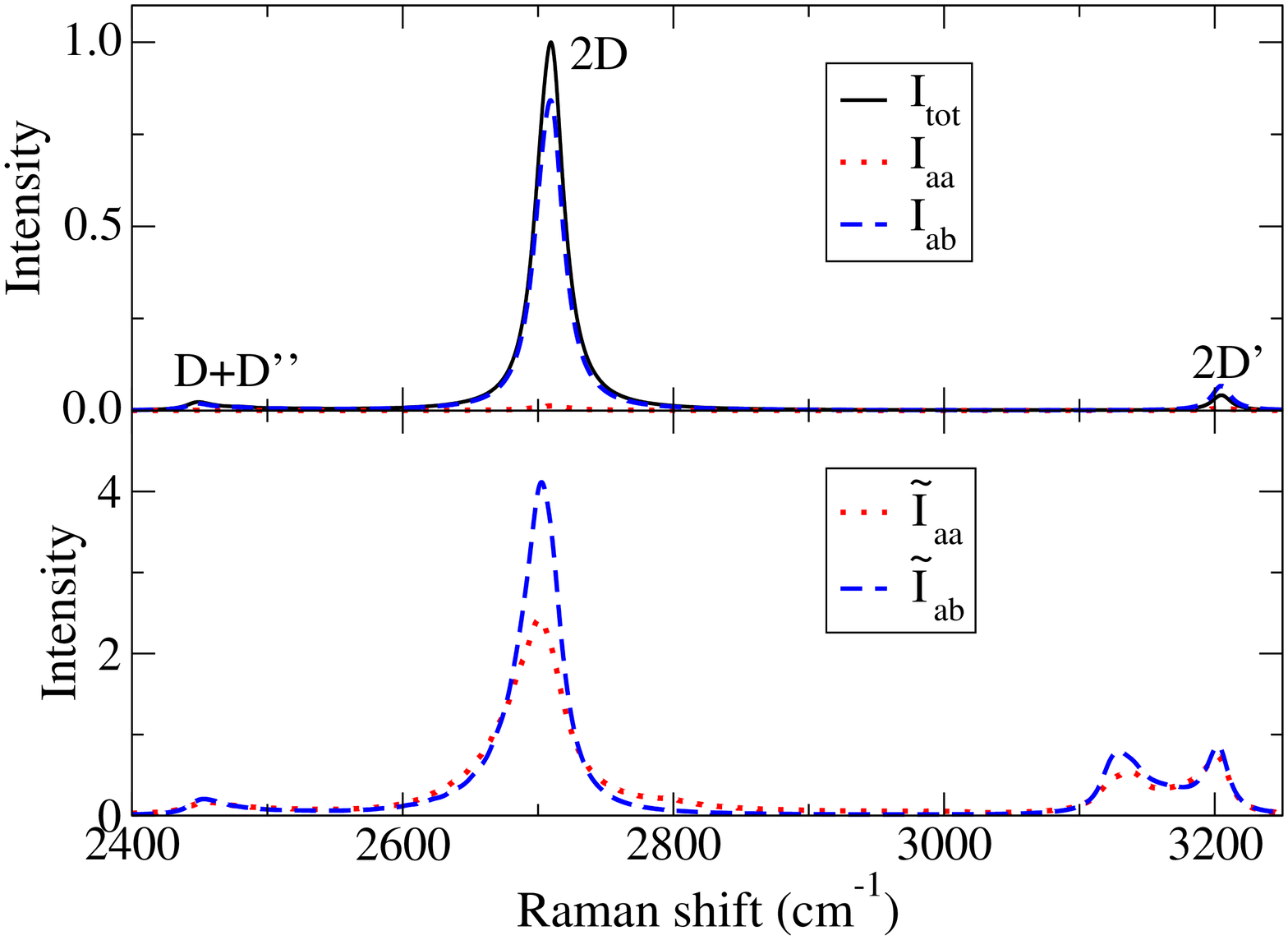}
\caption{(Color online)
The upper panels compare the calculated Raman spectrum ${\rm I}_{tot}$
with spectra determined considering only $aa$ processes, ${\rm I}_{aa}$,
or $ab$ processes, ${\rm I}_{ab}$.
More precisely, ${\rm I}_{tot}$ is determined considering all the processes
shown in Fig.~\ref{fig1};
I$_{aa}$ is computed by considering only $ee1$, $ee2$, $hh1$, and $hh2$ processes;
I$_{ab}$ is computed by considering only $eh1$, $eh2$, $he1$, and $he2$ processes
(see the text).
The lower panels display fictitious Raman intensities $\tilde {\rm I}$
obtained by substituting
to the DR scattering amplitudes $K$ in Eqs.~\ref{eq3} their modulus $|K|$
(see the text).
The two lines $\tilde{\rm I}_{aa}$ and $\tilde{\rm I}_{ab}$
are obtained by considering only $aa$ and $ab$ processes, as before.
Calculations are done using $\epsilon_L=2.4$~eV,
$\gamma^{tot}=84$~meV, and hopping defects with
$\alpha_{hopp}=6.4\times10^{13}$~eV$^2$cm$^{-2}$. All the
intensities are normalized to the $2D$ line maximum of ${\rm I}_{tot}$.
 }
\label{processes}
\end{figure}

In general, for all the simulations performed here, the $ab$ processes
are, by far, dominant over the $aa$ ones, that is, the largest part of
the Raman intensities are due to $ab$ processes.
This is true for both phonon-defect and two-phonon lines.
In general, among the $ab$ processes,
all the four processes $eh1$, $eh2$, $he1$, and $eh2$
are associated to intensities of the same order of magnitude.
Indeed, Fig.~\ref{processes} shows a typical Raman spectrum, in which
we compare the actual spectrum I$_{tot}$ with
two spectra obtained by including only $aa$ processes, I$_{aa}$, or
$ab$ ones, I$_{ab}$.
More precisely, 
I$_{tot}$ is the Raman intensity computed including all the processes;
I$_{aa}$ is computed by restricting the sums in $\alpha$ and $\beta$ in
Eqs.~\ref{eq3}
only to the $ee1$, $ee2$, $hh1$, and $hh2$ processes;
I$_{ab}$ is computed by restricting the sums in $\alpha$ and $\beta$ in
Eqs.~\ref{eq3}
only to the $eh1$, $eh2$, $he1$, and $he2$ processes.
In general, ${\rm I}_{tot}\neq{\rm I}_{aa}+{\rm I}_{ab}$.
From Fig.~\ref{processes}, ${\rm I}_{ab}\gg{\rm I}_{aa}$ for both
the $D$ and the $2D$ lines.

The dominance of the $ab$ processes is due to quantum interference effects.
In particular, from Eq.~\ref{eq3}, 
the Raman intensity for a given {\bf q}
results from a sum over {\bf k} of $K({\bf k})$ scattering
amplitudes, which are complex numbers.
The sum of these complex numbers can interfere in a constructive way,
as for the $ab$ processes, or in a destructive way, as for the $aa$ processes.
In particular,
the DR condition determines that for some resonant electronic wavevectors ${\bf k}_{r}$,
$|K({\bf k}_{r})|$ should have a maximum.
This maximum can be enhanced or suppressed by the interference of 
$K({\bf k}_{r})$ with the $K({\bf k})$ at wavevectors {\bf k} which are not exactly at
the resonance (this point is further discussed in App.~\ref{app_simple_model}).
It is important to remark that, according to the present calculations,
the DR scattering amplitudes $K$ are complex numbers in which the
real and imaginary parts are of the same order of magnitude even 
for the ${\bf k=k_r}$ wavevectors that satisfy the DR condition.

To quantify the importance of quantum interference,
we consider a fictitious Raman intensity $\tilde {\rm I}$, which is obtained
by substituting their modulus $|K|$
to the scattering amplitudes $K$
in Eqs.~\ref{eq3}.
As example, in Eqs.~\ref{eq3} we substitute
${\rm I}^{pp}_{\bf q} = \left|
\sum_{{\bf k},\beta}
K_\beta ({\bf k},{\bf q}) \right|^2/N_k$, with
$\tilde {\rm I}^{pp}_{\bf q} = \left|
\sum_{{\bf k},\beta}
|K_\beta ({\bf k},{\bf q})| \right|^2/N_k$.

Thus, within the intensities $\tilde {\rm I}$,
the presence of possible destructive interference effect
is cancelled.
Fig.~\ref{processes} shows a typical $\tilde {\rm I}$ spectrum, in which
we compare $\tilde {\rm I}_{aa}$ and $\tilde {\rm I}_{ab}$
obtained by solely including $aa$ or $ab$ processes.
The ratio $\tilde {\rm I}_{ab}/\tilde {\rm I}_{aa}$ is very different
from ${\rm I}_{ab}/{\rm I}_{aa}$ for both the $D$ and the $2D$ lines.
In particular, $\tilde {\rm I}_{ab}$ is no more dominant and it
is always comparable in intensity to $\tilde {\rm I}_{aa}$.
Thus, the fact that ${\rm I}_{ab}\gg{\rm I}_{aa}$ is indeed due to 
destructive interference effects.
Moreover, certain lines of the fictitious $\tilde {\rm I}$ spectrum, 
such as the $D'$ or the $2D'$, do not
appear as narrow and well defined lines as they are in the 
actual Raman spectrum, I.
Thus, interference effects also play a role in determining the shape of certain lines.

Notice that, often, when discussing the DR processes, it is used a
simplified argument which consists in finding the electronic and
phonon states which let two (or more) of the denominators in
Eq.~\ref{eq1} go to zero.  The assumption is that the physics is lead
only by those scattering amplitudes $K$ which satisfy the DR
condition.  This simplified approach, which we call the ``resonance
argument'', has been extensively used in literature with success
(e.g. to determine the momenta of the phonons associated to certain
lines), despite the fact that, within this approach, the possible role
of quantum interference is completely neglected.  The results of the
previous paragraph show that in certain specific situation the
``resonance argument'' can be very misleading.  For example, on the
basis of a ``resonance argument'' one would deduce that the intensity
associated $aa$ processes are of the same order of magnitude than that
associated to the $ab$ ones (indeed, $\tilde {\rm I}_{aa}\sim\tilde {\rm
I}_{ab}$ in Fig. ~\ref{processes}), which is not the case.

We remark that several authors describe the DR
by simply consider the $aa$ processes (usually the $ee$ processes in
Fig.~\ref{fig1},~\ref{fig2}),
as it is done in the seminal work by Thomsen and Reich~\cite{thomsen00}.
However, following the present conclusions, these processes cannot be used alone
to describe quantitatively the intensities of the $D$ and $2D$ lines.
The importance of interference effects in determining the shape of the
DR Raman lines has been already outlined by
Maultzsh {\it et al.} in ~\cite{maultzsch04prb}.
However, Ref.~\cite{maultzsch04prb} just consider $ee$ processes
and completely neglects the $ab$ ones, which are the most important.
The fact the $ab$ processes should be
dominant for the $2D$ line has been argued by Basko in Ref.~\cite{basko07}.
But, this conclusion is reached on the basis of
a ``resonance argument''.
Indeed, according to Ref.~\cite{basko07}, the $ab$ processes should be
dominant because within these process one can reach a condition
in which all the transitions are real (non virtual) and the three denominators
of Eq.~\ref{eq1} can be nullified simultaneously (triple resonance).
As already said, this kind of arguments cannot be applied to describe
the intensity of the $2D$ line
(basically, the conclusion is good but the argument is wrong).
The best way to understand this point is to
put to zero the phonon energies $\hbar\omega_{ph}$
in all the denominators of the Raman
scattering amplitudes $K$ (e.g. in Eqs.~\ref{eq4},~\ref{eq5}).
By doing this, the triple resonance condition of Basko applies also
to the $aa$ processes (not only to the $ab$). However,
actual calculations show that ${\rm I}_{ab}$ remains much
larger than ${\rm I}_{aa}$ even when $\hbar\omega_{ph}=0$.
Actually, the intensity and the shape of the $2D$ line are marginally affected
by including or not $\hbar\omega_{ph}$ in the denominators of the $K$s
(see Fig.~\ref{fig_phononzero} in App.~\ref{app_phononzero}).
We also remark that the triple resonance argument does not explain
why ${\rm I}_{ab}\gg {\rm I}_{aa}$ also for the $D$ line.
Finally, Ref.~\cite{basko09} argues that quantum interference {\it in real space}
plays a crucial role in enhancing the role of the $ab$ processes versus
the $aa$ ones, for the $D$ line.
However, the model of Ref.~\cite{basko09}, predicts a behavior which is
in contrast with the present calculations~\cite{note01}.
Notice that the model of ~\cite{basko09} was developed to describe extended
defects such as edges, while here we are considering point defects.

The main conclusion of this section is that the $ab$
processes ($eh1$, $eh2$, $he1$ and $eh2$ processes of
Figs.~\ref{fig1},~\ref{fig2}) are responsible for most of the Raman
intensity because of quantum interference.  We remark that
this conclusion is not due to the complex details of the present calculations
but can be deduced with a very simplified model in which the scattering matrix
elements in the numerator of Eq~\ref{eq1} are constant, the phonon
energies in the denominators (e.g. $\hbar\omega_{\bf q}^\nu$ in
Eqs.~\ref{eq4},~\ref{eq5}) are neglected, and in which the electronic
bands are conic.
This simple model can also be used to shed light on the role played by
quantum interference, see App.~\ref{app_simple_model}.

\subsubsection{\label{phon} Phonons wavevectors associated to the Raman lines}

\begin{figure}[ht]
\includegraphics[width=8.5cm]{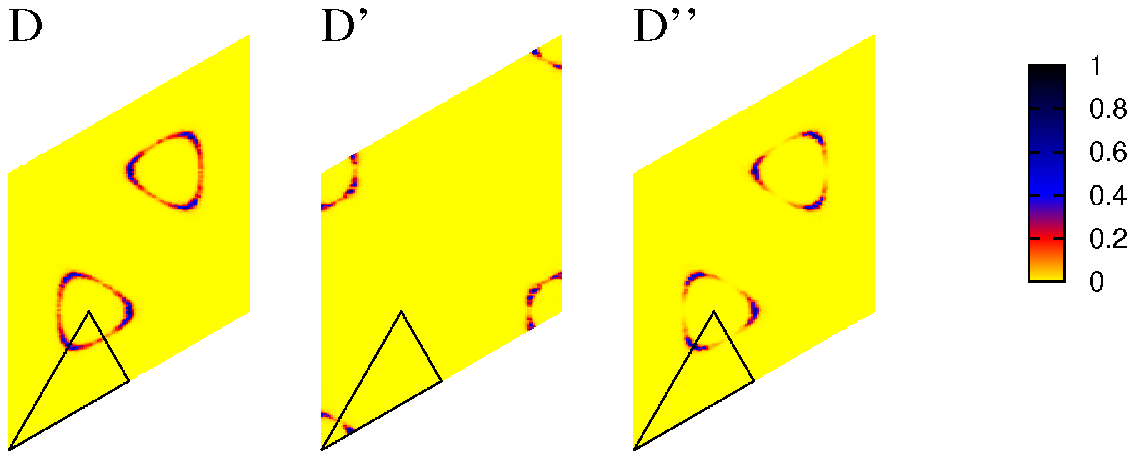}
\includegraphics[width=8.5cm]{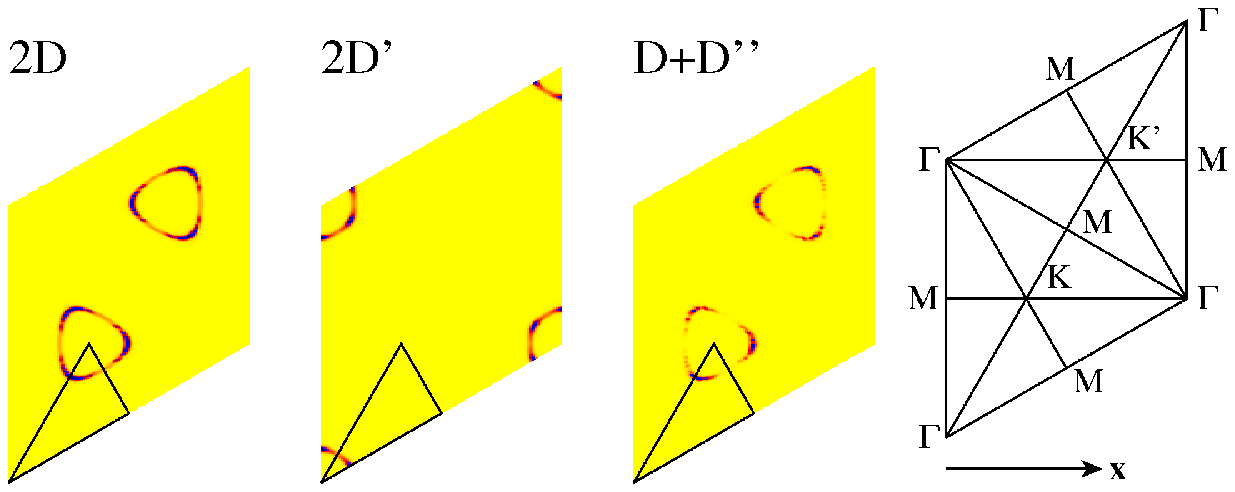}
\caption{(Color online) Decomposition the intensity of
the most important Raman bands into its components associated to
phonons with a given wavevector {\bf q}, ${\cal I}_{\bf q}$.
The rhombi are the
graphene first Brillouin zone.  For each band, we consider the
contribution to the Raman intensity in a window of frequencies
corresponding to that particular band~\cite{windows}.  The intensities
are normalized to the maximum of each band.  Calculations are done
using $\epsilon_L=2.4$~eV, $\gamma^{tot}=84$~meV, and hopping defects
with $\alpha_{hopp}=6.4\times10^{13}$~eV$^2$cm$^{-2}$.}
\label{maps}
\end{figure}

\begin{figure}[ht]
\includegraphics[width=7.5cm]{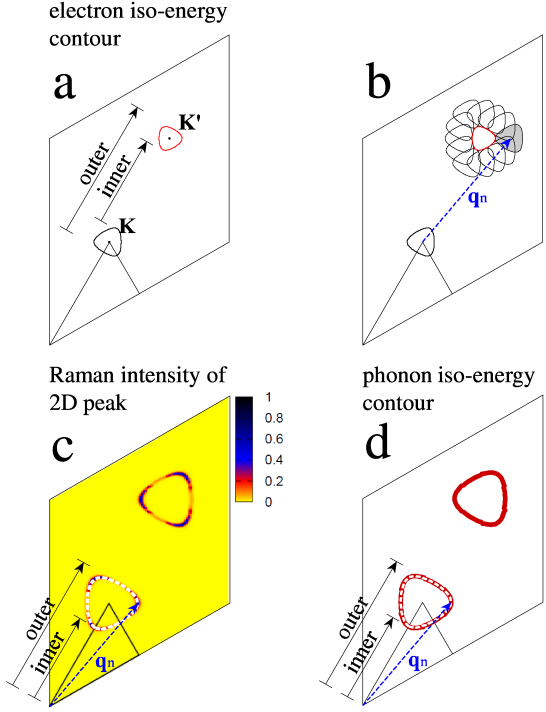}
\caption{(Color online) Electron and phonon
states relevant for the $2D$ line.  The rhombi are the graphene
Brillouin zone.  a) The triangularly distorted
contour around {\bf K} is obtained from
$\epsilon^{\pi^*}_{\bf k}-\epsilon^\pi_{\bf k}=2.4$~eV and represents
the electronic states near {\bf K} that are excited by a laser with
energy $\epsilon_L=2.4$~eV.  The contour around {\bf K}$'=2{\bf K}$ is
obtained from $\epsilon^{\pi^*}_{\bf k}-\epsilon^\pi_{\bf k}=2.06$~eV
and represents the electronic states near {\bf K}$'$ that are
deexcited by the emission of a quantum of light with energy
$\epsilon_L-2\omega_{ph}$~eV, with $\omega_{ph}=1354$~cm$^{-1}$ (half
the energy of the $2D$ line for $\epsilon_L=2.4$~eV).
b) {\bf q}$_n$ is one of the vectors such that the contour near {\bf
K} translated by {\bf q}$_n$ is tangent to the contour near {\bf K}$'$.
c) ${\cal I}_{\bf q}$ decomposition of the $2D$ intensity
(same figure as the $2D$ panel in Fig.~\ref{maps}).
The dashed closed line is defined by the ensemble of the {\bf q}$_n$ vectors.
d) The dashed line is the same as in c).
The
thick grey (red) line is the phonon iso-energy contour obtained from
$\omega_{\bf q}^\nu=1354$~cm$^{-1}$.  
The relevant phonon branch, thick grey line in
Fig.~\ref{fig2}, is disentangled form the other branches as in
Fig.2 of Ref.~\cite{gruneis09}.  
Notice that the iso-energy contours of electron states
(panels a, b) and phonons (panel c, d) have opposite trigonal warpings.
Notice also that phonon iso-energy contours in Fig.2 of
Ref.~\cite{gruneis09} are plotted with respect to the {\bf K}$'$ of
the present notation.
}
\label{fig_iso}
\end{figure}

\begin{figure}[ht]
\includegraphics[width=7.5cm]{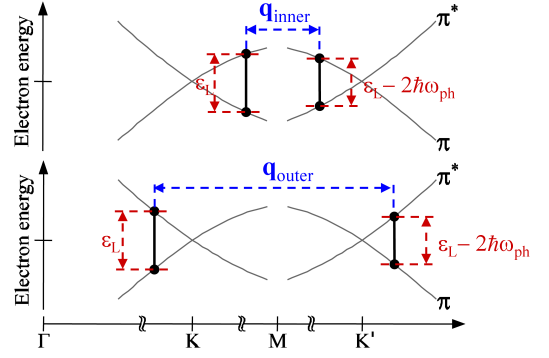}
\caption{(Color online)
Scheme of the double resonant process associated to the $2D$ line.
The momenta of the phonons mostly involved are indicated
as ``inner'' and ``outer''.}
\label{fig_inout}
\end{figure}

\begin{figure}[ht]
\includegraphics[width=8.0cm]{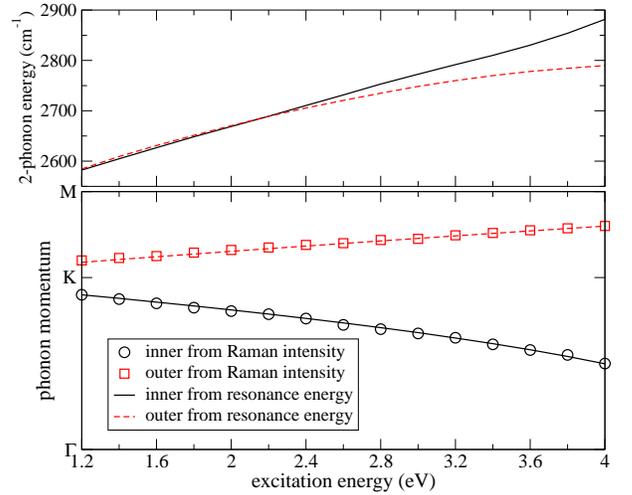}
\caption{(Color online) 
Lower panel: momenta of the inner and outer high symmetry phonons
which mostly contribute to the $2D$ band.
The lines are obtained
from the vectors connecting the isoenergy electronic contours
corresponding to that excitation energy.
For example, the values for $\epsilon_L=2.4$~eV are the 
moduli of the ``inner'' and ``outer'' vectors 
reported in Fig~\ref{fig_iso}.
The symbols are obtained from the maximum intensity in 
the ${\cal I}_{\bf q}$ plots (as those in Fig.~\ref{maps}
or in the left panels of Fig.~\ref{fig_2Dpm})
corresponding to that excitation energy.
Upper panel: frequency of the ``inner'' and ``outer'' phonons
reported in the lower panel.}
\label{fig_qk}
\end{figure}

We now discuss which phonons are responsible for the lines
presented in Figs.~\ref{fig3} and ~\ref{2D}.  In Fig.~\ref{maps}, we
consider the most important Raman lines and we decompose the Raman
intensity of a given band into its components associated to phonons
with a given wavevector {\bf q}.  For the defect-induced bands, $D$,
$D'$ and $D''$, we plot ${\cal I}_{\bf q}=\sum^*_\nu I^{pd}_{{\bf
q}\nu}$, and for the two-phonon bands, $2D$, $2D'$ and $D+D''$, we
plot ${\cal I}_{\bf q}=\sum^*_{\nu,\mu} I^{pp}_{{\bf q}\nu\mu}$, with
$I^{pd}_{{\bf q}\nu}$ and $I^{pp}_{{\bf q}\nu\mu}$ defined in Eq. 3
and the symbol $^*$ indicates that the summation is restricted to a
frequency window corresponding to a given Raman band (see
~\cite{windows}).  The {\bf q}-dependent intensity ${\cal I}_{\bf q}$
discloses which are the phonon wavevectors {\bf q} that mostly
contribute to a given Raman line.  The most remarkable result from
Fig.~\ref{maps} is that these phonons belong to limited regions of
the BZ consisting in very narrow (almost one-dimensional) lines.  
As expected, the $D$, $D''$,
$2D$ and $D+D''$ Raman bands originate from phonon {\bf q} wavevectors
belonging to a closed line around the {\bf K} and {\bf K'} high
symmetry points.  

In literature, the DR condition on the virtual transitions is
often used to determine the Raman-dominant phonon-wavevectors (see, e.g., 
\cite{thomsen00,saito01,kurti02,cancado02,ferrari,mafra}). To verify 
the validity of such a procedure,
we focus on the $2D$ line, which is
mostly due to $eh$ processes (Sec.~\ref{pr}) and consider an
excitation energy $\epsilon_L=2.4$~eV.  The DR consists in three
processes of excitation, phonon scattering, and recombination.  The
{\bf k} vectors of the electronic states which are excited by a laser
with energy $\epsilon_L$ form a triangularly-distorted
closed line, as the iso-energy contour surrounding the {\bf K} point in
Fig.~\ref{fig_iso}a.  The states involved in the emission of a quantum
of light with energy $\epsilon_L-2\hbar\omega_{\bf q}^\nu$
(recombination) form a second triangularly-distorted
closed line, as the iso-energy contour surrounding the {\bf K'} point in
Fig.~\ref{fig_iso}a.  These iso-energy contours are expected to give
the important contribution to the DR, although the energy is not
conserved in the intermediate virtual transitions.  The intermediate
DR processes are associated to a phonon {\bf q} and the important
processes are expected to be those associated to {\bf q} vectors that
connect the two triangles of Fig.~\ref{fig_iso}a. 
In particular, let us translate the {\bf K} triangle by {\bf q} and let 
us consider the
nesting vectors ({\bf q}$_n$) for which the {\bf K} triangle becomes
tangent to the ${\bf K'}$ one, as in Fig. ~\ref{fig_iso}b.
These phonon-wavevectors are expected to dominate the Raman spectra, since
for such nesting vectors there is a high density of electronic transitions satisfying the DR mechanism~\cite{cancado02,kurti02}.
The {\bf q}$_n$ vectors are shown in Fig.~\ref{fig_iso}c
as a dashed white line which is compared with the Raman intensity ${\cal
I}_{\bf q}$ from our most precise calculation (as in Fig.~\ref{maps}).
Within the scale of the figure, the nesting vectors reproduce very
well the maximum of the ${\cal I}_{\bf q}$, meaning that the simple
picture of Fig.~\ref{fig_iso}b provides a quantitative prediction of
the relevant phonon momenta.  

To generalize the analysis to an arbitrary laser excitation-energy,
we now consider, the isoenergy electronic contours as those of
Fig.~\ref{fig_iso}a for different values of $\epsilon_L$.  For each
$\epsilon_L$, we determined the phonon {\bf q}$_n$ vectors that are
nesting the corresponding contours. Among these points, we consider
only the vectors along high symmetry lines. In this case the
nesting vectors, ${\bf q}_{\rm inner}$ and ${\bf q}_{\rm outer}$, can be easily extracted from the one-dimensional
electronic-band dispersion along the high symmetry line,
as schematically shown in Fig.~\ref{fig_inout}. 
In the lower panel of Fig.~\ref{fig_qk} we report ${\bf q}_{\rm inner}$ and ${\bf q}_{\rm outer}$
obtained by the DR condition of Fig.~\ref{fig_inout} as a function of $\epsilon_L$.
In Fig.~\ref{fig_qk}, we also report the corresponding vectors obtained
by finding the maximum intensity in the ${\cal I}_{\bf q}$ plots (as
those in Fig.~\ref{maps}) corresponding to that excitation energy.
The sets of {\bf q} vectors obtained with these two different
procedures nicely coincide.  

\begin{figure}[ht]
\includegraphics[width=8.5cm]{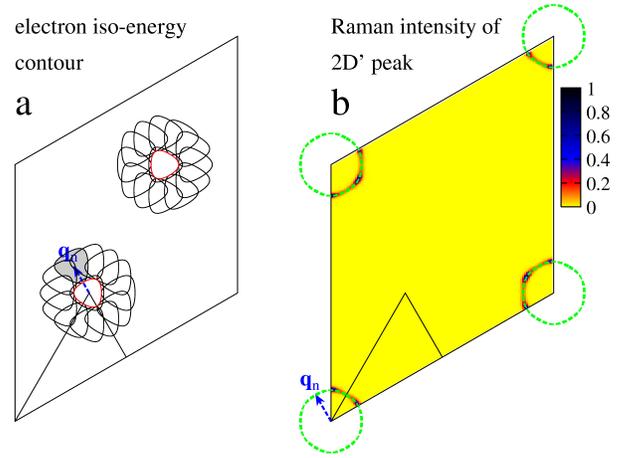}
\caption{
(Color online) Electron and phonon states relevant for the $2D'$ line.
a) The triangularly distorted contours around {\bf K} are obtained from
$\epsilon^{\pi^*}_{\bf k}-\epsilon^\pi_{\bf k}=2.4$~eV
and $\epsilon^{\pi^*}_{\bf k}-\epsilon^\pi_{\bf k}=2.0$~eV.
They represents the electronic states that are excited by a laser with
energy $\epsilon_L=2.4$~eV and those that are
deexcited by the emission of a quantum of light with energy
$\epsilon_L-2\omega_{ph}$~eV, with $\omega_{ph}=1602$~cm$^{-1}$ (half
the energy of the $2D'$ line for $\epsilon_L=2.4$~eV).
{\bf q}$_n$ is one of the vectors such that the excited-states contour
translated by {\bf q}$_n$ is tangent to the second contour.
The analogous construction arond {\bf K'} is also shown.
b) ${\cal I}_{\bf q}$ decomposition of the $2D'$ intensity
(same figure as the $2D'$ panel in Fig.~\ref{maps}).
The two dashed (green) closed lines (almost indistinguishable in the scale of the figure)
are defined by the ensemble of the nesting {\bf q}$_n$ vectors
among {\bf K} states or among {\bf K'} states.
}
\label{fig_isonew}
\end{figure}

We remark that the simplified scheme of Figs.
~\ref{fig_iso}b and ~\ref{fig_inout}
is used
for the $2D$ line, and that its validity comes ``a posteriori''
after the comparison with our most precise calculations.
The analogous construction for the $2D'$ line works equally well,
as can be seen in Fig.~\ref{fig_isonew}b,
by comparing the nesting vector profile (dashed line)
with the ${\cal I}_{\bf q}$ decomposition of the $2D'$ intensity.

\subsubsection{\label{angularsub} Dominant directions of the Raman phonon-wavevectors}

\begin{figure}[ht]
\includegraphics[width=7.5cm]{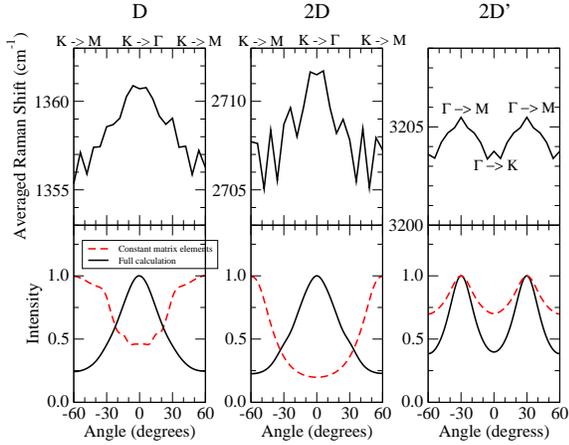}
\caption{(Color online)
Angular dependence of the intensity (lower panels) and of the weighted
average Raman shift (upper panels) for the $D$, $2D$ and $2D'$ bands.
The angles are measured taking the horizontal direction in
Fig.~\ref{maps} as reference.  Thus, for the $D$ and $2D$ bands, zero
degrees is the {\bf K}$\rightarrow$${\bm \Gamma}$ direction in the BZ,
while $\pm$60 degrees are the {\bf K}$\rightarrow${\bf M} one.  For
the $2D'$ band, zero degrees is the ${\bm \Gamma}$$\rightarrow${\bf K}
direction, while, $\pm$30 degrees are the ${\bm
\Gamma}$$\rightarrow${\bf M} direction.  In the lower panels, the
solid lines correspond to our most precise calculation.  Dashed lines
correspond to an approximated simulations in which the electron-light,
electron-phonon, and electron-defect scattering matrix elements are
kept constant (see the text).  Calculations are done using
$\epsilon_L=2.4$~eV, $\gamma^{tot}=84$~meV, and hopping defects with
$\alpha_{hopp}=6.4\times10^{13}$~eV$^2$cm$^{-2}$.
 }
\label{angle}
\end{figure}

A close look at Fig.~\ref{maps}
reveals that the most intense contributions of $D$, $D''$,
$2D$ and $D+D''$  are due to {\bf q} points
along the high symmetry directions {\bf K}$\rightarrow$${\bm \Gamma}$
and {\bf K'}$\rightarrow$${\bm \Gamma}$.  The $D'$
and $2D'$ bands originate from a closed line around ${\bm \Gamma}$ and
the most intense contributions are due to {\bf q} points along the
high symmetry ${\bm \Gamma}$$\rightarrow${\bf M} direction.

To analyze the results we consider the following definitions.  The
intensities ${\cal I}_{\bf q}$ (Fig.~\ref{maps}) form,
basically, a closed profile surrounding one high symmetry point ({\bf
K} for the $D$ and $2D$ lines, and ${\bm \Gamma}$ for the $2D'$).
Taking the high symmetry point as the reference, we consider how the
intensity of a given Raman band varies as a function of the direction
${\hat q}$ of the vector {\bf q}. Thus, in the lower panel of
Fig.~\ref{angle} we plot ${\cal I}_{\hat q}=\int^{\overline q}_0 qdq
{\cal I}_{\bf q}$, where the integral is done in a region containing
the most intense contribution.  It is also interesting to consider the
intensity weighted average phonon frequency associated to a given
Raman band and to a given {\bf q} point, $\langle\omega_{\bf
q}\rangle$.  As example, for the two-phonon lines $\langle\omega_{\bf
q}\rangle=[\sum^*_{\nu,\mu} I^{pp}_{{\bf q}\nu\mu} (\omega_{\bf
q}^\nu+\omega_{\bf q}^{\mu})]/[\sum^*_{\nu,\mu} I^{pp}_{{\bf
q}\nu\mu}]$, where the summation is restricted to the corresponding
frequency window~\cite{windows}.  This
quantity, basically, gives the frequency of the phonons associated to
that Raman band.  In analogy to ${\cal I}_{\hat q}$, we define
$\langle\omega_{\hat q}\rangle$ as the average of $\langle\omega_{\bf
q}\rangle$ along a direction ${\hat q}$ of the vector {\bf q}. Here,
also, the origin of ${\hat q}$ is {\bf K} for the $D$ and $2D$ lines,
and ${\bm \Gamma}$ for the $2D'$.  Fig. ~\ref{angle} shows the angular
dependence of the averaged phonon frequency $\langle\omega_{\hat q}\rangle$ for
the $D$, $2D$, and $2D'$ lines (actually, the shifts in the upper
panel of Fig. ~\ref{angle} are obtained after an average on a small
angle interval from $\theta - \Delta\theta$ to $\theta + \Delta\theta$).

Let us consider the $D$ and $2D$ bands.  From Fig.~\ref{angle}, the
phonons along the {\bf K}$\rightarrow$${\bm \Gamma}$ directions (in
literature these are usually called ``inner'' phonons, Fig.~\ref{fig_inout})
provide a
contribution which is almost four times higher than the one from the
{\bf K}$\rightarrow${\bf M} ones (``outer''
phonons).  Contrary to the present findings, in literature it is usually
assumed~\cite{thomsen00,kurti02,ferrari} that the phonons which mostly
contribute to the $D$ and $2D$ lines are outer phonons (along {\bf
K}$\rightarrow${\bf M}).
Only very recently some authors have
outlined the possible importance of the inner phonons ({\bf
K}$\rightarrow$${\bm \Gamma}$) ~\cite{daniela,mohr,huang10,frank11,yoon11}.
The present finding is
counter-intuitive and stems from the complex behavior of the scattering
matrix elements in the numerators of Eq.~\ref{eq1}.  
To understand this point, in Fig.~\ref{angle} we show the
results of calculations in which the numerators in Eq.~\ref{eq1} are
taken as a constant (that is, independent form {\bf k} and {\bf q} as,
e.g., in Eqs.~\ref{eq4},~\ref{eq5}).  Within this simplified approach
(which completely neglects, for example, the dependence
electron-phonon scattering matrix elements on {\bf q})
the outer phonons become dominant (in Fig.~\ref{angle}, lower panel,
the intensity has the maximum along the {\bf K}$\rightarrow${\bf M}
direction for both $D$ and $2D$), 
in agreement with the simplified models previously used in literature,
but in disagreement with our most precise calculations.
Concluding, inner processes are dominant for both $D$ and $2D$
lines. A proper description of the electronic
scattering matrix elements (in particular of the electron-phonon
coupling) is crucial to obtain this result.

\subsubsection{The width of the Raman bands}
\label{sec_2Dwidth}

\begin{figure}[ht]
\includegraphics[width=8.0cm]{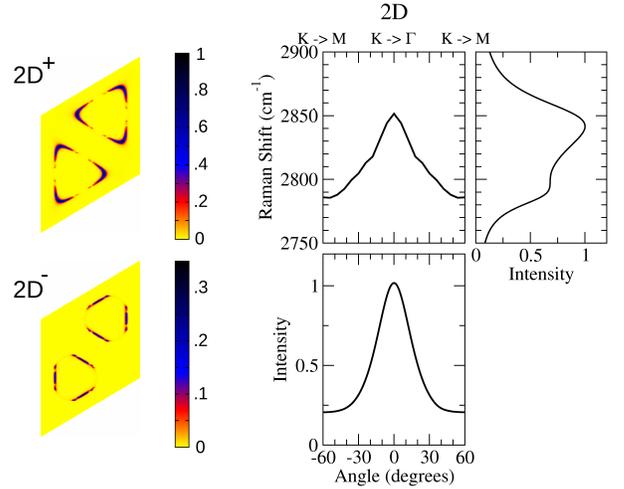}
\caption{(Color online) 
Calculated $2D$ line for the excitation $\epsilon_L=3.8$~eV
and $\gamma^{tot}=170$~meV.
Top right panel: intensity vs. Raman shift.
The line appears as a broad band with two maxima near
2790~cm$^{-1}$ ($2D^-$) and 2840~cm$^{-1}$ ($2D^+$).
Left panels:
mapping of the Raman intensity in the Brillouin zone
(as in Fig.~\ref{maps})
of the two components $2D^-$ $2D^+$ obtained by integrating in the
corresponding frequency windows~\cite{windows}.
Central panels:
angular dependence of the weighted average Raman shift and of the
intensity, as in Fig.~\ref{angle}
}
\label{fig_2Dpm}
\end{figure}

One of the most interesting feature of the simulated Raman spectra of
Figs.~\ref{fig3} and ~\ref{2D} is the narrow width of the bands, which
reproduces the measured spectra.  The narrow width of the $D$ and $2D$
lines is indeed surprising since already at
$\epsilon_L=2.4$~eV the electronic states involved in the Raman
process display an important trigonal warping (i.e. the electron
isoenergy contour are triangularly distorted as in
Fig.~\ref{fig_iso}a).  In the presence of trigonal warping one should
expect the excited phonons to have energies distributed in a broad
range.  Indeed, previous calculations\cite{kurti02,park} did not
reproduce narrow lineshape of the DR lines.  The present improved
description of the electronic scattering matrix elements partially
explains such narrow lines.
The most important role is played by the phonon energy dispersions.
The upper panels of Fig.~\ref{angle} show that,
for the $D$, $2D$ and $2D'$ lines at $\epsilon_L=2.4$~eV, 
the excited phonons have almost the same energy (within $\sim$5 cm$^{-1}$),
despite the strong electron trigonal warping.
This fact explains the small width of the DR Raman lines and it is
due to the details of the phonon dispersion we used.
Indeed, with a reasonable description of the electronic trigonal warping
and using a rough description of the phonon energies,
larger dispersions in frequencies (and broader Raman lines) are found~\cite{park}.
Ref.~\cite{gruneis09} has clearly demonstrated that the phonon trigonal
warping is important and that it is opposite to the electronic one.
The present results show that,
as already argued in Ref.~\cite{gruneis09}, the interplay between the
electronic and phononic trigonal warping provides a sort of
cancellation.  This results in the small dispersion of the phonon
frequencies of the upper panel of Fig.~\ref{angle} and, consequently,
in the small width of the associated Raman lines.

To illustrate the concept of trigonal warpings cancellation,
Fig.~\ref{fig_iso}d compares the line of the nesting vectors {\bf q}$_n$
(white dashed line, see Fig.~\ref{fig_iso}b and Sec.~\ref{phon}) 
with the iso-energy contour of the
phonons having half the energy of the $2D$ (thick red lines).  
The two lines nicely
resemble each other, meaning that all the nesting phonons have nearly
the same energy and, as a consequence, the $2D$ line width is small.
If the phonon
isoenergy contour was different, the two lines would not superimpose
and the $2D$ line would have a broader shape.
The perfect cancellation of electronic and phononic trigonal warping
breaks down for laser energy in the UV range.
Indeed in the upper panel of 
Fig.~\ref{fig_qk}, we report, as a function of $\epsilon_L$, 
the frequency associated with the inner and outer phonons.
At $\epsilon_L=2.4$~eV, the frequencies associated to inner
and outer phonons almost coincide. However, for higher $\epsilon_L$,
the two frequencies become different, meaning that for a sufficiently
high $\epsilon_L$ the $2D$ line is expected to become broader.

Indeed, according to our most precise calculations, at
$\epsilon_L=3.8$~eV the Raman $2D$ band appears much broader than the
one at $\epsilon_L=2.4$~eV and displays two maxima at 2790 cm$^{-1}$
and 2840 cm$^{-1}$ (Fig.~\ref{spectra}).  
At $\epsilon_L=3.8$~eV (Fig.~\ref{fig_2Dpm})  the angular dependence
of the average frequency shift is more dispersive than in the
$\epsilon_L=2.4$~eV case. The inner phonons correspond to the highest
frequency components, $2D^+$ at $\sim$2840 cm$^{-1}$, and the outer
phonons to the lowest one, $2D^-$ at $\sim$2790 cm$^{-1}$.  In
Fig.~\ref{fig_2Dpm} we also show the {\bf q} vectors decomposition of
the intensities of the $2D^+$ and $2D^-$ components.  For the $2D^+$,
the shape is triangularly distorted and the maximum corresponds to the
inner phonons, while for the $2D^-$ the maximum corresponds to the
outer phonons.

\section{Conclusions}
\label{sec_conclusions}

We calculated the double resonant Raman spectrum of graphene with a
computational method which tries to overcome the most common
approximations used in literature.  Calculations are done using the
standard approach based on the golden rule generalized to the fourth
perturbative order~\cite{thomsen00} (Eq.~\ref{eq1}).  We determined the
Raman lines associated to both phonon-defect processes
(defect-induced excitations of {\bf q}$\neq${\bf 0} phonons, such as
in the $D$, $D'$, and $D''$ Raman lines) and two-phonons processes
(excitations in a defect-free sample of a {\bf -q} and a {\bf q}
phonons, such as in the $2D$, $2D'$, or $D+D''$ lines).  The
lowest-order processes (excitation of a {\bf q=0} phonon, such in the
$G$ line) and higher-order processes (such as in the $D+D'$ line at
$\sim$2900~cm$^{-1}$, which is usually attributed to a defect-induced
excitation of two phonons $\bf q$ and $\bf q'$ with {\bf
q+q}$'$$\neq${\bf 0}) are not described by the present approach.

The electronic summation is performed all over the two dimensional
Brillouin zone and all the possible phonons (with any wavevector) are
considered.  Electronic bands are obtained from a 5-neighbors tight
binding (TB) approach in which the parameters are fitted to reproduce
ab-initio calculations based on density functional theory (DFT)
corrected with GW.  This procedure provides a Fermi velocity (the
slope of the Dirac cone) in good agreement with measurements and a
good description of the trigonal warping.  The resulting electron/hole
asymmetry is not negligible.  The phonon dispersion is obtained from
fully ab-initio DFT calculations corrected with GW.  This procedure is
necessary to obtain a good description of the slope of the phonon
branch associated with the $D$ and $2D$ lines, near {\bf K}.  The
electron-phonon, electron-light, and electron-defect scattering matrix
elements are obtained within the TB approach.  The defect-induced Raman
processes are simulated by considering three different kinds of model
defects: i) {\it on-site defects}, obtained by changing the on-site TB
parameter; ii) {\it hopping defects}, obtained by changing one of the
first-neighbors hopping TB parameters; iii) {\it Coulomb defects},
corresponding to charged impurities adsorbed at a given distance from
the graphene sheet, which interact with graphene through a Coulomb potential.

The electronic linewidth (the inverse of the electronic lifetime),
which turns out to be a very relevant parameter, is calculated explicitly
considering the contributions from electron-phonon and electron-impurity scattering.
To give an idea,
for $\epsilon_L=2.4$~eV, in the absence of defects and for zero doping,
the sum of the electron and hole linewidths is $\gamma^{tot}=84$~meV
(which is roughly two times the FWHM of the electron spectral function).

By looking at the overall shape of the typical Raman spectra, 
for an excitation energy of $\epsilon_L=2.4$~eV, the
agreement between calculations and measurements is very good.  In
particular, all the Raman lines observed experimentally, even the
small intensity ones, are present in the calculated spectra and the
relative intensities among two-phonon lines (such as $2D$, $2D'$, or
$D+D''$ lines) or among phonon-defect lines (such as the $D$ and the $D'$ lines)
are correctly reproduced (being the hopping defect the best model to
study defect-induced Raman processes).  The most remarkable agreement
between theory and measurements relates to the line widths. Indeed,
the present calculations reproduce very well the measured small widths
of the $D$, $D'$, $2D$ and $2D'$ lines.  Moreover, calculations
reproduce quite well the symmetric Lorentzian shapes of the $2D$ and
$2D'$ lines and the asymmetric shape of $D+D''$ band. We remark that,
in the present model, the only parameter used to fit Raman
measurements, $\alpha_{hopp}$, determines the ratio of the $D$
vs. $2D$ intensities but does not affect the relative intensities
among phonon-defect or among two-phonon lines, the width of the lines,
and their shape.

We determined how the Raman spectra change by changing the laser
excitation energy $\epsilon_L$ from 1.2 to 4.0 eV, which are
the energies mainly used experimentally.
All the visible lines change in position, intensity and shape.
In particular, the $2D$ line has a small-width
Lorentzian shape for $\epsilon_L\leq 2.4$~eV
and it is asymmetric and broader at $\epsilon_L=3.8$~eV.
The measured shift of the Raman line position as a function of $\epsilon_L$
is well reproduced for all the available measurements.
The calculated spectra also display some small intensity bands associated
to acoustic phonons. Some of them, such as the
$D'+D^3$ and the $D'+D^4$ (in the 1800, 2000 cm$^{-1}$ range)
are actually visible in the measured spectra~\cite{cong,rao11}.
Finally, for high energy excitations, e.g. $\epsilon_L=3.8$~eV,
the most intense Raman lines ($2D$ and $D$)
change shape and intensity as a function of the polarization of the light.
This finding is remarkable since it could lead to measurable effects.

We determined how the intensity of the main DR Raman lines is
affected by various parameters such as the electronic linewidth,
the excitation energy, and the defect concentration.
The absolute intensity of the double resonant Raman lines is strongly
affected by the actual value of the electronic linewidth, $\gamma^{tot}$.
In general, the intensity of a DR Raman line decreases when the electronic
linewidth increases (at fixed defect concentration) because
the electronic broadening tends to kill the double resonance condition.
According to the present findings, also the ratio of the intensities 
of the $2D$ and $2D'$ lines depends on $\gamma^{tot}$.
This result is particularly appealing since the measurement of this ratio
(followed by the comparison with the present calculations)  could be used to determine experimentally the electron/hole
linewidth $\gamma^{tot}$ and, in particular, its components due to
defects and/or to electron-electron scattering in doped samples.
We determined how the intensity ratio among various Raman lines
change as a function of the excitation energy of the laser.
In particular, we determined the evolution of
A($2D'$)/A($2D$), A($D+D''$)/A($2D$), A($D''$)/A($D$), and A($D'$)/A($D$)
[where A($X$) is the integrated area under the $X$ Raman line]
as a function of the excitation energy.
All these ratios considerably change in the range of excitation energies
available experimentally, however
measurements to compare with are not presently available.

We studied the dependence of the $D$ and $2D$ lines intensity on
the defect concentration, comparing to recent measurements~\cite{erlon,lucchese}.
We first remind that the electronic linewidth 
$\gamma^{tot}$ is given by the sum of an intrinsic component $\tilde\gamma^{(ep)}$
(due to the electron-phonon interaction) and an extrinsic defect-induced component
$\tilde\gamma^{(D)}$ which increases linearly by increasing the defect concentration.
The intensity of the $2D$ line
monotonously decreases by increasing the defect concentration $n_d$.
Indeed, the $2D$ line (which is a two-phonon process) depends on $n_d$
only through the electronic linewidth $\gamma^{tot}$, which, in turn, increases
by increasing $n_d$.
The intensity of the $D$ line has a non-monotonic behavior.
The $D$ line (which is a defect induced process) depends on $n_d$ through two
distinct mechanisms: first there is a proportionality factor between the Raman intensity
and $n_d$, second, the linewidth $\gamma^{tot}$ depends on $n_d$ as for the $2D$ line.
For small $n_d$, $\gamma^{tot}\sim\tilde\gamma^{(ep)}$ and the $D$ intensity increases linearly with $n_d$.
For high $n_d$, the dependence of $\gamma^{tot}$ on $n_d$
becomes the dominant mechanism, leading to a decrease of the intensity, as for the $2D$ line.
The maximum of the $D$ intensity is reached for the defect concentration
corresponding to the condition $\tilde\gamma^{(D)}\sim\tilde\gamma^{(ep)}$.

We have compared Raman spectra calculated with the three different
model defects. The intensity ratio
between the defect-induced $D$ and $D'$ lines
strongly depends on the kind of model defect,
suggesting that this ratio could possibly be tuned in actual
experiments by selecting special kind of impurities on the sample.
Charged impurities outside the graphene plane (Coulomb defects) could
be an important source of scattering during electronic transport.
However, according to the present calculations, they should provide an almost
undetectable contribution to the Raman signal, the $D$ line being
completely absent and the $D'$ having an intensity orders of magnitude
smaller than the $2D$ line.

Finally, the analysis of the results has focused on certain specific
issues currently debated.

Among the different possible DR processes, the electron-hole ones
(processes in which both electronic and hole states are involved in the
scattering, $ab$ in the text)
are responsible for most of the Raman intensity of both the $D$ and the $2D$ lines.
Several authors (e.g. ~\cite{thomsen00}) describe the DR by simply considering
electron-electron or hole-hole processes
(processes in which only electrons or only holes are involved in the
scattering, $aa$ in the text)
which, according to the present findings, give a negligible contribution to the Raman intensity.
The dominance of the electron-hole processes stems from the presence of a
destructive quantum interference that
kills the contribution of the electron-electron and hole-hole ones.
This conclusion is not due to the complex details of the present calculations
but can be deduced with a very simplified model, easy to implement.

The most intense contribution to both the $D$ and $2D$ lines
is due to phonons along the high symmetry directions
{\bf K}$\rightarrow{\bm \Gamma}$ ($inner$ phonons).
This is contrary to the common assumption
~\cite{thomsen00,kurti02,ferrari} that the phonons which mostly
contribute to the $D$ and $2D$ lines belong to the {\bf
K}$\rightarrow${\bf M} direction ($outer$ phonons).
The present result (the dominance of the inner phonons) is
counterintuitive and stems from the complex behavior of the electronic
scattering matrix elements in the numerator of the double resonance
scattering amplitude.

The observed small width of the $2D$ line at $\epsilon_L=2.4$~eV
is explained as a consequence of the interplay between the opposite
trigonal warpings of the electron and phonon dispersions:
the excited electronic states form a triangularly distorted
profile having vertex along the {\bf K}$\rightarrow${\bf M} direction, while
the phonon isoenergy contour is a triangularly distorted
profile having vertex along the {\bf K}$\rightarrow{\bm \Gamma}$ direction.
Because of this, the excited phonons (both the inner and the outer ones)
have almost the same energy and, as a consequence, the $2D$ line-width is small.
At higher excitation energies this condition is no more verified
and the $2D$ line becomes broader and asymmetric.
For instance at $\epsilon_L=3.8$~eV the calculated spectrum
displays two maxima corresponding to a main
component at $\sim$2840~cm$^{-1}$ (due to inner phonons) and to 
a less intense one at $\sim$2790~cm$^{-1}$ (due to outer phonons).

\section*{Acknowledgments}
We thank D. Basko and P. Gava for usefull discussions.
P. V. has received financial support from the
Conselho Nacional de Desenvolvimento Cient{\'\i}fico e Tecnol{\'o}gico (CNPq), Brazil.
Part of the calculations were performed at IDRIS (France), proj. 096128.

\appendix

\section{Raman double-resonant scattering amplitudes}
\label{app_DR}

Explicit expressions are now given for all the double resonant scattering
amplitudes $K^{pd}({\bf k},{\bf q},\nu)$ and
$K^{pp}({\bf k},{\bf q},\nu,\mu)$,
which have been included in the sums of Eq.~\ref{eq3}.
The following processes are described diagrammatically in Fig.~\ref{fig1}.
The arguments {\bf k}, {\bf q}, $\nu$, and $\mu$ are dropped for simplicity.
The sign $\pm$ before each $K$
is determined by the fermionic statistics
of the carriers.
The broadening energies $\gamma_{\bf k}$ in the denominators of
the DR scattering amplitudes $K$
are the sum of the broadenings of the corresponding electronic states
(see Sec.~\ref{sec_el-life}).
As examples, in $K^{pd}_{ee1}$
$\gamma^A_{\bf k} = \gamma^{\pi^*}_{\bf k} + \gamma^\pi_{\bf k}$,
$\gamma^B_{\bf k} = \gamma^{\pi^*}_{\bf k+q} + \gamma^\pi_{\bf k}$,
$\gamma^C_{\bf k} = \gamma^{\pi^*}_{\bf k} + \gamma^\pi_{\bf k}$.
In $K^{pd}_{eh1}$	   	 		     
$\gamma^A_{\bf k} = \gamma^{\pi^*}_{\bf k} + \gamma^\pi_{\bf k}$,
$\gamma^B_{\bf k} = \gamma^{\pi^*}_{\bf k+q} + \gamma^\pi_{\bf k}$,
$\gamma^C_{\bf k} = \gamma^{\pi^*}_{\bf k+q} + \gamma^\pi_{\bf k+q}$.

\begin{widetext}

\centerline{There are eight phonon-defect (pd) processes.}
Process $ee1$: the electron is first scattered by a phonon and then by a defect,
\[
K_{ee1}^{pd} =
\frac{
\langle {\bf k}    \pi   |D_{out}                 | {\bf k}    \pi^* \rangle
\langle {\bf k}    \pi^* |H_D                     | {\bf k+q}, \pi^* \rangle
\langle {\bf k+q}, \pi^* |\Delta H_{{\bf q},\nu} | {\bf k}    \pi^* \rangle
\langle {\bf k}    \pi^* |D_{in}                  | {\bf k}    \pi   \rangle
}
{
( \epsilon_L-\epsilon^{\pi^*}_{\bf k}   + \epsilon^{\pi}_{\bf k} - \hbar\omega_{{\bf -q}}^\nu
-i\frac{\gamma^C_{\bf k}}{2})
( \epsilon_L-\epsilon^{\pi^*}_{\bf k+q} + \epsilon^{\pi}_{\bf k} - \hbar\omega_{{\bf -q}}^\nu
-i\frac{\gamma^B_{\bf k}}{2})
( \epsilon_L-\epsilon^{\pi^*}_{\bf k}   + \epsilon^{\pi}_{\bf k}
-i\frac{\gamma^A_{\bf k}}{2})
}.
\]

Process $ee2$: the electron is first scattered by a defect and then by a phonon,
\[
K_{ee2}^{pd} =
\frac{
\langle {\bf k}    \pi   |D_{out}                 | {\bf k}    \pi^* \rangle
\langle {\bf k}    \pi^* |\Delta H_{{\bf q},\nu} | {\bf k-q}, \pi^* \rangle
\langle {\bf k-q}, \pi^* |H_D                     | {\bf k}    \pi^* \rangle
\langle {\bf k}    \pi^* |D_{in}                  | {\bf k}    \pi   \rangle
}
{
( \epsilon_L-\epsilon^{\pi^*}_{\bf k}   + \epsilon^{\pi}_{\bf k} - \hbar\omega_{{\bf -q}}^\nu
-i\frac{\gamma^C_{\bf k}}{2})
( \epsilon_L-\epsilon^{\pi^*}_{\bf k-q} + \epsilon^{\pi}_{\bf k}
-i\frac{\gamma^B_{\bf k}}{2})
( \epsilon_L-\epsilon^{\pi^*}_{\bf k}   + \epsilon^{\pi}_{\bf k}
-i\frac{\gamma^A_{\bf k}}{2})
}.
\]

Process $hh1$: the hole is first scattered by a phonon and then by a defect,
\[
K_{hh1}^{pd} =
\frac{
\langle {\bf k}    \pi   |D_{out}                 | {\bf k}    \pi^* \rangle
\langle {\bf k-q}, \pi   |H_D                     | {\bf k}    \pi   \rangle
\langle {\bf k}    \pi   |\Delta H_{{\bf q},\nu} | {\bf k-q}, \pi   \rangle
\langle {\bf k}    \pi^* |D_{in}                  | {\bf k}    \pi   \rangle
}
{
( \epsilon_L-\epsilon^{\pi^*}_{\bf k}   + \epsilon^{\pi}_{\bf k} - \hbar\omega_{{\bf -q}}^\nu
-i\frac{\gamma^C_{\bf k}}{2})
( \epsilon_L-\epsilon^{\pi^*}_{\bf k}   + \epsilon^{\pi}_{\bf k-q} - \hbar\omega_{{\bf -q}}^\nu
-i\frac{\gamma^B_{\bf k}}{2})
( \epsilon_L-\epsilon^{\pi^*}_{\bf k}   + \epsilon^{\pi}_{\bf k}
-i\frac{\gamma^A_{\bf k}}{2})
}.
\]

Process $hh2$: the hole is first scattered by a defect and then by a phonon,
\[
K_{hh2}^{pd} =
\frac{
\langle {\bf k}    \pi   |D_{out}                 | {\bf k}    \pi^* \rangle
\langle {\bf k+q}, \pi   |\Delta H_{{\bf q},\nu} | {\bf k}    \pi   \rangle
\langle {\bf k}    \pi   |H_D                     | {\bf k+q}, \pi   \rangle
\langle {\bf k}    \pi^* |D_{in}                  | {\bf k}    \pi   \rangle
}
{
( \epsilon_L-\epsilon^{\pi^*}_{\bf k}   + \epsilon^{\pi}_{\bf k} - \hbar\omega_{{\bf -q}}^\nu
-i\frac{\gamma^C_{\bf k}}{2})
( \epsilon_L-\epsilon^{\pi^*}_{\bf k}   + \epsilon^{\pi}_{\bf k+q}
-i\frac{\gamma^B_{\bf k}}{2})
( \epsilon_L-\epsilon^{\pi^*}_{\bf k}   + \epsilon^{\pi}_{\bf k}
-i\frac{\gamma^A_{\bf k}}{2})
}.
\]

Process $eh1$: first the electron is scattered by a phonon and then the hole by a defect,
\[
K_{eh1}^{pd} = -
\frac{
\langle {\bf k+q}, \pi   |D_{out}                 | {\bf k+q}, \pi^* \rangle
\langle {\bf k}    \pi   |H_D                     | {\bf k+q}, \pi   \rangle
\langle {\bf k+q}, \pi^* |\Delta H_{{\bf q},\nu} | {\bf k}    \pi^* \rangle
\langle {\bf k}    \pi^* |D_{in}                  | {\bf k}    \pi   \rangle
}
{
( \epsilon_L-\epsilon^{\pi^*}_{\bf k+q} + \epsilon^{\pi}_{\bf k+q} - \hbar\omega_{{\bf -q}}^\nu
-i\frac{\gamma^C_{\bf k}}{2})
( \epsilon_L-\epsilon^{\pi^*}_{\bf k+q} + \epsilon^{\pi}_{\bf k} - \hbar\omega_{{\bf -q}}^\nu
-i\frac{\gamma^B_{\bf k}}{2})
( \epsilon_L-\epsilon^{\pi^*}_{\bf k}   + \epsilon^{\pi}_{\bf k}
-i\frac{\gamma^A_{\bf k}}{2})
}.
\]

Process $eh2$: first the electron is scattered by a defect and then the hole by a phonon,
\[
K_{eh2}^{pd} = -
\frac{
\langle {\bf k-q}, \pi   |D_{out}                 | {\bf k-q}, \pi^* \rangle
\langle {\bf k}    \pi   |\Delta H_{{\bf q},\nu} | {\bf k-q}, \pi   \rangle
\langle {\bf k-q}, \pi^* |H_D                     | {\bf k}    \pi^* \rangle
\langle {\bf k}    \pi^* |D_{in}                  | {\bf k}    \pi   \rangle
}
{
( \epsilon_L-\epsilon^{\pi^*}_{\bf k-q} + \epsilon^{\pi}_{\bf k-q} - \hbar\omega_{{\bf -q}}^\nu
-i\frac{\gamma^C_{\bf k}}{2})
( \epsilon_L-\epsilon^{\pi^*}_{\bf k-q} + \epsilon^{\pi}_{\bf k}
-i\frac{\gamma^B_{\bf k}}{2})
( \epsilon_L-\epsilon^{\pi^*}_{\bf k}   + \epsilon^{\pi}_{\bf k}
-i\frac{\gamma^A_{\bf k}}{2})
}.
\]

Process $he1$: first the hole is scattered by a phonon and then the electron by a defect,
\[
K_{he1}^{pd} = -
\frac{
\langle {\bf k-q}, \pi   |D_{out}                 | {\bf k-q}, \pi^* \rangle
\langle {\bf k-q}, \pi^* |H_D                     | {\bf k}    \pi^* \rangle
\langle {\bf k}    \pi   |\Delta H_{{\bf q},\nu} | {\bf k-q}, \pi   \rangle
\langle {\bf k}    \pi^* |D_{in}                  | {\bf k}    \pi   \rangle
}
{
( \epsilon_L-\epsilon^{\pi^*}_{\bf k-q} + \epsilon^{\pi}_{\bf k-q} - \hbar\omega_{{\bf -q}}^\nu
-i\frac{\gamma^C_{\bf k}}{2})
( \epsilon_L-\epsilon^{\pi^*}_{\bf k}   + \epsilon^{\pi}_{\bf k-q} - \hbar\omega_{{\bf -q}}^\nu
-i\frac{\gamma^B_{\bf k}}{2})
( \epsilon_L-\epsilon^{\pi^*}_{\bf k}   + \epsilon^{\pi}_{\bf k}
-i\frac{\gamma^Q_{\bf k}}{2})
}.
\]

Process $he2$: first the hole is scattered by a defect and then the electron by a phonon,
\[
K_{he2}^{pd} = -
\frac{
\langle {\bf k+q}, \pi   |D_{out}                 | {\bf k+q}, \pi^* \rangle
\langle {\bf k+q}, \pi^* |\Delta H_{{\bf q},\nu} | {\bf k}    \pi^* \rangle
\langle {\bf k}    \pi   |H_D                     | {\bf k+q}, \pi   \rangle
\langle {\bf k}    \pi^* |D_{in}                  | {\bf k}    \pi   \rangle
}
{
( \epsilon_L-\epsilon^{\pi^*}_{\bf k+q} + \epsilon^{\pi}_{\bf k+q} - \hbar\omega_{{\bf -q}}^\nu
-i\frac{\gamma^C_{\bf k}}{2})
( \epsilon_L-\epsilon^{\pi^*}_{\bf k}   + \epsilon^{\pi}_{\bf k+q}
-i\frac{\gamma^B_{\bf k}}{2})
( \epsilon_L-\epsilon^{\pi^*}_{\bf k}   + \epsilon^{\pi}_{\bf k}
-i\frac{\gamma^A_{\bf k}}{2})
}.
\]

\centerline {There are eight two-phonon (pp) processes.}
Process $ee1$: the electron is first scattered by the {\bf -q}$\nu$ phonon and then by the {\bf q}$\mu$ one,
\[
K_{ee1}^{pp} = 
\frac{
\langle {\bf k}    \pi   |D_{out}                  | {\bf k}    \pi^* \rangle
\langle {\bf k}    \pi^* |\Delta H_{{\bf -q},\mu} | {\bf k+q}, \pi^* \rangle
\langle {\bf k+q}, \pi^* |\Delta H_{{\bf q},\nu}  | {\bf k}    \pi^* \rangle
\langle {\bf k}    \pi^* |D_{in}                   | {\bf k}    \pi   \rangle
}
{
( \epsilon_L - \epsilon^{\pi^*}_{\bf k}   + \epsilon^{\pi}_{\bf k} - \hbar\omega_{{\bf -q}}^\nu - \hbar\omega_{{\bf q}}^\mu
-i\frac{\gamma^C_{\bf k}}{2})
( \epsilon_L - \epsilon^{\pi^*}_{\bf k+q} + \epsilon^{\pi}_{\bf k} - \hbar\omega_{{\bf -q}}^\nu
-i\frac{\gamma^B_{\bf k}}{2})
( \epsilon_L - \epsilon^{\pi^*}_{\bf k}   + \epsilon^{\pi}_{\bf k}
-i\frac{\gamma^A_{\bf k}}{2})
}.
\]

Process $ee2$: the electron is first scattered by the {\bf q}$\mu$ phonon and then by the {\bf -q}$\nu$ one,
\[
K_{ee2}^{pp} = 
\frac{
\langle {\bf k}    \pi   |D_{out}                  | {\bf k}    \pi^* \rangle
\langle {\bf k}    \pi^* |\Delta H_{{\bf q},\nu}  | {\bf k-q}, \pi^* \rangle
\langle {\bf k-q}, \pi^* |\Delta H_{{\bf -q},\mu} | {\bf k}    \pi^* \rangle
\langle {\bf k}    \pi^* |D_{in}                   | {\bf k}    \pi   \rangle
}
{
( \epsilon_L - \epsilon^{\pi^*}_{\bf k}   + \epsilon^{\pi}_{\bf k} - \hbar\omega_{{\bf -q}}^\nu - \hbar\omega_{{\bf q}}^\mu
-i\frac{\gamma^C_{\bf k}}{2})
( \epsilon_L - \epsilon^{\pi^*}_{\bf k-q} + \epsilon^{\pi}_{\bf k} - \hbar\omega_{{\bf q}}^\mu
-i\frac{\gamma^B_{\bf k}}{2})
( \epsilon_L - \epsilon^{\pi^*}_{\bf k}   + \epsilon^{\pi}_{\bf k}
-i\frac{\gamma^A_{\bf k}}{2})
}.
\]

Process $hh1$: the hole is first scattered by the {\bf -q}$\nu$ phonon and then by the {\bf q}$\mu$ one,
\[
K_{hh1}^{pp} = 
\frac{
\langle {\bf k}    \pi   |D_{out}                  | {\bf k}    \pi^* \rangle
\langle {\bf k-q}, \pi   |\Delta H_{{\bf -q},\mu} | {\bf k}    \pi   \rangle
\langle {\bf k}    \pi   |\Delta H_{{\bf q},\nu}  | {\bf k-q}, \pi   \rangle
\langle {\bf k}    \pi^* |D_{in}                   | {\bf k}    \pi   \rangle
}
{
( \epsilon_L - \epsilon^{\pi^*}_{\bf k}   + \epsilon^{\pi}_{\bf k} - \hbar\omega_{{\bf -q}}^\nu - \hbar\omega_{{\bf q}}^\mu
-i\frac{\gamma^C_{\bf k}}{2})
( \epsilon_L - \epsilon^{\pi^*}_{\bf k}   + \epsilon^{\pi}_{\bf k-q} - \hbar\omega_{{\bf -q}}^\nu
-i\frac{\gamma^B_{\bf k}}{2})
( \epsilon_L - \epsilon^{\pi^*}_{\bf k}   + \epsilon^{\pi}_{\bf k}
-i\frac{\gamma^A_{\bf k}}{2})
}.
\]

Process $hh2$: the hole is first scattered by the {\bf q}$\mu$ phonon and then by the {\bf -q}$\nu$ one,
\[
K_{hh2}^{pp} = 
\frac{
\langle {\bf k}    \pi   |D_{out}                  | {\bf k}    \pi^* \rangle
\langle {\bf k+q}, \pi   |\Delta H_{{\bf q},\nu}  | {\bf k}    \pi   \rangle
\langle {\bf k}    \pi   |\Delta H_{{\bf -q},\mu} | {\bf k+q}, \pi   \rangle
\langle {\bf k}    \pi^* |D_{in}                   | {\bf k}    \pi   \rangle
}
{
( \epsilon_L - \epsilon^{\pi^*}_{\bf k}   + \epsilon^{\pi}_{\bf k} - \hbar\omega_{{\bf -q}}^\nu - \hbar\omega_{{\bf q}}^\mu
-i\frac{\gamma^C_{\bf k}}{2})
( \epsilon_L - \epsilon^{\pi^*}_{\bf k}   + \epsilon^{\pi}_{\bf k+q} - \hbar\omega_{{\bf q}}^\mu
-i\frac{\gamma^B_{\bf k}}{2})
( \epsilon_L - \epsilon^{\pi^*}_{\bf k}   + \epsilon^{\pi}_{\bf k}
-i\frac{\gamma^A_{\bf k}}{2})
}.
\]

Process $eh1$: first the electron is scattered by the {\bf -q}$\nu$ phonon and then the hole by the {\bf q}$\mu$ one,
\[
K_{eh1}^{pp} = -
\frac{
\langle {\bf k+q}, \pi   |D_{out}                  | {\bf k+q}, \pi^* \rangle
\langle {\bf k}    \pi   |\Delta H_{{\bf -q},\mu} | {\bf k+q}, \pi   \rangle
\langle {\bf k+q}, \pi^* |\Delta H_{{\bf q},\nu}  | {\bf k}    \pi^* \rangle
\langle {\bf k}    \pi^* |D_{in}                   | {\bf k}    \pi   \rangle
}
{
( \epsilon_L - \epsilon^{\pi^*}_{\bf k+q} + \epsilon^{\pi}_{\bf k+q} - \hbar\omega_{{\bf -q}}^\nu - \hbar\omega_{{\bf q}}^\mu
-i\frac{\gamma^C_{\bf k}}{2})
( \epsilon_L - \epsilon^{\pi^*}_{\bf k+q} + \epsilon^{\pi}_{\bf k}   - \hbar\omega_{{\bf -q}}^\nu
-i\frac{\gamma^B_{\bf k}}{2})
( \epsilon_L - \epsilon^{\pi^*}_{\bf k}   + \epsilon^{\pi}_{\bf k}
-i\frac{\gamma^A_{\bf k}}{2})
}.
\]

Process $eh2$: first the electron is scattered by the {\bf q}$\mu$ phonon and then the hole by the {\bf -q}$\nu$ one,
\[
K_{eh2}^{pp} = -
\frac{
\langle {\bf k-q}, \pi   |D_{out}                  | {\bf k-q}, \pi^* \rangle
\langle {\bf k}    \pi   |\Delta H_{{\bf q},\nu}  | {\bf k-q}, \pi   \rangle
\langle {\bf k-q}, \pi^* |\Delta H_{{\bf -q},\mu} | {\bf k}    \pi^* \rangle
\langle {\bf k}    \pi^* |D_{in}                   | {\bf k}    \pi   \rangle
}
{
( \epsilon_L - \epsilon^{\pi^*}_{\bf k-q} + \epsilon^{\pi}_{\bf k-q} - \hbar\omega_{{\bf -q}}^\nu - \hbar\omega_{{\bf q}}^\mu
-i\frac{\gamma^C_{\bf k}}{2})
( \epsilon_L - \epsilon^{\pi^*}_{\bf k-q} + \epsilon^{\pi}_{\bf k}   - \hbar\omega_{{\bf q}}^\mu
-i\frac{\gamma^B_{\bf k}}{2})
( \epsilon_L - \epsilon^{\pi^*}_{\bf k}   + \epsilon^{\pi}_{\bf k}
-i\frac{\gamma^A_{\bf k}}{2})
}.
\]

Process $he1$: first the hole is scattered by the {\bf -q}$\nu$ phonon and then the electron by the {\bf q}$\mu$ one,
\[
K_{he1}^{pp} = -
\frac{
\langle {\bf k-q}, \pi   |D_{out}                  | {\bf k-q}, \pi^* \rangle
\langle {\bf k-q}, \pi^* |\Delta H_{{\bf -q},\mu} | {\bf k}    \pi^* \rangle
\langle {\bf k}    \pi   |\Delta H_{{\bf q},\nu}  | {\bf k-q}, \pi   \rangle
\langle {\bf k}    \pi^* |D_{in}                   | {\bf k}    \pi   \rangle
}
{
( \epsilon_L - \epsilon^{\pi^*}_{\bf k-q} + \epsilon^{\pi}_{\bf k-q} - \hbar\omega_{{\bf -q}}^\nu - \hbar\omega_{{\bf q}}^\mu
-i\frac{\gamma^C_{\bf k}}{2})
( \epsilon_L - \epsilon^{\pi^*}_{\bf k}   + \epsilon^{\pi}_{\bf k-q} - \hbar\omega_{{\bf -q}}^\nu
-i\frac{\gamma^B_{\bf k}}{2})
( \epsilon_L - \epsilon^{\pi^*}_{\bf k}   + \epsilon^{\pi}_{\bf k}
-i\frac{\gamma^A_{\bf k}}{2})
}.
\]

Process $he2$: first the hole is scattered by the {\bf q}$\mu$ phonon and then the electron by the {\bf -q}$\nu$ one,
\[
K_{he2}^{pp} = -
\frac{
\langle {\bf k+q}, \pi   |D_{out}                  | {\bf k+q}, \pi^* \rangle
\langle {\bf k+q}, \pi^* |\Delta H_{{\bf q},\nu}  | {\bf k}    \pi^* \rangle
\langle {\bf k}    \pi   |\Delta H_{{\bf -q},\mu} | {\bf k+q}, \pi   \rangle
\langle {\bf k}    \pi^* |D_{in}                   | {\bf k}    \pi   \rangle
}
{
( \epsilon_L - \epsilon^{\pi^*}_{\bf k+q} + \epsilon^{\pi}_{\bf k+q} - \hbar\omega_{{\bf -q}}^\nu - \hbar\omega_{{\bf q}}^\mu
-i\frac{\gamma^C_{\bf k}}{2})
( \epsilon_L - \epsilon^{\pi^*}_{\bf k}   + \epsilon^{\pi}_{\bf k+q} - \hbar\omega_{{\bf q}}^\mu
-i\frac{\gamma^B_{\bf k}}{2})
( \epsilon_L - \epsilon^{\pi^*}_{\bf k}   + \epsilon^{\pi}_{\bf k}
-i\frac{\gamma^A_{\bf k}}{2})
}.
\]
\end{widetext}

\section{The Tight-Binding Model}
\label{app_TB}

Here we describe the tight-binding model which is used
to calculate the electronic structure,
the electron-phonon,
the electron-light and
the electron-defect scattering matrix elements.

\subsection{Electronic structure}
\label{app_el-struct}

Let us call
$|l,s\rangle$ the orthonormalized
$p_z$ orbital of the $s$ atom (in graphene $s=1,2$),
in the position $\tau_s$,
in the cell identified by the lattice vectors {\bf R}$_l$ ($l=1,\infty$).
Let us consider the wavefunction (normalized in the unit cell)
\[
|{\bf k},s\rangle = \sum_l e^{i{\bf k}\cdot({\bf R}_l+\tau_s)} |l,s\rangle.
\]
Given the tight-binding Hamiltonian $H$,
$H_{{\bf k},s,s'} = \langle {\bf k},s|H|{\bf k},s'\rangle/N$
($N$ is the number of cells in the crystal)
is the $2\times2$ matrix:
\begin{equation}
H_{\bf k} =
\begin{pmatrix}
g({\bf k}) & f({\bf k}) \\
f^*({\bf k}) & g ({\bf k})
\end{pmatrix},
\end{equation}
where
\begin{eqnarray}
&&f(\mathbf{k})=
-t_1\sum_{i=1,3}e^{i\mathbf{k}\cdot\mathbf{C_i^1}}
-t_3\sum_{i=1,3}e^{i\mathbf{k}\cdot\mathbf{C_i^3}}
-t_4\sum_{i=1,6}e^{i\mathbf{k}\cdot\mathbf{C_i^4}} \nonumber \\
&&g(\mathbf{k})=
-t_2\sum_{i=1,6}e^{i\mathbf{k}\cdot\mathbf{C_i^2}}
-t_5\sum_{i=1,6}e^{i\mathbf{k}\cdot\mathbf{C_i^5}}=g^*({\bf k}).
\end{eqnarray}
Here, $t_i$ is the i-th neighbor hopping parameter.
$\mathbf{C_i^1}$ are the three vectors connecting the $s=1$ atom
with its three nearest neighbors ($i=1,3$).
More in general, $\mathbf{C_i^j}$ are the vectors
connecting the $s=1$ atom with the i-th atom in the j-th neighborhood.

By diagonalizing $H_{{\bf k},s,s'}$,
\[
\sum_{s'=1,2} H_{{\bf k},s,s'} a^\alpha_{{\bf k}s'} = \epsilon^{\alpha}_{{\bf k}} a^\alpha_{{\bf k}s},
\]
one obtains the eigenvalues $\epsilon^{\alpha}_{{\bf k}}$ ($\alpha=\pi$,$\pi^*$)
and the eigen wavefunctions
$|{\bf k},\alpha\rangle =\sum_{s} a^\alpha_{{\bf k}s} |{\bf k},s\rangle$:
\begin{eqnarray}
\epsilon^{\pi^*}_{\bf k} = g(\mathbf{k})+ |f(\mathbf{k})|
&~~,~~&
a^{\pi^*}_{\bf k} =\frac{1}{\sqrt{2}}
\begin{pmatrix}
1\\
\phi(\mathbf{k})
\end{pmatrix} \nonumber \\
\epsilon^{\pi}_{\bf k} = g(\mathbf{k})-|f(\mathbf{k})|
&~~,~~&
a^{\pi}_{\bf k} =\frac{1}{\sqrt{2}}
\begin{pmatrix}
1\\
-\phi(\mathbf{k})
\end{pmatrix},
\label{eqb3}
\end{eqnarray}
where $\phi(\mathbf{k})= f^*(\mathbf{k})/|f(\mathbf{k})|$.

Finally, here the overlap matrix is the identity
because of the use of orthonormal $p_z$ orbitals.
In alternative, a precise description of the bands can also be obtained
by using  pristine (non-orthonormal) $p_z$ orbital
with only  three neighbors interaction parameters
at the expense of using a non-diagonal overlap matrix
(see e.g.~\cite{reich02,grueneis08}).

\subsection{Electron-phonon scattering}
\label{app_el-phon}

Given a phonon mode {\bf q}$\nu$,
with pulsation $\omega_{{\bf q}\nu}$ and
polarization $\epsilon^{s,c}_{{\bf q},\nu}$
($s=1,2$ is an atomic index and $c=1,3$ is a Cartersian coordinate index,
$\epsilon^{s,c}_{{\bf q},\nu}$ is normalized to 1 in the unit cell,
corresponding to a displacement
$\epsilon^{s,c}_{{\bf q},\nu}e^{i{\bf q}\cdot({\bf R}_l+{\bm \tau}_s)}$
of the $s$ atom in the $l$ unit-cell),
the electron-phonon scattering matrix element is
\begin{eqnarray}
&&\langle {\bf k+q},\alpha|\Delta H_{{\bf q},\nu}|{\bf k},\beta\rangle =
\sqrt{\frac{\hbar}{2M\omega_{{\bf q},\nu}}}
\sum_{s,c}
\epsilon^{s,c}_{{\bf q},\nu} \nonumber \\
&&\times~(a^{\alpha}_{\bf k+q})^\dagger
\Delta H_{{\bf k+q},{\bf k}}^{s,c}
a^{\beta}_{\bf k},
\label{eqb4}
\end{eqnarray}
where $M$ is the carbon mass.
All the unit cells give the same contribution and the bra-ket integration is done on the unit cell
(with this choice the numerators of the scattering amplitudes are
independent from the number of cells of the crystal).
The 2$\times$2 matrix $\Delta H_{{\bf k+q},{\bf k}}^{s,c}$ is the derivative of the
TB Hamiltonian with respect to a periodic displacement (with periodicity {\bf q})
of the atom $s$ along the $c$ Cartesian coordinate.
By defining $\eta_1$ as the derivative of the nearest-neighbor hopping parameter
$t_1$ with respect to the bond length,
\begin{eqnarray}
\Delta H_{{\bf k+q},{\bf k}}^{1,c} &=& \sqrt{3} \eta_1
\begin{pmatrix}
0 & h_c({\bf k}) \\
h_c^*({\bf k+q}) & 0
\end{pmatrix}
\nonumber \\
\Delta H_{{\bf k+q},{\bf k}}^{2,c} &=& -\sqrt{3} \eta_1
\begin{pmatrix}
0 & h_c({\bf k+q}) \\
h_c^*({\bf k}) & 0
\end{pmatrix}
\nonumber \\
h_c({\bf k}) &=& \sum_{i=1,3} e^{i{\bf k}\cdot{\bf C^1_i}} C^1_{i,c}/a_0~~~~~,
\label{eqb5}
\end{eqnarray}
where $C^1_{i,c}$ is the Cartesian component along the $c$ direction of
${\bf C^1_i}$, and $a_0$ is the graphene lattice spacing.

\subsection{Electron-light scattering}
\label{app_el-light}

The electron-light interaction is calculated as
\begin{eqnarray}
\langle {\bf k} \pi^*|D_{in}  |{\bf k}\pi\rangle&=&
\frac
{e\vec{P_{in}}\cdot
(a^\pi_{\bf k})^\dagger \vec{\nabla} H({\bf k})a^{\pi^*}_{\bf k}}
{\epsilon_L} \nonumber \\
\langle {\bf k} \pi|D_{out}  |{\bf k}\pi^*\rangle&=&
\frac
{e\vec{P_{out}}\cdot
(a^\pi_{\bf k})^\dagger \vec{\nabla} H({\bf k})a^{\pi^*}_{\bf k}}
{\epsilon_L^{out}},
\end{eqnarray}

where $\vec{P_{in}}$ and $\vec{P_{out}}$ are the polarizations of the incident and
scattered radiation, $\vec{\nabla} H({\bf k})$ is the gradient of the TB Hamiltonian
and is a 2$\times$2 matrix.
$\epsilon_L$ is the incident laser energy
and $\epsilon_L^{out}$ is the scattered radiation energy
($\epsilon_L^{out}=\epsilon_L-\hbar\omega_{\bf -q}^\nu$
for a $K^{pd}({\bf q},\nu)$ process and
$\epsilon_L^{out}=\epsilon_L-\hbar\omega_{\bf -q}^\nu-\hbar\omega_{\bf q}^\mu$
for a $K^{pp}({\bf q},\nu,\mu)$ process).

\subsection{Electron-defect scattering}
\label{app_el-def}

We consider three distinct kind of defects.
The electron-defect scattering operator is defined accordingly.

i) The {\it on-site defect} changes the on-site TB parameter of the atom $\tau_1$
by $\delta V_0$, in this case we will use the notation $H_D=V_{on}$ and
\begin{equation}
\langle {\bf k}\alpha|V_{on}|{\bf k'}\alpha \rangle =
\frac{\delta V_0}{2}.
\end{equation}
$\alpha=\pi$ or $\pi^*$. Here we have considered $\tau_1$ in the origin
and here the bra-ket integration is done all over the space.
ii) The {\it hopping defect}
changes the hopping parameter of two nearest-neighbor atoms
connected by the vector $C_i^1$ by $\delta t_1$.
$H_D=V_{hopp}$ and
\begin{equation}
\langle {\bf k}\alpha|V_{hopp}|{\bf k'}\alpha \rangle =
\frac{\delta t_1}{2}
[\phi^*(\mathbf{k})e^{-i{\bf k}\cdot{\bf C_i^1}}+
\phi(\mathbf{k'})e^{i{\bf k'}\cdot{\bf C_i^1}}],
\label{eq_b8}
\end{equation}
where $\phi$ is defined as in Eq.~\ref{eqb3}.
In the calculations of the Raman scattering probability
averages among the three different $C_i^1$ vectors are taken.

iii) The {\it Coulomb defect} is a Coulomb impurity with charge $e$, placed at a distance
$h$ from the graphene sheet. In this case, $H_D=V_{Coul}$.
The Coulomb potential
in the position $\mathbf{r}$ in the graphene's plane is
\begin{equation}
V_{Coul}(\mathbf{r})
=\frac{e^2}{4\pi\epsilon_0\kappa}\frac{1}{\sqrt{r^2+h^2}}
=\frac{e^2}{4\pi\epsilon_0\kappa}\int d^2k~\frac{e^{-kh}}{k}e^{i\mathbf{k}\cdot\mathbf{r}}
\end{equation}
where
$\epsilon_0$ the vacuum permittivity,
$\kappa$ an environment dielectric constant,
and the integral is performed on all the reciprocal space.
By assuming that the $p_z$ orbitals are localized with respect to $a_0$ and $h$
(this is done to avoid the introduction of new parameters in the model),
\begin{eqnarray}
\langle {\bf k}\alpha|V_{Coul}|{\bf k'}\alpha \rangle =
\frac{e^2}{2\epsilon_0\kappa A_0}
\sum_\mathbf{G}
\frac{e^{-|\mathbf{k}-\mathbf{k'}+\mathbf{G}|h}}{|\mathbf{k}-\mathbf{k'}+\mathbf{G}|}
&&
\nonumber \\
\times
\left[
1+e^{i(\mathbf{k}-\mathbf{k'}+\mathbf{G})\cdot\mathbf{C_1^1}}\phi^*(\mathbf{k})\phi(\mathbf{k'})
\right]
&&
\label{eq_b10}
\end{eqnarray}
where the sum is done on the reciprocal lattice vectors $\mathbf{G}$
and $A_0$ is the unit-cell area.

Note that in the three cases the Raman intensity is calculated by
Eqs.~\ref{eq2} and~\ref{eq3}. As a consequence, for the cases of
on-site and hopping defects the intensity is proportional to
$\alpha_{on} = n_d \delta V_0^2$ and $\alpha_{hopp} = n_d
\delta t_1^2$, respectively, being $n_d$ the impurity concentration.
On the other hand, for the Coulomb impurities, the intensity is
proportional to $n_d$, but it also depends on the
impurity-graphene distance, $h$, as in Eq.(B10) above.

\section{Role of the phonon energies in the DR}
\label{app_phononzero}

\begin{figure}[ht]
\includegraphics[width=7.8cm]{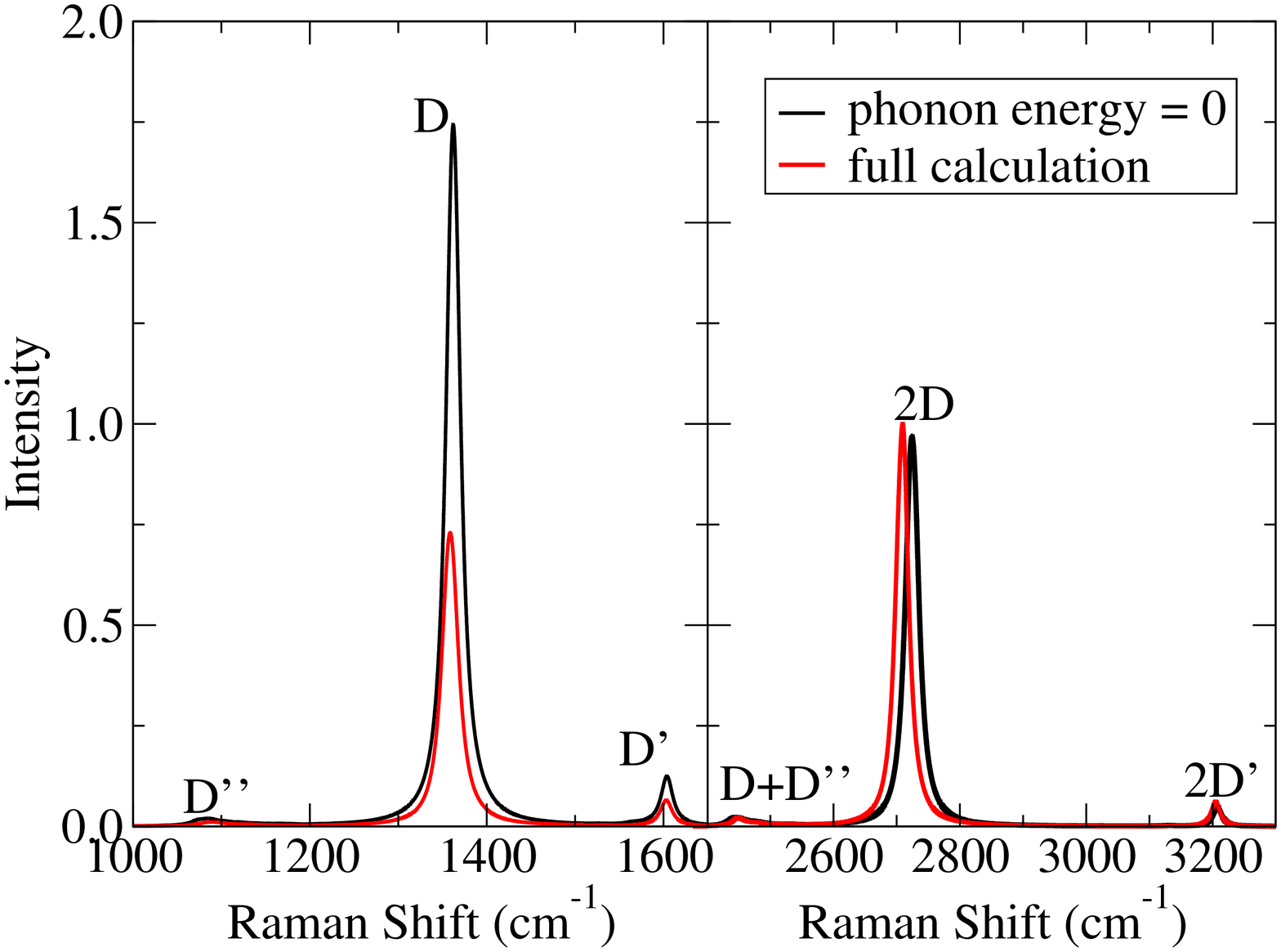}
\caption{(Color online) 
Comparison of a typical Raman spectrum (full calculation)
with a test calculation in which the phonon energies
in the denominators of the DR scattering amplitudes $K$
are considered zero.
Calculations are done using $\epsilon_L=2.4$~eV,
$\gamma^{tot}=96$~meV, and hopping defects with
$\alpha_{hopp}=6.4\times10^{13}$~eV$^2$cm$^{-2}$. All the
intensities are normalized to the $2D$ line maximum value
of the full calculation.
 }
\label{fig_phononzero}
\end{figure}

The Raman spectra depend on the phonon frequencies $\omega_{\bf q}^\nu$
through the energy conservation between the initial and the final states
(expressed in the $\delta$ functions in Eq.~\ref{eq2})
and through the denominators of the
DR scattering amplitudes $K$ (e.g. in Eqs.~\ref{eq4},~\ref{eq5}).
We performed a serie of test calculations in which we consider the
phonon energies $\omega_{\bf q}^\nu=0$ in all the denominators of the
amplitudes $K$ (e.g. $\omega_{\bf -q}^\nu=\omega_{\bf q}^\mu=0$
in Eqs.~\ref{eq4},~\ref{eq5}).
It turns out that, qualitatively, the Raman spectra are not affected.
For example, the $2D$ line intensity is basically unchanged, 
while the $D$ one remains of the same orders of magnitude
(Fig.~\ref{fig_phononzero}).
We also checked that the results of Sec.~\ref{pr} are not affected by the
actual value of $\omega_{\bf q}^\nu$ in the denominators.
Using the notation of Sec.~\ref{pr}, by letting $\omega_{\bf q}^\nu=0$
in the $K$ denominators,
${\rm I}_{ab}\gg {\rm I}_{aa}$ and
$\tilde {\rm I}_{aa}\sim\tilde {\rm I}_{ab}$
for both the $2D$ and the $D$ lines.
That is, the $ab$ processes are still, by far, the dominant ones.


\section{A Simple model}
\label{app_simple_model}

In Sec.~\ref{pr} we have shown that the largest part of the DR Raman spectrum
is due to the processes involving the scattering of both one electron and one hole
($ab$ processes). We now show that the same conclusions are reached by considering
a simple model in which the scattering matrix
elements in the numerator of Eq~\ref{eq1} are constant, the phonon
energies in the denominators (e.g. $\hbar\omega_{\bf q}^\nu$ in
Eqs.~\ref{eq4},~\ref{eq5}) are neglected
(see discussion in App.~\ref{app_phononzero}), and in which the electronic
bands are conic: $\epsilon_{\bf k}^{\pi^*/\pi}=\pm \hbar v_F |{\bf k}|$, where
$v_F$ is the Fermi velocity and {\bf k=0} corresponds to the high symmetry
{\bf K} point.

For a given excitation energy $\epsilon_L$, the scattering cross section
associated
to a phonon of momentum {\bf q} are ${\rm I}_{aa}({\bf q},\epsilon_L)$
and ${\rm I}_{ab}({\bf q}, \epsilon_L)$. As usual,
$aa$ refers to the $ee1$, $ee2$, $hh1$, and $hh2$ processes, and
$ab$ to the $eh1$, $eh2$, $he1$, and $he2$ ones.
By using the equations of App. ~\ref{app_DR}, one obtains,
\begin{widetext}
\begin{eqnarray}
{\rm I}_{aa}({\bf q}, \epsilon_L)&=&\left| \int
\frac{d^2 {\bf k}}{(2\pi)^2} K_{aa}({\bf k}, {\bf q}, \epsilon_L)
\right|^2
~~,~~~~
{\rm I}_{ab}({\bf q}, \epsilon_L)=\left| \int
\frac{d^2 {\bf k}}{(2\pi)^2} K_{ab}({\bf k}, {\bf q}, \epsilon_L)
\right|^2, \nonumber \\
K_{aa}({\bf k}, {\bf q}, \epsilon_L)&=&
\frac{1}
{\left(\epsilon_L-2\hbar v_Fk-i\frac{\gamma}{2} \right)
 \left(\epsilon_L-\hbar v_F|{\bf k+q}|-\hbar v_Fk-i\frac{\gamma}{2} \right)
 \left(\epsilon_L-2\hbar v_Fk-i\frac{\gamma}{2} \right)}, \nonumber \\
K_{ab}({\bf k}, {\bf q}, \epsilon_L)&=&
\frac{1}
{\left(\epsilon_L-2\hbar v_F|{\bf k+q}|-i\frac{\gamma}{2} \right)
 \left(\epsilon_L-\hbar v_F|{\bf k+q}|-\hbar v_Fk-i\frac{\gamma}{2} \right)
 \left(\epsilon_L-2\hbar v_Fk-i\frac{\gamma}{2} \right) }.
\label{eqd1}
\end{eqnarray}
\end{widetext}

\begin{figure}
\includegraphics[width=7.8cm]{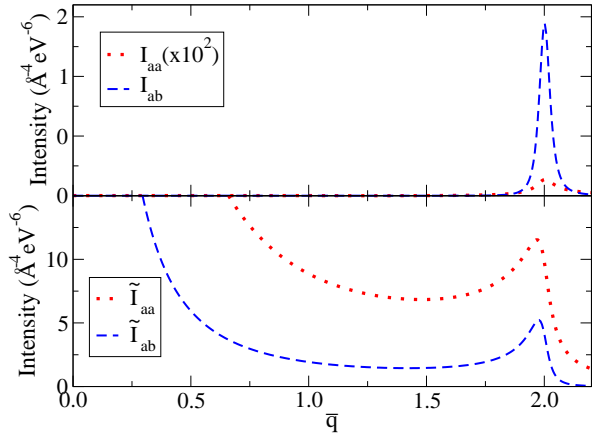}
\caption{(Color online)
Numerical solution of Eqs.~\ref{eqd1} using
$\epsilon_L=2.4$~eV, $\gamma=84$~meV, and $\hbar v_F=6.49$~eV\AA.
$\overline{\rm q}=2q\hbar v_F/\epsilon_L$ is an adimensional
momentum and ${\overline q}=2$ corresponds to the double resonance
condition.
${\rm I}_{aa}$ is magnified by 10$^2$ for clarity.
$\tilde{\rm I}_{aa}$ and $\tilde{\rm I}_{ab}$ are 
intensities in which quantum interference has been artificially
suppressed (see the text).
}
\label{fig_integral}
\end{figure}

\begin{figure}
\includegraphics[width=7.8cm]{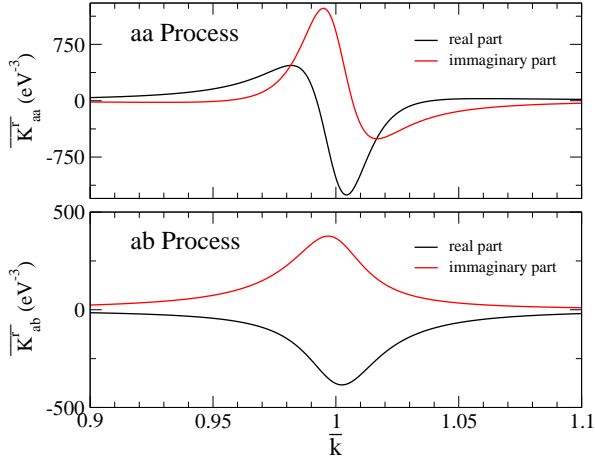}
\caption{(Color online)
DR scattering amplitudes $\overline{K^r}$
as defined in Eq.~\ref{eqd2}
for the $aa$ and $ab$ processes, as a function of the adimensional
momentum $\overline{\rm k}=2k\hbar v_F/\epsilon_L$.
The real and imaginary part of the complex number $\overline{K^r}$
are plotted as two different lines. 
Calculations are done using
$\epsilon_L=2.4$~eV, $\gamma=84$~meV, and $\hbar v_F=6.49$~eV\AA.
}
\label{fig_modelIVSk}
\end{figure}

In analogy to Sec.~\ref{pr},
$\tilde{\rm I}_{aa}$ and $\tilde{\rm I}_{ab}$
are obtained by considering only the modulus of the integrand,
e.g. $\tilde {\rm I}_{aa}= \left| \int d^2{\bf k}/(2\pi)^2|K_{aa}|\right|^2 $.
Fig.~\ref{fig_integral} reports the intensities ${I}$ thus obtained 
for a fixed value of $\epsilon_L$, as a function of $q$
(the results do not depend on the direction of {\bf q}).
As expected from the DR picture, ${\rm I}(q)$
has a maximum at $q=\epsilon_L/(\hbar v_F)$.
Even with this simplified model, one recover the result that
$ab$ processes are by far dominant: ${\rm I}_{ab}\gg{\rm I}_{aa}$
from Fig.~\ref{fig_integral}.
The importance of quantum interference effects is understood by
considering that the intensities $\tilde{\rm I}_{ab}$ and $\tilde{\rm I}_{aa}$
(in which quantum interference effects are artificially canceled,
Sec.~\ref{pr}) are very different from ${\rm I}_{ab}$ and ${\rm I}_{aa}$.
In particular, $\tilde{\rm I}_{ab}$ and $\tilde{\rm I}_{aa}$ have
the same order of magnitude.
As already noticed in ~\cite{maultzsch04prb}
the shapes of ${\rm I}(q)$  and $\tilde{\rm I}(q)$ are very different,
thus the fact that ${\rm I}(q)$
is associated to a well defined narrow line is a direct consequence
of quantum interferece.
Notice that, however, the authors of ~\cite{maultzsch04prb} consider only
the $aa$ processes.

To further explain the concept of quantum interference
we consider that
for a fixed value of $\epsilon_L$ the resonance condition
$q_r=\epsilon_L/(\hbar v_F)$ ($\overline{q}=2$ in Fig.~\ref{fig_integral}),
implies that the maximum of the intensities are
\begin{equation}
{\rm I}_{\alpha}(q_r,\epsilon_L)
= \left| \int_0^\infty \frac{kdk}{2\pi} \overline{K^r_{\alpha}}(k)
\right|^2,
\label{eqd2}
\end{equation}
where the label $\alpha=aa$ or $ab$, and
$\overline{K^r_{\alpha}}(k)$ are the $K$ scattering amplitudes
of Eqs.~\ref{eqd1} calculated at $\epsilon_L$ and $q_r$,
averaged over the angular dependence of {\bf k}.

Fig.~\ref{fig_modelIVSk} shows 
$\overline{K^r_{aa}}(k)$ and $\overline{K^r_{ab}}(k)$
for realistic values of the parameters $\epsilon_L$, $\gamma$ and $v_F$.
Both $\overline{K^r_{aa}}(k)$ and $\overline{K^r_{ab}}(k)$
have a maximum near $k=\epsilon_L/(2\hbar v_F)$
which corresponds to the DR condition ($\overline{k}=1$ in
Fig.~\ref{fig_modelIVSk}).
First we remark that, for realistic values of $\gamma$,
the real, ${\rm Re}$,
and imaginary parts, ${\rm Im}$, of the $\overline{K^r}$
amplitudes are of the same order of magnitude.
Thus, the $\overline{K^r}$ cannot be approximated as purely real
or purely imaginary numbers.
Second we notice that
${\rm Re}(\overline{K^r_{ab}})$ and
${\rm Im}(\overline{K^r_{ab}})$
do not change their sign when plotted as a function of $k$.
On the contrary,
${\rm Re}(\overline{K^r_{aa}})$ and
${\rm Im}(\overline{K^r_{aa}})$ change their sign
(Fig.~\ref{fig_modelIVSk}).
Because of this, the $\overline{K^r_{ab}}(k)$ inside the integral
of Eq.~\ref{eqd2} add coherently, while the $\overline{K^r_{aa}}(k)$
interfere in a destructive way.
As a consequence, 
${\rm I}_{ab}\gg{\rm I}_{aa}$, despite the fact that
$\overline{K^r_{ab}}$ and $\overline{K^r_{aa}}$
are of the same order of magnitude.


\begin{thebibliography}{23}

\bibitem{ferrari} A.C. Ferrari, J.C. Meyer, V. Scardaci, C. Casiraghi, M. Lazzeri,
F. Mauri, S. Piscanec, D. Jiang, K.S. Novoselov, S. Roth , and A. K. Geim,
 Phys. Rev. Let. {\bf 97}, 187401 (2006).

\bibitem{gupta06}
A. Gupta, G. Chen, P. Joshi, S. Tadigadapa, and P.C. Eklund,
Nano Lett. {\bf 6}, 2667 (2006).

\bibitem{lazzeri06PRL}
M. Lazzeri and F. Mauri,
Phys. Rev. Lett. {\bf 97}, 266407 (2006).

\bibitem{pisana07}
S. Pisana, M. Lazzeri, C. Casiraghi, K.S. Novoselov, A.K. Geim, A.C. Ferrari, and F. Mauri,
Nature Materials {\bf 6}, 198 (2007).

\bibitem{yan07}
J. Yan, Y. Zhang, P. Kim, and A. Pinczuk, Phys. Rev. Lett. {\bf 98}, 166802 (2007).

\bibitem{chen09}
J.H. Chen, W.G. Cullen, C. Jang, M.S. Fuhrer, and E.D. Williams,
Phys. Rev. Lett. {\bf 102}, 236805 (2009).

\bibitem{lucchese} M.M. Lucchese, F. Stavale, E.H. Martins Ferreira, C. Vilani,
M.V.O. Moutinho, R.B. Capaz, C.A. Achete and A. Jorio, Carbon {\bf 48}, 1592 (2010).

\bibitem{ni10}
Z.H. Ni, L.A. Ponomarenko, R.R. Nair, R. Yang, S. Anissimova, I.V. Grigorieva, F. Schedin, Z.X. Shen, E.H. Hill, K.S. Novoselov, and A.K. Geim,
Nano Lett. {\bf 10}, 3868 (2010).

\bibitem{mafra} D. L. Mafra, G. Samsonidze, L. M. Malard, D. C. Elias, J. C. Brant, F. Plentz,
E. S. Alves, and M. A. Pimenta,
 Phys. Rev. B {\bf 76}, 233407 (2007). In this work, the 2D and D+D'' bands were called as
G' and G*, respectively.

\bibitem{thomsen00} C. Thomsen and S. Reich,
Phys. Rev. Lett. {\bf 85}, 5214 (2000).

\bibitem{erlon} E. H. Martins Ferreira, M. V. O. Moutinho, F. Stavale, M. M. Lucchese, R. B. Capaz,
C. A. Achete and A. Jorio, Phys. Rev. B {\bf 82}, 125429 (2010).

\bibitem{berciaud} S. Berciaud, S. Ryu, L. E. Brus, and T. F. Heinz,
NanoLett.  {\bf 9}, 346 (2009).

\bibitem{martinfalicov}R. M. Martin and L. M. Falicov,
in {\it Light Scattering in Solids I}, edited by M. Cardona,
 Topics in Applied Physics Vol.8 (Springer, Berlin, 1983), p. 79.

\bibitem{kurti02} J. Kurti, V. Zolyomi, A. Gruneis, and H. Kuzmany,
Phys. Rev. B {\bf 65}, 165433 (2002).

\bibitem{narula} R. Narula and S. Reich,
Phys. Rev. B {\bf 78}, 165422 (2008).

\bibitem{basko} D. M. Basko,
Phys. Rev. B {\bf 78}, 125418 (2008).

\bibitem{park} J. S. Park, A. Reina, R. Saito, J. Kong, G. Dresselhaus, and
M. S. Dresselhaus, Carbon {\bf 47}, 1303 (2009).

\bibitem{daniela}
D.L. Mafra, E.A. Moujaes, S.K. Doorn, H. Htoon, R.W. Nunes and M.A. Pimenta,
Carbon, {\bf 49}, 1511 (2011).

\bibitem{mohr}M. Mohr, J. Maultzsch, and C. Thomsen,
Phys. Rev. B {\bf 82}, 201409(R) (2010).

\bibitem{huang10}
M. Huang, H. Yan, T.F. Heinz, and J. Hone, Nano Lett. {\bf 10}, 4074 (2010).

\bibitem{frank11}
O. Frank, M. Mohr, J. Maultzsch, C. Thomsen, I. Riaz, R. Jalil, K.S. Novoselov, G. Tsoukleri, J. Parthenios, K. Papagelis, L. Kavan, and C. Galiotis,
ACS Nano {\bf 5}, 2231 (2011)

\bibitem{yoon11}
D. Yoon, Y.W. Son, and H. Cheong,
Phys. Rev. Lett. {\bf 106}, 155502 (2011).

\bibitem{basko07}
D.M. Basko, Phys. Rev. B {\bf 76}, 081405(R) (2007).

\bibitem{paola}P. Gava, M. Lazzeri, A. M. Saitta and F. Mauri, Phys. Rev. B
{\bf 79}, 165431 (2009).

\bibitem{gruneis} A. Gruneis, C. Attaccalite, L. Wirtz, H. Shiozawa,
R. Saito, T. Pichler, and A. Rubio, Phys. Rev. B {\bf 78}, 205425 (2008).


\bibitem{dfpt}
S. Baroni, S. de~Gironcoli, A. Dal~Corso, and P. Giannozzi,
Rev. Mod. Phys. {\bf 73}, 515 (2001).

\bibitem{michele} M. Lazzeri, C. Attaccalite, L. Wirtz and F. Mauri,
 Phys. Rev. B {\bf 78}, 081406 (2008).

\bibitem{gruneis09}
A. Gr\"uneis, J. Serrano, A. Bosak, M. Lazzeri, S.L. Molodtsov, L. Wirtz, C. Attaccalite, M. Krisch, A. Rubio, F. Mauri, and T. Pichler,
Phys. Rev. B {\bf 80}, 085423 (2009).

\bibitem{maultzsch04prb}
J. Maultzsch, S. Reich, and C. Thomsen,
Phys. Rev. B {\bf 70}, 155403 (2004).

\bibitem{maultzsch04}
J. Maultzsch, S. Reich, C. Thomsen, H. Requardt, and P. Ordejon,
Phys. Rev. Lett. {\bf 92}, 075501 (2004).

\bibitem{mohr07}
M. Mohr, J. Maultzsch, E. Dobardzic, S. Reich, I. Milosevic, M. Damnjanovic, A. Bosak, M. Krisch, and C. Thomsen,
Phys. Rev. B {\bf 76}, 035439 (2007).

\bibitem{piscanec} S. Piscanec, M. Lazzeri, F. Mauri, A. C. Ferrari, and J. Robertson
Phys. Rev. Lett. {\bf 93}, 185503 (2004).

\bibitem{dassarma} E. H. Hwang and S. Das Sarma, Phys. Rev. B {\bf 77}, 195412 (2008).

\bibitem{basko2} D. M. Basko, S. Piscanec and A. C. Ferrari,
 Phys. Rev. B {\bf 80}, 165413 (2009).

\bibitem{calizo}I. Calizo, I. Bejenari, M. Rahman, G. Liu, and
A. A. Balandinc, J. Appl. Phys. {\bf 106}, 043509 (2009).

\bibitem{cong} 
C. Cong, T. Yu, R. Saito, G. F. Dresselhaus, and M. S. Dresselhaus,
ACS Nano {\bf 5}, 1600 (2011).

\bibitem{rao11}
R. Rao, R. Podila, R. Tsuchikawa, J. Katoch, D. Tishler, A.M. Rao, and M. Ishigami,
ACS Nano, 2011, {\bf 5}, 1594 (2011).

\bibitem{reichptrs} S. Reich and C. Thomsen, Phil. Trans. R. Soc. A {\bf 362}, 2271 (2010).

\bibitem{yoon08}
D. Yoon, H. Moon, Y.W. Son, G. Samsonidze, B.H. Park, J.B. Kim, Y.P. Lee, and H. Cheong,
Nano Lett. {\bf 8}, 4270 (2008).

\bibitem{attaccalite10}
C. Attaccalite, L. Wirtz, M. Lazzeri, F. Mauri, and A. Rubio,
Nano Letters {\bf 10}, 1172 (2010).

\bibitem{alzina}F. Alzina, H. Tao, J. Moser, Y. Garcia, A. Bachtold and
C. M. Sotomayor-Torres,  Phys. Rev. B {\bf 82}, 075422  (2010).

\bibitem{ferrari00}
A.C. Ferrari, and J. Robertson, Phys. Rev. B {\bf 61}, 14095 (2000).

\bibitem{chen08}
J. H. Chen, C. Jang, S. Adam, M. S. Fuhrer, E. D. Williams, and M. Ishigami,
Nature Physics {\bf 4}, 377 (2008).

\bibitem{basko09}
D. M. Basko, Phys. Rev. B {\bf 79}, 205428 (2009).

\bibitem{note01}
Ref.~\cite{basko09} predicts (last sentence os Sec. II B)
that for the $D$ line, the $aa$ processes should be weaker by a factor
$\hbar\omega_{ph}/\epsilon_L$, where $\omega_{ph}$ is the {\bf K} phonon
pulsation.
We verified by direct calculations
that this relation does not apply to the present results.
Indeed, by considering $\omega_{ph}=0$ in the denominators of the Raman
scattering matrix elements $K$ (e.g. in Eqs.~\ref{eq4},~\ref{eq5})
the ratio ${\rm I}_{aa}/{\rm I}_{ab}$ increases by $25\%$ instead of
decreasing to zero as predicted by~\cite{basko09}.

\bibitem{windows}  In Fig.~\ref{maps},
the mapping of the Raman intensity
in the first BZ, is done by integrating in the
following frequency windows:
[1040 cm$^{-1}$ , 1180  cm$^{-1}$] for the $D''$ line;
[1200 cm$^{-1}$ , 1520  cm$^{-1}$] for  $D$    ;
[1520 cm$^{-1}$ , 1720  cm$^{-1}$] for  $D'$   ;
[2380 cm$^{-1}$ , 2550  cm$^{-1}$] for  $D+D''$ ;
[2550 cm$^{-1}$ , 3000  cm$^{-1}$] for  $2D$   ;
[3120 cm$^{-1}$ , 3300  cm$^{-1}$] for  $2D'$.
In Fig.~\ref{fig_2Dpm}, the mapping is done by integrating
in the windows:
[2760 cm$^{-1}$ , 2793  cm$^{-1}$] for $2D^-$ ;
[2793 cm$^{-1}$ , 3060  cm$^{-1}$] for $2D^+$.

\bibitem{saito01}
R. Saito, A. Jorio, A. G. Souza Filho, G. Dresselhaus, M. S. Dresselhaus and M. A. Pimenta,
Phys. Rev. Lett. {\bf 88}, 027401 (2001)

\bibitem{cancado02}
L. G. Can\c{c}ado, M. A. Pimenta, R. Saito, A. Jorio, L. O. Ladeira, A. Grueneis, A. G. Souza-Filho, G. Dresselhaus, and M. S. Dresselhaus
Phys. Rev. {\bf B 66}, 035415 (2002)

\bibitem{reich02}
S. Reich, J. Maultzsch, C. Thomsen, and P. Ordejon
Phys. Rev. B {\bf 66}, 035412 (2002).

\bibitem{grueneis08}
A. Gr\"uneis, C. Attaccalite, L. Wirtz, H. Shiozawa, R. Saito, T. Pichler, and A. Rubio,
Phys. Rev. B {\bf 78}, 205425 (2008).

\end{thebibliography}
\end{document}